\documentclass[12pt]{article}
\usepackage{graphicx}

\font\grande=cmr9.5 scaled \magstep4
\font\medio=cmr9.5 scaled \magstep2
\outer\def\beginsection#1\par{\medbreak\bigskip
      \message{#1}\leftline{\bf#1}\nobreak\medskip
\vskip-\parskip
      \noindent}

\begin{document}
\bibliographystyle {unsrt}

\titlepage

\begin{flushright}
CERN-PH-TH/2009-117
\end{flushright}

\vspace{15mm}
\begin{center}
{\grande Dark energy, integrated Sachs-Wolfe effect}\\
\vspace{3mm}
{\grande and large-scale magnetic fields}\\
\vskip1.cm
 
Massimo Giovannini

\vskip1cm
{\sl Department of Physics, Theory Division, CERN, 1211 Geneva 23, Switzerland}
\vskip 0.2cm
{\sl INFN, Section of Milan-Bicocca, 20126 Milan, Italy}
\end{center}
\vskip 2cm
\centerline{\medio  Abstract}
The impact of large-scale magnetic fields on the interplay between the ordinary and integrated Sachs-Wolfe effects is investigated in the presence of a fluctuating dark energy component.  The modified initial conditions of  the Einstein-Boltzmann hierarchy allow for the simultaneous inclusion of dark energy perturbations and of large-scale magnetic fields. 
The temperature and polarization angular power spectra are compared with the results obtained in the magnetized version
of the (minimal) concordance model. Purported compensation effects arising at large scales  are specifically investigated.  
The fluctuating dark energy component modifies, in a computable manner, the shapes of the $1$- and $2$-$\sigma$ contours in the parameter space of the magnetized background. 
The allowed spectral indices and magnetic field intensities turn out to be
 slightly larger than those determined in the framework of the magnetized concordance model where the dark energy fluctuations are absent.  
\noindent

\vspace{5mm}

\vfill
\newpage
\renewcommand{\theequation}{1.\arabic{equation}}
\setcounter{equation}{0}
\section{Motivations and goals}
\label{sec1}
Defining $z_{{\mathrm{rec}}} = 1090.51 \pm 0.95$ as the typical redshift of recombination according to the 
WMAP data alone (see \cite{WMAP3a,WMAP3b} for the 3yr data release
and \cite{WMAP51,WMAP52,WMAP53,WMAP54,WMAP55} for the 5yr data release) 
the temperature anisotropies for multipoles $\ell < \sqrt{z_{{\mathrm{rec}}}}$ are 
practically unaffected by the thickness of the visibility function.
The resulting angular power of the temperature inhomogeneities 
is determined by the ordinary Sachs-Wolfe effect (SW in what follows), by the 
integrated Sachs-Wolfe effect (ISW effect in what follows) 
and by their mutual correlation.  The origin of the SW and of the ISW  contributions 
is physically distinct. When the opacity drops suddenly, the SW term 
is determined by the density contrast of the photons and by the 
curvature perturbations around $z_{{\mathrm{rec}}}$. Conversely the ISW effect depends upon an integral over the conformal time coordinate $\tau$ extending between $\tau_{{\mathrm{rec}}}$ (corresponding to $z_{{\mathrm{rec}}}$) and $\tau_{0}$ (i.e. the present time when 
$z=0$). The integration path determining the ISW passes through $z_{\Lambda}$, i.e. the redshift at which the dark energy contribution
is of the same order of the total density of matter and the geometry starts accelerating.  The explicit contribution of the dark energy to the SW and ISW effects (as well as to their cross-correlation) has been scrutinized, for the first time, in \cite{KS1} (see also \cite{KS2}) where the effects 
of the late dominance of a cosmological term have been taken into account in a consistent semi-analytic treatment. 

The SW contribution typically 
peaks for comoving wavenumbers $k \simeq 0.0002 \, \mathrm{Mpc}^{-1}$ while the ISW effects contributes between $k_{\mathrm{min}} =0.001\, \mathrm{Mpc}^{-1}$ and $k_{\mathrm{max}}= 0.01\, \mathrm{Mpc}^{-1}$. For comparison recall that 
the angular power spectra of the Cosmic Microwave Background (CMB in what follows) fluctuations are customarily assigned at a pivot scale $k_{\mathrm{p}} =0.002\,\,
\mathrm{Mpc}^{-1}$ which corresponds, for the best fit parameters of the WMAP 5yr alone (and in the light of the $\Lambda$CDM scenario\footnote{$\Lambda$ 
stays for the dark energy component (i.e. just a cosmological constant) and CDM stands for the cold dark matter component.}) to $\ell_{\mathrm{p}} \simeq 30$.
Even if  both contributions are reasonably separated in scales, the SW and ISW effects may partially compensate.
An effective compensation would imply a suppression 
of the lowest multipoles (and in particular of the quadrupole) in the angular power spectrum of the temperature 
autocorrelations. In the context 
of the $\Lambda$CDM scenario with purely adiabatic fluctuations the compensation is partial and  
uneffective for the suppression of the quadrupole. If non-adiabatic fluctuations in the dark-energy sector are 
consistently included, the physical situation is different and a quadrupole suppression 
cannot be excluded. This possibility has been investigated especially in the context of the (observed) low quadrupole of the temperature autocorrelations \cite{hugor, elga,karwan} (see also \cite{BDVS}  for a contending explanations of the same class of phenomena).

The relative contribution of the SW and ISW terms does depend upon the evolution of the background geometry between the recombination epoch and the present time. Furthermore it does also depend upon the potential presence of non-adiabatic fluctuations as well as upon the specific way dark energy is parametrized. In the $\Lambda$CDM paradigm   there are, by definition, no fluctuations in the dark energy sector. The rationale for such a statement stems directly from the value of the barotropic index 
of the dark energy (i.e. $w_{\mathrm{de}} = -1 $). The natural extension of the $\Lambda$CDM paradigm to the case when the 
dark energy fluctuations are dynamical is represented by the $w$CDM model where $w$ stands for the barotropic index 
of the dark energy component \footnote{To be more clear, since also other barotropic indices will intervene, we will denote 
the barotropic index of dark energy as $w_{\mathrm{de}}$. At the same time, since it is conventional to talk about the 
$w$CDM paradigm we will adhere to this convention and refrain from writing $w_{\mathrm{de}}$CDM.}.
In the $w$CDM scenario the dark energy fluctuations affect virtually all CMB observables and, more specifically, the best 
fit parameters will differ from the ones determined, for instance, within the standard $\Lambda$CDM scenario. 

A  legitimate question concerns, in this context, the role of the large-scale magnetic fields. 
Indeed the $\Lambda$CDM 
paradigm served as the simplest framework for the systematic scrutiny of magnetized CMB anisotropies
(see, e.g., \cite{maxa1,maxa2} and \cite{max1,max2,max3} for more recent developments).
The purpose of the present investigation is to extend  the program of \cite{maxa1,maxa2} by relaxing the hypotheses of \cite{max1,max2,max3} on the dark energy component.
What happens if the dominant energy density of the background is not parametrized  in terms of a cosmological constant and, simultaneously, a magnetized background 
is present? For large angular separations the interplay between the SW 
and the ISW effect answers satisfactorily to the previous question, but what 
happens for higher multipoles?  What are the changes in the parameter space 
of the magnetized background induced by a fluctuating dark energy component?
 
 While there are no doubts that large-scale magnetic fields exist today in nearly all gravitationally 
bound systems there has been 
mounting evidence from diverse observations in galaxies \cite{gal1,gal2}, clusters \cite{cl1,cl2}, 
superclusters \cite{scl1}, high-redshift 
quasars \cite{q1,q2} that magnetic fields could also have pretty large correlation scales (see \cite{maxrev} for a review on the subject).   The evolution of electromagnetic fields in a plasma 
is subject to daily test in terrestrial laboratories and in astrophysical 
observations (see, e.g. \cite{pb1,pb2} for historic monographs on the 
subject).  When plasma physics is used to interpret or even justify 
some astrophysical observations the governing equations are exactly the 
ones used in terrestrial experiments. This approximation is 
reasonable for sufficiently small redshift. However, as we move towards higher 
redshift, the evolution of the space-time curvature 
cannot be neglected anymore.
The spirit of the approach summarized in \cite{max1,max2,max3} is to translate the plasma dynamics in flat space-time to a curved background which is the one dictated, for simplicity, by the $\Lambda$CDM paradigm and by its neighboring extensions. Specific attention must be paid, in this respect, to the accurate treatment of the relativistic fluctuations of the geometry and to their initial conditions \cite{maxa1,maxa2}.

The content of the forthcoming sections can be summarized, in short, as follows.
In section \ref{sec2} the governing equations of the system will be introduced.  
In section \ref{sec3} the ordinary and integrated SW contributions will be analyzed. 
Section \ref{sec4} deals with the initial conditions for the magnetized Einstein-Boltzmann hierarchy in the presence of a fluctuating dark-energy component.  In section \ref{sec5}  the numerical aspects of the whole analysis have been collected. Section \ref{sec6} contains our concluding considerations. Various technical results have been reported in the appendix to avoid excessive digressions from the main logical line of each section.  

\renewcommand{\theequation}{2.\arabic{equation}}
\setcounter{equation}{0}
\section{Governing equations}
\label{sec2}
The governing equations of the pre-decoupling plasma can be written in terms 
of the global magnetohydrodynamical variables, i.e. the total current, the 
charge density and the baryon velocity (see, e.g. \cite{max4}).  
In most of the astrophysical applications the space-time curvature hardly 
plays any role in the evolution of the plasma. Prior to decoupling, as already emphasized 
in section \ref{sec1}, the space-time metric as well its relativistic inhomogeneities are two
necessary ingredients for gauging the effects of large-scale magnetism on the CMB observables. 
The reduction from the two-fluid  system to the one-fluid system is 
discussed in \cite{max1,max2,max3} (see, in particular,  \cite{max4}) with special attention to the 
effects of the space-time curvature.  The magnetohydrodynamical reduction  holds, 
of course under various approximations which are not that dissimilar from the ones usually 
employed in the case of flat space-time. There are, however, also notable differences which can be summarized as follows:
\begin{itemize}
\item{} since electrons and ions are non-relativistic (i.e. the plasma is cold) 
the system is not invariant under Weyl rescaling of the space-time metric and this 
implies that the corresponding evolution equations will be qualitatively 
different from the case of a non-expanding space-time;
\item{} even if the charge carriers are non-relativistic, the fluctuations 
of the geometry cannot be treated in the standard Newtonian 
approximation since the relevant modes of the gravitational 
field have wavelengths which are 
larger than the Hubble radius before matter-radiation equality \footnote{ 
The numerical value of the redshift of matter-radiation equality, denoted by $z_{\mathrm{eq}}$ is given by $z_{\mathrm{eq}} +1 = 3228.91 [h_{0}^2 \Omega_{\mathrm{M}0}/(0.134)]$.}.
\end{itemize}
With these specifications in mind, the relevant evolution equations will be written prior to decoupling when the electron-photon and electron-ion coupling is still strong. 

\subsection{Background variables}
In a spatially flat background metric of the type 
$\overline{g}_{\mu\nu}(\tau) = a^2(\tau) \eta_{\mu\nu}$ (where $\eta_{\mu\nu}$ is the Minkowski metric) the Friedmann-Lema\^itre equations take the form 
\begin{equation}
{\mathcal H}^2 = \frac{8 \pi G a^2}{3} \rho_{\mathrm{t}}, \qquad 
{\mathcal H}^2 - {\mathcal H}' = 4 \pi G a^2 (p_{\mathrm{t}} + \rho_{\mathrm{t}}),\qquad {\mathcal H} = \frac{a'}{a},
\label{FL1}
\end{equation}
where the prime denotes a derivation with respect to the conformal time coordinate $\tau$.
The explicit form of the energy density and of the enthalpy density appearing in Eq. (\ref{FL1}) 
is given by:
\begin{eqnarray}
&& \rho_{\mathrm{t}} = \rho_{\mathrm{e}} + \rho_{\mathrm{i}} +  \rho_{\mathrm{c}} +
\rho_{\gamma} + \rho_{\nu} + \rho_{\mathrm{de}},
\label{FL2}\\
&& \rho_{\mathrm{t}} + p_{\mathrm{t}} = \frac{4}{3} (\rho_{\nu} + \rho_{\gamma}) + \rho_{\mathrm{c}} +  \rho_{\mathrm{e}} + \rho_{\mathrm{i}} + 
(w_{\mathrm{de}} + 1) \rho_{\mathrm{de}}.
\label{FL3}
\end{eqnarray}
In the one-fluid limit the electron and ion energy densities 
form a single physical entity, i.e. the baryonic matter density 
$\rho_{\mathrm{b}} = \rho_{\mathrm{e}} + \rho_{\mathrm{i}}$
where $\rho_{\mathrm{e}} = m_{\mathrm{e}} n_{0}$ and $\rho_{\mathrm{i}}= 
m_{\mathrm{i}} n_{0}$ are, respectively, the electron and ion matter densities. The comoving concentrations of electrons and ions (i.e. $n_{0} =
a^3 \tilde{n}_{0}$) coincide because of the electric neutrality of the plasma
\footnote{The plasma is globally neutral $n_{\mathrm{i}} = n_{\mathrm{e}} = n_{0}$ 
and the common value of the electron and ion concentrations can be expressed as 
$n_{\mathrm{0}} = \eta_{\mathrm{b}} n_{\gamma}$ where $n_{\gamma}$ is the 
comoving concentration of photons, $\eta_{\mathrm{b}} =  
6.219 \times 10^{-10} [h_{0}^2\Omega_{\mathrm{b}}/(0.02773)]  [T_{\gamma 0}/(2.725\,\mathrm{K})]^{-3}$ and $\Omega_{\mathrm{b}}$ is, as usual, the critical fraction of baryonic matter.}. 
The remaining energy densities in Eqs. (\ref{FL2}) and (\ref{FL3})  
parametrize the contributions of neutrinos (with subscript $\nu$), 
of photons (with subscript $\gamma$), of cold dark matter particles (with subscript 
c); finally $\rho_{\mathrm{de}}$ denotes the dark energy density while $w_{\mathrm{de}}$, as already mentioned, is the barotropic index of dark energy.
\subsection{Fluctuations of the geometry}
Diverse gauges lead to slightly different (but mathematically equivalent) physical pictures of the effects 
discussed in the present investigation.
To avoid a pedantic presentation,  the relevant equations will be swiftly introduced in the 
conformally Newtonian gauge with the proviso that the very same 
equations will be expressed, when needed, in the 
synchronous frame. The recipe to move between these two gauges 
will now be given together with the appropriate definitions of the perturbed line 
elements. 
In the conformally Newtonian gauge the perturbed entries of the metric are given by:
\begin{equation}
\delta^{(\mathrm{cn})}_{\mathrm{s}} g_{00}(k, \tau) = 2 a^2(\tau) \phi(k,\tau),\qquad 
\delta^{(\mathrm{cn})}_{\mathrm{s}} g_{ij}(k, \tau) = 2 a^2(\tau) \psi(k,\tau) \delta_{ij};
\label{gauge1}
\end{equation}
while in the synchronous gauge the perturbed metric can be written as: 
\begin{equation}
\delta^{(\mathrm{S})}_{\mathrm{s}} g_{ij}(k, \tau) = a^2(\tau) \biggl[ \hat{k}_{i} \hat{k}_{j}\, h(k,\tau) 
+ 6\, \xi(k,\tau) \,\biggl(\hat{k}_{i} \hat{k}_{j} - \frac{\delta_{ij}}{3} \biggr)\biggr],
\label{gauge2}
\end{equation}
where $\hat{k}_{i} = k_{i}/|\vec{k}|$. The parametrizations of Eqs. (\ref{gauge1}) and (\ref{gauge2}) are related 
by the appropriate coordinate transformations, i.e.  
\begin{eqnarray}
\psi(k,\tau) &=& - \xi(k,\tau) + \frac{{\mathcal H}}{2 k^2} [h(k,\tau) + 6 \xi(k,\tau)]',
\nonumber\\
\phi(k,\tau) &=& - \frac{1}{2k^2} \{[h(k,\tau) + 6 \xi(k,\tau)]'' + {\mathcal H} [h(k,\tau) + 6 \xi(k,\tau)]'\}.
\label{gauge3}
\end{eqnarray}
Consider, for instance, the curvature perturbations on comoving 
orthogonal hypersurfaces, i.e. ${\mathcal R}(k,\tau)$, 
within  the description provided by the conformally Newtonian gauge:
\begin{equation}
{\mathcal R}(k,\tau) = - \psi - \frac{{\mathcal H} ({\mathcal H} \phi + \psi')}{{\mathcal H}^2 - {\mathcal H}'} \to \xi + \frac{{\mathcal H} \xi'}{{\mathcal H}^2 - {\mathcal H}'},
\label{gauge4}
\end{equation}
where the  arrow simply labels the coordinate 
transformation from the conformally Newtonian gauge to the 
synchronous coordinate system; the last expression in Eq. 
(\ref{gauge4}) is  obtained by shifting metric fluctuations 
according to Eq. (\ref{gauge3}). From the last equality in Eq. (\ref{gauge4}) 
it is also apparent that  when ${\mathcal R}' =0$, $\xi(k) = {\mathcal R}(k)$.
In more general terms, by solving Eq. (\ref{gauge4}) in terms of $\xi$,
it can be shown, after integration by part, that 
\begin{equation}
\xi(k,\tau) = {\mathcal R}(k,\tau) - \frac{{\mathcal H}(\tau)}{a(\tau)} \int_{0}^{\tau} \frac{a(\tau_{1})}{{\mathcal H}(\tau_{1})} {\mathcal R}'(k,\tau_{1}) \, 
d\tau_{1},
\label{gauge5}
\end{equation}
where $\tau_{1}$ is an integration variable and where the prime denotes, as usual, a derivation with respect to $\tau$.
Both in analytical and numerical calculations the normalization of the 
curvature perturbations is customarily expressed in terms of ${\mathcal R}$; for this 
reason Eqs. (\ref{gauge4})  and (\ref{gauge5}) turn out 
to be particularly useful in the explicit estimates.  The  same transformations of Eq. (\ref{gauge3}) can be used to 
gauge-transform the governing equations from the conformally Newtonian 
farme to the synchronous coordinate system\footnote{Not only the metric fluctuations will change under coordinate transformations but also the inhomogeneities of the sources. In particular, it can be 
easily shown that $\delta^{(\mathrm{cn})}\rho = 
\delta^{(\mathrm{S})}\rho - \rho' [(h' + 6 \xi')/(2 k^2)]$. From the latter 
relation it also follows that a fluctuation of the energy density 
does not transform if its homogeneous background value is constant in time.}.
 Equation (\ref{gauge5}) also justifies, a posteriori, the perturbed form of the line 
 element introduced in Eq. (\ref{gauge2}): the factor of $6$ in front of $\xi$ is essential 
 if we want $\xi(k,\tau)$ to coincide with ${\mathcal R}(k,\tau)$ at least in the case of the adiabatic mode when ${\mathcal R}_{*}(k)$ does not depend on time prior to equality and for wavelengths 
 larger than the Hubble radius. 
 
The present conventions as well as the whole 
approach slightly differ from the treatment of, for 
instance, Ref. \cite{bertschinger1} (see also \cite{PV1}). In the present approach 
 the curvature perturbations on comoving 
orthogonal hypersurfaces (i.e. ${\mathcal R}$) are simply related, in the 
large-scale limit, to the curvature perturbations on uniform 
density hypersurfaces (see, e.g. \cite{bardeen} and also 
\cite{hwang1,hwang2}). From the latter observations, 
most of the synchronous gauge results needed in the present analysis
 follow  by making judicious use of Eq. (\ref{gauge5}).  The need for the 
 synchronous description is also motivated since the code used for present numerical analysis is based, originally, on  
 COSMICS \cite{bertschinger2} and on CMBFAST \cite{zalda1,zalda2}. The numerical code used here is an extension 
 of what has been described and exploited in \cite{max1,max2,max3}.
 It is also worth mentioning, for completeness, that there exist 
approaches to cosmological perturbations which are fully covariant \cite{EB} and which have been 
also applied to the case of large-scale magnetic fields \cite{cov} without leading, in the latter case, to any explicit estimate or of the temperature and polarization angular power spectra neither in the 
$\Lambda$CDM paradigm nor in its neighboring extensions such as the ones analyzed in the present paper. 
 
\subsection{Evolution equations for the inhomogenities} 
The Hamiltonian and momentum constraints stemming, respectively, from the $(00)$ and $(0i)$ (perturbed) Einstein equations are given, in real space, by 
\begin{eqnarray}
&& \nabla^2 \psi - 3 {\mathcal H} ({\mathcal H} \phi + \psi') = 4\pi G a^2[ \delta_{\mathrm{s}} \rho_{\mathrm{f}} + 
\delta_{\mathrm{s}} \rho_{\mathrm{de}} + \delta_{\mathrm{s}} \rho_{\mathrm{B}} + \delta_{\mathrm{s}} \rho_{\mathrm{E}}],
\label{ham1}\\
&& \vec{\nabla} ({\mathcal H} \phi + \psi') = - 4\pi G a^2 \biggl[ (p_{\mathrm{t}} + \rho_{\mathrm{t}}) \vec{v}_{\mathrm{t}}
+ \frac{\vec{E} \times \vec{B}}{4\pi} \biggr],
\label{mom1}
\end{eqnarray}
where, $\delta_{\mathrm{s}} \rho_{\mathrm{f}} = \delta_{\mathrm{s}} \rho_{\mathrm{b}} + \delta_{\mathrm{s}} \rho_{\mathrm{c}}+\delta_{\mathrm{s}} \rho_{\nu} + \delta_{\mathrm{s}} \rho_{\gamma}$ 
is the density fluctuation of the fluid sources in the longitudinal gauge.
In Eq. (\ref{mom1}) the total velocity field  $\vec{v}_{\mathrm{t}}$ obeys \footnote{
Following exactly the same conventions established in
 Eqs. (\ref{FL2}) and (\ref{FL3}), the various subscripts denote the velocities 
of the different fluid components.}
\begin{equation}
(p_{\mathrm{t}} + \rho_{\mathrm {t}}) \vec{v}_{\mathrm{t}} = \frac{4}{3} \rho_{\nu} \vec{v}_{\nu} + 
\frac{4}{3} \rho_{\gamma} ( 1 + R_{\mathrm{b}}) \vec{v}_{\gamma\mathrm{b}} + \rho_{\mathrm{c}} \vec{v}_{\mathrm{c}} + 
( w_{\mathrm{de}} +1) \rho_{\mathrm{de}} \vec{v}_{\mathrm{de}};
\label{mom2}
\end{equation}
as already mentioned $w_{\mathrm{de}}$ is the barotropic index of 
the dark energy component i.e.
\begin{equation}
w_{\mathrm{de}} = \frac{p_{\mathrm{de}}}{\rho_{\mathrm{de}}}, \qquad c_{\mathrm{s\,de}}^2 = w_{\mathrm{de}} - 
\frac{w_{\mathrm{de}}'}{3 {\mathcal H} (w_{\mathrm{de}}+1)},
\label{mom3}
\end{equation}
where the sound speed of dark energy has been also introduced. 
In Eq. (\ref{mom2}) 
$R_{\mathrm{b}} = 3 \rho_{\mathrm{b}}/(4 \rho_{\gamma})$ denotes the  baryon to photon ratio.  Prior to photon decoupling the baryon and photon velocities 
effectively coincide with $\vec{v}_{\gamma\mathrm{b}}$
\begin{equation}
\vec{v}_{\gamma\mathrm{b}} \simeq \vec{v}_{\gamma} \simeq \vec{v}_{\mathrm{b}}, \qquad \vec{v}_{\mathrm{b}} = 
\frac{m_{\mathrm{e}} \vec{v}_{\mathrm{e}} + m_{\mathrm{i}} \vec{v}_{\mathrm{i}}}{m_{\mathrm{e}} + m_{\mathrm{i}}}.
\label{mom4}
\end{equation}
In Eqs. (\ref{ham1}) and (\ref{mom1}) the gravitational effects 
of the large-scale electromagnetic fields have been included 
in terms of the comoving electric and magnetic fields 
$\vec{E}(\vec{x},\tau) = a^2(\tau) \vec{{\mathcal E}}(\vec{x},\tau)$ and $\vec{B}(\vec{x},\tau) = a^2(\tau) 
\vec{{\mathcal B}}(\vec{x},\tau)$, i.e. 
\begin{equation}
\delta_{\mathrm{s}} \rho_{\mathrm{B}} = \frac{B^2}{8\pi a^4},\qquad 
\delta_{\mathrm{s}} \rho_{\mathrm{E}} =  \frac{E^2}{8\pi a^4},\qquad 
\delta_{\mathrm{s}} p_{\mathrm{B}} = \frac{\delta_{\mathrm{s}}\rho_{\mathrm{B}}}{3},\qquad \delta_{\mathrm{s}} p_{\mathrm{E}} = \frac{\delta_{\mathrm{s}}\rho_{\mathrm{E}}}{3},
\label{s5}
\end{equation}
where $B^2 = |\vec{B}|^2$ and $E^2 = |\vec{E}|^2$.
The evolution equations for $\vec{E}$ and $\vec{B}$ are: 
\begin{eqnarray}
&&\vec{\nabla}\cdot \vec{B} =0,\qquad  \vec{\nabla}\cdot \vec{E} = 4 \pi 
\rho_{\mathrm{q}},
\label{s1}\\
&& \vec{\nabla} \times \vec{E} + \vec{B}' =0, \qquad 
\vec{\nabla} \times \vec{B} = 4 \pi \vec{J} + \vec{E}'
\label{s2}\\
&& \rho_{\mathrm{q}}'  + \vec{\nabla}\cdot \vec{J} =0, 
\qquad \rho_{\mathrm{q}} = e (n_{\mathrm{i}} - n_{\mathrm{e}}),\qquad 
\vec{J} = e (n_{\mathrm{i}} \vec{v}_{\mathrm{i}} - n_{\mathrm{e}} \vec{v}_{\mathrm{e}}).
\label{s3}
\end{eqnarray}
The total current $\vec{J}$ has been expressed in terms of the two 
fluid variables, however, as in the case of flat-space 
magnetohydrodynamics, $\vec{J}$ can be related to the 
electromagnetic fields by means of the generalized Ohm law \cite{max4}:
\begin{equation}
\vec{J} =\sigma  \biggl(\vec{E} + \vec{v}_{\mathrm{b}} \times \vec{B} + \frac{\vec{\nabla} p_{\mathrm{e}}}{e\,n_{0}}- \frac{\vec{J}\times \vec{B}}{n_{0} e}\biggr), \qquad \sigma =  \frac{\omega_{\mathrm{pe}}^2 }{4\pi\{a \Gamma_{\mathrm{ie}} + (4/3)[\rho_{\gamma}/(n_{0}m_{\mathrm{e}})] \Gamma_{\mathrm{e}\gamma}\}},
\label{s4}
\end{equation}
where $\sigma$ is the conductivity; $\Gamma_{\mathrm{ie}}$ and $\Gamma_{\gamma\mathrm{e}}$ are, respectively, 
the  electron-ion and electron-photon interaction rates. The three terms appearing in the 
Ohm's law are, besides the electric field, the drift term (i.e. 
$\vec{v}_{\mathrm{b}} \times \vec{B}$), the thermoelectric term (containing 
the gradient of the electron pressure\footnote{In Eq. (\ref{s4}), following \cite{max4}, what appears is the comoving electron pressure 
given by $p_{\mathrm{e}} = n_{\mathrm{e}} T_{\mathrm{e}}$ where $n_{\mathrm{e}}$ and $T_{\mathrm{e}}$ are, respectively, the comoving concentration and the comoving temperature of the electrons.}) and the Hall term (i.e. 
$\vec{J}\times \vec{B}$). It can be shown \cite{max4} that for frequencies 
much smaller than the (electron) plasma frequency and for typical length-scales much larger than the  Debye screening length the Hall and thermoelectric terms are subleading for the purposes of the present analysis. 
The electromagnetic pressure as well as the anisotropic stresses enter the perturbed $(ij)$ components of the Einstein equations:
\begin{eqnarray}
&& \psi'' + {\mathcal H} (\phi' + 2\psi') + ({\mathcal H}^2 + 2 {\mathcal H}') \phi 
\nonumber\\
&&+ \frac{1}{3} \nabla^2 ( \phi - \psi) = 
4\pi G a^2[ \delta_{\mathrm{s}} p_{\mathrm{f}} + \delta_{\mathrm{s}} p_{\mathrm{de}} + \delta_{\mathrm{s}} p_{\mathrm{B}}
+ \delta p_{\mathrm{E}}], 
\label{ij1}\\
&& \nabla^4(\phi - \psi) = 16 \pi G a^2 \biggl[ \rho_{\nu} \nabla^2 \sigma_{\nu} + \rho_{\gamma}\nabla^2 \sigma_{\mathrm{B}}
+ \rho_{\gamma} \nabla^2 \sigma_{\mathrm{E}} + \frac{3}{4} \rho_{\mathrm{de}} (w_{\mathrm{de}} +1) \nabla^2 \sigma_{\mathrm{de}}\biggr],
\label{ij2}
\end{eqnarray}
where $\delta_{\mathrm{s}} p_{\mathrm{f}} = (\delta_{\mathrm{s}} \rho_{\gamma} + \delta_{\mathrm{s}} \rho_{\nu})/3$,
in analogy with $\delta_{\mathrm{s}} \rho_{\mathrm{f}}$, denotes the fluctuation of the pressure of the fluid components.
The total anisotropic stress is given, as usual,  by 
\begin{equation}
\Pi_{i\,\mathrm{t}}^{j} = \Pi_{i\,\nu}^{j}  +  \Pi_{i\,\mathrm{B}}^{j}  + \Pi_{i\,\mathrm{E}}^{j} + \Pi_{i\,\mathrm{de}}^{j};
\label{anis1}
\end{equation}
the various subscripts denote the respective components and, in 
particular the electromagnetic contribution:
\begin{equation}
  \Pi_{i\,\mathrm{B}}^{j} = \frac{1}{4 \pi a^4} \biggl[ B_{i} B^{j} - \frac{\delta_{i}^{j}}{3} B^2 \biggr],\qquad \Pi_{i\,\mathrm{E}}^{j} = \frac{1}{4 \pi a^4} \biggl[ E_{i} E^{j} - \frac{\delta_{i}^{j}}{3} E^2 \biggr].
\label{anisBE}
\end{equation}
In Eq. (\ref{ij2}) the  notation
\begin{equation}
\partial_{i}\partial_{j} \Pi^{ij}_{\mathrm{t}} = \frac{4}{3} \rho_{\nu} \nabla^2 \sigma_{\nu} + \frac{4}{3} \rho_{\gamma} 
\nabla^2 \sigma_{\mathrm{B}} + \frac{4}{3} \rho_{\gamma} \nabla^2\sigma_{\mathrm{E}} +  \rho_{\mathrm{de}} (w_{\mathrm{de}} +1) \nabla^2 \sigma_{\mathrm{de}},
\label{anis2}
\end{equation}
has been adopted.
The various species of the plasma either interact strongly 
with the plasma (like the elctrons, the ions and the photons) 
or they only feel the effects of the geometry (like the CDM 
component and the dark-energy). For large-scales 
the baryon-photon system obeys 
\begin{eqnarray}
&& \delta_{\mathrm{b}}' = 3 \psi' - \theta_{\mathrm{b}},
\label{BP1}\\
&& \theta_{\mathrm{b}}' + {\mathcal H} \theta_{\mathrm{b}}' = \frac{\vec{\nabla} \cdot [ \vec{J} \times \vec{B}]}{a^4 \rho_{\mathrm{b}}} - \nabla^2 \phi + \frac{4}{3} \frac{\rho_{\gamma}}{\rho_{\mathrm{b}}} \epsilon' (\theta_{\gamma} - \theta_{\mathrm{b}}),
\label{BP2}\\
&&  \theta_{\gamma}' = - \frac{1}{4} \nabla^2 \delta_{\gamma} - \nabla^2 \phi + 
\epsilon' (\theta_{\mathrm{b}} - \theta_{\gamma}),
\label{BP3}\\
&& \delta_{\gamma}' = 4 \psi' - \frac{4}{3} \theta_{\gamma},
\label{BP4}
\end{eqnarray}
where $\epsilon'$ is the differential optical depth; furthermore, defining as 
$\vec{v}_{X}$ the velocity of the species $X$, $\theta_{X} = \vec{\nabla} \cdot \vec{v}_{X}$.
The CDM and the neutrino components will evolve, respectively, as 
\begin{eqnarray}
&&\delta_{\mathrm{c}}' = 3 \psi' - \theta_{\mathrm{c}},
\label{C1}\\
&& \theta_{\mathrm{c}}' + {\mathcal H} \theta_{\mathrm{c}} = - \nabla^2 \phi,
\label{C2}
\end{eqnarray}
and as 
\begin{eqnarray}
&&\theta_{\nu}' = - \frac{1}{4} \nabla^2 \delta_{\nu} + \nabla^2 \sigma_{\nu} - \nabla^2 \phi,
\label{N1}\\
&& \delta_{\nu}' = 4 \psi' - \frac{4}{3} \theta_{\nu}, 
\label{N2}\\
&& \sigma_{\nu}' = \frac{4}{15} \theta_{\nu} - \frac{3}{10} {\mathcal F}_{\nu 3},
\label{N3}
\end{eqnarray}
where ${\mathcal F}_{\nu 3}$ reminds of the coupling of the monopole and of the dipole 
to the higher multipoles of the neutrino phase space distribution. The latter term will 
be set to zero
in the class of initial conditions discussed in the present paper but it can be relevant 
when magnetized non-adiabatic modes are consistently included.
The fluctuations of the dark energy density obey, in the $w$CDM scenario, the following 
equation:
\begin{equation}
\delta_{\mathrm{s}} \rho_{\mathrm{de}}' + 3 {\mathcal H} ( \delta_{\mathrm{s}} \rho_{\mathrm{de}} + 
\delta p_{\mathrm{de}}) - 3 \psi' (p_{\mathrm{de}} + \rho_{\mathrm{de}}) 
+ (\rho_{\mathrm{de}} + p_{\mathrm{de}
}) \vec{\nabla} \cdot \vec{v}_{\mathrm{de}} =0,
\label{DE1}
\end{equation}
where, as already mentioned, $\vec{v}_{\mathrm{de}}$ is the velocity of the dark energy 
in the conformally Newtonian frame whose evolution equation can be written as 
\begin{eqnarray}
&& (p_{\mathrm{de}} + \rho_{\mathrm{de}}) \vec{v}_{\mathrm{de}}' + 
[ p_{\mathrm{de}}' + \rho_{\mathrm{de}}' + 4 {\mathcal H} (p_{\mathrm{de}} + \rho_{\mathrm{de}}) ]\vec{v}_{\mathrm{de}}
+ \vec{\nabla} \delta p_{\mathrm{de}} 
\nonumber\\
&&+ (p_{\mathrm{de}} + \rho_{\mathrm{de}}) \vec{\nabla} \phi - 
(p_{\mathrm{de}} + \rho_{\mathrm{de}}) \vec{\nabla} \sigma_{\mathrm{de}}=0.
\label{DE2}
\end{eqnarray}
In Eq. (\ref{DE2}) $\sigma_{\mathrm{de}}$ accounts for the possible presence 
of an anisotropic stress for the dark energy contribution. This contribution as argued in \cite{stress1} (see also, in a related 
perspective, \cite{stress2}) can be rather 
interesting in connection with dark energy models. 
The contribution of the dark energy anisotropic stress has been included, so far, 
to have sufficiently general equations.  In the standard version of the  {\it w}CDM scenario
 the dark energy has no anisotropic stress and therefore, for practical reasons, the anisotropic stress 
 will be set to zero. It might be interesting, in the future, to relax this assumption by following, for instance, the
 approach of \cite{stress1}.
By introducing the density contrast for the dark energy component, Eqs. (\ref{DE1}) and (\ref{DE2}) 
can also be written as 
\begin{eqnarray}
&& \delta_{\mathrm{de}}' + 3 {\mathcal H}( c_{\mathrm{s\, de}}^2 - w_{\mathrm{de}}) \delta_{\mathrm{de}} + 
(w_{\mathrm{de}} + 1) \theta_{\mathrm{de}} - 3 (w_{\mathrm{de}} + 1 ) \psi'=0,
\label{DE3}\\
&& \theta_{\mathrm{de}}' + {\mathcal H} ( 1 - 3 c_{\mathrm{s\,de}}^2) \theta_{\mathrm{de}} + 
\frac{c_{\mathrm{s\,de}}^2\nabla^2 \delta_{\mathrm{de}}}{(w_{\mathrm{de}} +1) }  + \nabla^2 \phi - \nabla^2 \sigma_{\mathrm{de}}=0,
\label{DE4}
\end{eqnarray}
where $\theta_{\mathrm{de}} = \vec{\nabla}\cdot \vec{v}_{\mathrm{de}}$ and $\delta_{\mathrm{s}} \rho_{\mathrm{de}} = \rho_{\mathrm{de}}\delta_{\mathrm{de}}$.
The governing equations 
can be easily translated in the synchronous coordinate system
 (see, in particular, Eqs. (\ref{gauge1}), (\ref{gauge2}) and (\ref{gauge3})--(\ref{gauge5})). For instance the synchronous 
form of Eq. (\ref{ham1}) can be obtained by using, directly, Eqs. 
(\ref{gauge1}) and (\ref{gauge2}) and by appreciating that  
\begin{equation}
\delta^{(\mathrm{cn})}_{\mathrm{s}} \rho_{\mathrm{f}} =
\delta^{(\mathrm{S})}_{\mathrm{s}} \rho_{\mathrm{f}} - \rho_{\mathrm{f}}' 
\frac{(h'+ 6 \xi')}{2 k^2}, \qquad \delta^{(\mathrm{cn})}_{\mathrm{s}} \rho_{\mathrm{de}} =
\delta^{(\mathrm{S})}_{\mathrm{s}} \rho_{\mathrm{de}} - \rho_{\mathrm{de}}' 
\frac{(h'+ 6 \xi')}{2 k^2}.
\label{gauge6} 
\end{equation}
Since $\rho_{\mathrm{f}} = (\rho_{\gamma} + \rho_{\nu} + \rho_{\mathrm{c}} + \rho_{\mathrm{b}})$ and $\rho_{\mathrm{de}}$ are 
separately conserved (i.e. $\rho_{\mathrm{f}}' = -3 {\mathcal H}(\rho_{\mathrm{f}} + p_{\mathrm{f}})$ and $\rho_{\mathrm{de}}' = -3 {\mathcal H}(\rho_{\mathrm{de}} + p_{\mathrm{de}})$) 
the result will be, as expected: 
\begin{equation}
2 k^2 \xi - {\mathcal H} h' = 8\pi G a^2 [ \delta^{(\mathrm{S})}_{\mathrm{s}} \rho_{\mathrm{f}} + \delta^{(\mathrm{S})}_{\mathrm{s}} \rho_{\mathrm{de}} + \delta^{(\mathrm{S})}_{\mathrm{s}} \rho_{\mathrm{B}} + \delta^{(\mathrm{S})}_{\mathrm{s}} \rho_{\mathrm{E}}].
\label{synch1}
\end{equation}
In the same way all the other governing equations reported in this sections can be 
translated in the synchronous frame. 

\subsection{Large-scale evolution}
The Hamiltonian constraint of Eq. (\ref{ham1}) can be expressed as 
\begin{equation}
\zeta = {\mathcal R} +
 \frac{\nabla^2 \psi}{12 \pi Ga^2 (p_{\mathrm{t}} + \rho_{\mathrm{t}})},
\label{LSS1}
\end{equation}
where ${\mathcal R}$ has been introduced in Eq. (\ref{gauge4}) 
and $\zeta$ \cite{max3} (see also \cite{hwang1,hwang2})
is a further gauge-invariant variable which measures 
 the density contrast on uniform curvature hypersurfaces. 
The constraint (\ref{LSS1}) can also be used  
as an explicit definition of $\zeta$ whose evolution depends also upon the 
gravitating effects of the electromagnetic fields and of the Ohmic current \cite{max4}
\begin{eqnarray}
\frac{\partial \zeta}{\partial\tau} &=& - \frac{{\mathcal H}\, \delta p_{\mathrm{nad}}}{p_{\mathrm{t}} + \rho_{\mathrm{t}}}
+ \frac{{\mathcal H} ( 3 \, c_{\mathrm{st}}^2 -1)}{3 (p_{\mathrm{t}} + \rho_{\mathrm{t}})} \delta_{\mathrm{s}} \rho_{\mathrm{B}}
\nonumber\\
&+& \frac{ \vec{E}\cdot\vec{J}}{3 a^4 (p_{\mathrm{t}} + \rho_{\mathrm{t}})} 
+ \frac{\delta_{\mathrm{s}}\rho_{\mathrm{E}}' + 3 (c_{\mathrm{st}}^2 +1) {\mathcal H} \delta_{\mathrm{s}} \rho_{\mathrm{E}}}{3 (p_{\mathrm{t}} + \rho_{\mathrm{t}})} - \frac{\theta_{\mathrm{t}}}{3},
\label{LSS2}\\
\frac{\partial\vec{J}}{\partial \tau} &=&  -  \biggl({\mathcal H} + 
a\Gamma_{\mathrm{ie}} -  \frac{4 \rho_{\gamma}\Gamma_{\mathrm{e}\gamma}}{ 3 n_{0} \,m_{\mathrm{e}}}\biggr) \vec{J} 
\nonumber\\
&+& \frac{\omega_{\mathrm{pe}}^2 }{4\pi} \biggl(\vec{E} + \vec{v}_{\mathrm{b}} \times \vec{B} + \frac{\vec{\nabla} p_{\mathrm{e}}}{e\,n_{0}}- \frac{\vec{J}\times 
\vec{B}}{e n_{0}}\biggr) + \frac{4 e \rho_{\gamma} \Gamma_{\mathrm{e}\gamma}}{3 m_{\mathrm{e}}} (\vec{v}_{\mathrm{b}} - \vec{v}_{\gamma}).
\label{LSS3}
\end{eqnarray}
In Eq. (\ref{LSS2}) the contribution of non-adiabatic 
pressure fluctuations has been included for sake of completeness even if $\delta p_{\mathrm{nad}}$ 
will not play a specific role. One of the possible extensions of the present work 
could indeed be the inclusion of non-adiabatic modes in the 
dark-energy sector (as argued, for instance, in \cite{hugor} in connection 
with the problem of the quadrupole suppression).
In connection with Eqs. (\ref{LSS1}), (\ref{LSS2}) and (\ref{LSS3}) few remarks are in order:
\begin{itemize}
\item{} since both $\zeta$ and ${\mathcal R}$ are invariant under 
gauge transformations, they do coincide, in any frame, in the limit
$k \ll {\mathcal H}$;
\item{} the right hand side of Eq. (\ref{LSS2})  contains three 
qualitatively different contributions: the purely electric contribution (proportional to 
$\delta_{\mathrm{s}} \rho_{\mathrm{E}}$ and its derivative), the purely magnetic contribution (proportional to $\delta_{\mathrm{s}} \rho_{\mathrm{B}}$) and the Ohmic contribution (containing explicitly the Ohmic current $\vec{J}$);
\item{} the factor $\theta_{\mathrm{t}}$ at the right hand side of Eq. (\ref{LSS2}) 
is of order $k^2/{\mathcal H}$ and it is subleading  for typical length-scales larger than the Hubble radius;
\item{} Eq. (\ref{LSS4}) reduces, at large scales, to 
Eq. (\ref{s4}): this statement can be directly verified by taking into account 
that the rate of electron-ion interactions (i.e. $\Gamma_{\mathrm{ie}}$) is always much larger than the rate of electron-photon interaction (i.e. $\Gamma_{\gamma\mathrm{e}}$); prior to decoupling both $\Gamma_{\mathrm{ie}}$ and $\Gamma_{\gamma\mathrm{e}}$ are larger than ${\mathcal H}$.
\end{itemize}
The physical hierarchies between the electric,  magnetic and Ohmic contributions follow from the typical 
length-scales of the problem and also from the frequency range dictated by the magnetohydrodynamical approximation 
(see \cite{max4} and, in particular, the left plot of Fig.1):
\begin{equation}
\frac{\delta_{\mathrm{s}} \rho_{\mathrm{E}}'}{(p_{\mathrm{t}} + \rho_{\mathrm{t}})} < {\mathcal H} \delta_{\mathrm{s}} \rho_{\mathrm{E}} \ll \frac{\vec{E} \cdot \vec{J}}{a^4 (p_{\mathrm{t}} + \rho_{\mathrm{t}})}
\ll \frac{ {\mathcal H} \delta_{\mathrm{s}} \rho_{\mathrm{B}}}{(p_{\mathrm{t}} + \rho_{\mathrm{t}})}.
\label{LSS4}
\end{equation}
The magnetic contribution appearing at the right hand side of Eq. (\ref{LSS2}) can also be written as
\begin{equation}
\frac{{\mathcal H} (3 c_{\mathrm{st}}^2 -1) }{3 (1 + w_{\mathrm{t}})}  \frac{\rho_{\mathrm{R}}}{\rho_{\mathrm{t}}} R_{\gamma} \Omega_{\mathrm{B}},\qquad R_{\gamma} + R_{\nu} =1, 
\label{LSS5}
\end{equation}
where $R_{\nu} = 0.405$ and $R_{\gamma} = 0.595$ are, respectively, the photon and neutrino fractions 
present in the radiation plasma and 
\begin{equation}
\Omega_{\mathrm{B}}(\vec{x},\tau) = \frac{\delta_{\mathrm{s}} \rho_{\mathrm{B}}(\vec{x}, \tau)}{\rho_{\gamma}(\tau)}.
\label{LSS6}
\end{equation}
Recalling the parametrization of the anisotropic stress $\sigma_{\mathrm{B}}$, the divergence 
of the magnetohydrodynamical Lorentz force, i.e. $\vec{\nabla} \cdot [\vec{J} \times \vec{B}]$ 
can be expressed as a combination of $\nabla^2 \sigma_{\mathrm{B}}$ and $\nabla^2 \Omega_{\mathrm{B}}$. 
These vector identities are important in various analytic estimates (see \cite{max1,max2,max3} and discussions 
therein). In the case where the dark energy is simply given by the cosmological 
constant the total barotropic index and the total sound speed can be written as:
\begin{equation}
w_{\mathrm{t}} =  \frac{p_{\mathrm{t}}}{\rho_{\mathrm{t}}} = \frac{1 - 3 \frac{\alpha^4}{\alpha_{\Lambda}^{3}}}{3(
 1 + \alpha  + \frac{\alpha^4}{\alpha_{\Lambda}^3})},
\qquad c_{\mathrm{st}}^2 = \frac{p_{\mathrm{t}}'}{\rho_{\mathrm{t}}'} = w_{\mathrm{t}}  - \frac{\alpha}{3(1 + w_{\mathrm{t}})} \frac{\partial w_{\mathrm{t}}}{\partial \alpha} = 
\frac{4 }{3( 3\alpha + 4)},
\label{LONG7}
\end{equation}
where $\alpha = a/a_{\mathrm{eq}}$ and 
\begin{equation}
\alpha_{\Lambda} = \frac{a_{\Lambda}}{a_{\mathrm{eq}}} = 2246.81 \biggl(\frac{h_{0}^2 \Omega_{\mathrm{M}0}}{0.1326}
\biggr)^{4/3} \biggl(\frac{h_{0}^2 \Omega_{\Lambda}}{0.3835}\biggr),
\qquad \alpha_{0} = 3195.18\, \biggl(\frac{h_{0}^2 \Omega_{\mathrm{M}0}}{0.1326}\biggr).
\label{LONG8}
\end{equation}
Using Eqs. (\ref{LONG7}) into Eq. (\ref{LSS5}) we also have, in Fourier space,
\begin{equation}
\frac{{\mathcal H} (3 c_{\mathrm{st}}^2 -1) }{3 (1 + w_{\mathrm{t}})}  \frac{\rho_{\mathrm{R}}}{\rho_{\mathrm{t}}} R_{\gamma} \Omega_{\mathrm{B}}(k) = - \frac{3 \alpha}{(3 \alpha + 4)^2} R_{\gamma} \Omega_{\mathrm{B}}(k).
\label{LONG9}
\end{equation}

\renewcommand{\theequation}{3.\arabic{equation}}
\setcounter{equation}{0}
\section{Large-scale compensations}
\label{sec3}
The evolution of the brightness perturbations can be written as \footnote{As usual, $\mu = \hat{k}\cdot \hat{n}$ denotes the projection of the Fourier mode on the direction of propagation of the CMB photon.}:
\begin{equation}
 \Delta_{\mathrm{I}}' + (i k \mu + \epsilon') \Delta_{\mathrm{I}} = \psi' - i k \mu \phi
 +  \epsilon' \biggl[\Delta_{\mathrm{I}0} + \mu v_{\mathrm{b}} + \frac{(1 - 3 \mu^2)}{4}S_{\mathrm{P}}(k,\tau)\biggl],
\label{HTA}
\end{equation}
where $S_{\mathrm{P}}$ can be expressed as the sum of the 
quadrupole of the intensity, of the monopole of the polarization and 
of the quadrupole of the  polarization, i.e.,   respectively, $S_{\mathrm{P}}(k,\tau) = 
(\Delta_{\mathrm{I}2} + \Delta_{\mathrm{P}0} + \Delta_{\mathrm{P}2})$. The multipole expansion of the 
brightness perturbation reads, within the present conventions, 
\begin{equation}
\Delta_{\mathrm{I}}(k,\mu,\tau_{0}) = \sum_{\ell} (-i)^{\ell} ( 2 \ell +1 ) \Delta_{\mathrm{I}\ell}(k,\tau_{0}) P_{\ell}(\mu),
\label{INT2}
\end{equation}
where $P_{\ell}(\mu)$ are the standard Legendre polynomials. The line of sight solution of Eq. (\ref{HTA}) can be written
\begin{eqnarray}
\Delta_{\mathrm{I}}(k, \mu, \tau_{0}) &=& \int_{0}^{\tau_{0}} {\mathcal K}(\tau) \biggl[ \Delta_{\mathrm{I}0} + \phi + \mu v_{\mathrm{b}}  + \frac{(1 - 3 \mu^2)}{4} S_{\mathrm{P}}\biggr] e^{- i \mu x(\tau)}
\nonumber\\
&+& \int_{0}^{\tau_{0}} d\tau e^{- \epsilon(\tau,\tau_{0})} ( \phi' + \psi') e^{-i \mu x(\tau)} d\tau,\qquad {\mathcal K}(\tau) = \epsilon' 
e^{- \epsilon(\tau,\tau_{0})},
\label{LONG1}
\end{eqnarray}
where  the term $-i k \mu\phi$ has been integrated by parts\footnote{Of course one might also wish to continue with the integrations by parts and integrate all the 
$\mu$-dependent terms. Such a step will however produce various time 
derivatives of the ${\mathcal K}(\tau)$ which are difficult to evaluate 
explicitly when the visibility function is infinitely thin. See also, in this respect, the discussion 
of appendix \ref{APPA}.} and where ${\mathcal K}(\tau)$ denotes the visibility function whose 
explicit form can be approximated by a Gaussian profile \cite{zeld1,wyse,pav1,pav2} 
with different methods. At large scales the Gaussian can be considered effectively 
a Dirac delta function approximately centered at recombination.

For typical multipoles $\ell \leq \sqrt{z_{\mathrm{rec}}}$ the finite width 
of the visibility function is immaterial. This means that for sufficiently small $\ell$ 
everything goes as if the opacity suddenly drops at recombination. This 
implies that the visibility function presents a sharp (i.e. infinitely thin peak at the recombination time).  Thus, since ${\mathcal K}(\tau)$ is proportional to a Dirac delta 
function and $e^{- \epsilon(\tau,\tau_{0})}$ is proportional to an Heaviside theta function.  Under the latter approximations, 
 Eq. (\ref{LONG1}) leads to the wanted separation between SW and ISW contributions:
\begin{eqnarray}
&& \Delta_{\mathrm{I}}(k,\mu,\tau_{0}) = \Delta_{\mathrm{I}}^{(\mathrm{SW})}(k,\mu,\tau_{0})  + \Delta_{\mathrm{I}}^{(\mathrm{ISW})}(k,\mu,\tau_{0}), 
\label{HT14a}\\
&& \Delta_{\mathrm{I}}^{(\mathrm{SW})}(k,\mu,\tau_{0}) = \biggl[ \frac{\delta_{\gamma}}{4} + \phi \biggr]_{\tau_{\mathrm{rec}}} e^{- i \mu y_{\mathrm{rec}}},
\label{HT14}\\
&& \Delta_{\mathrm{I}}^{(\mathrm{ISW})}(k,\mu,\tau_{0}) = \int_{\tau_{\mathrm{rec}}}^{\tau_{0}} 
(\phi' +\psi') e^{- i \mu x(\tau)} \, d\tau.
\label{HT15}
\end{eqnarray}
where it has been used that the $\delta_{\gamma}$ obeying, for instance, Eq. (\ref{BP4}), is also related to $\Delta_{\mathrm{I}0}$ as 
$\delta_{\gamma}(k,\tau) = 4\Delta_{\mathrm{I}0}(k,\tau)$ . 
Equations (\ref{HT14}) and (\ref{HT15}) can be evaluated within three complementary approximation schemes:
\begin{itemize}
\item{[{\it a}]} in the first approximation we can assume that $a_{\mathrm{rec}}/a_{\mathrm{eq}} \gg 1 $ and that, 
simultaneously, the phase appearing in Eq. (\ref{HT15}) is $\tau$-independent (i.e. 
$i\mu x(\tau) \simeq i k \mu (\tau_{0} - \tau_{\mathrm{rec}})$);
\item{[{\it b}]} in a more accurate perspective the assumption that 
$a_{\mathrm{rec}}/a_{\mathrm{eq}} \gg 1 $ can be dropped; after all 
\begin{equation}
\alpha_{\mathrm{rec}} = \frac{a_{\mathrm{rec}}}{a_{\mathrm{eq}}} = \frac{z_{\mathrm{eq}} + 1}{z_{\mathrm{rec}}+1} = 3.04 \biggl( \frac{h_{0}^2 \Omega_{\mathrm{M}0}}{0.134}\biggr);
\label{APP2}
\end{equation}
since $\alpha_{\mathrm{rec}} \simeq  3$ is not extremely larger then $1$, significant corrections can be expected;
\item{[{\it c}]} finally, Eq. (\ref{APP2}) can be taken into account in conjunction with  
the $\tau$ dependence of the phase in the integrand of Eq. (\ref{HT15}).
\end{itemize}
In the absence of large-scale magnetic fields, it is sometimes practical to estimate the 
large-scale temperature autocorrelations by only keeping the ordinary 
SW contribution evaluated in the limit 
$\alpha_{\mathrm{rec}} \gg 1$. The same approximation (with the same 
level of accuracy) can also be used when large-scale magnetic fields are included in the analysis. 
It would be incorrect to use different levels of accuracy for the curvature contribution and for the 
magnetized contribution.
\subsection{Simplified analysis  of the compensation}
In the case labeled by $[{\it a}]$ in the above list of items $|a_{\mathrm{rec}}/a_{\mathrm{eq}}| \gg1$
and the ordinary SW contribution turns out to be\footnote{The time labeled by $\tau_{*}$ 
is such that $\tau \ll \tau_{\mathrm{eq}}$ and, simultaneously, $k/{\mathcal H}_{*} \simeq k \tau_{*} \ll 1$.}:
\begin{equation}
 \biggl[ \frac{\delta_{\gamma}(k,\tau)}{4} + \psi(k,\tau) \biggr]_{\tau_{\mathrm{rec}}} = 
 2 \psi(k,\tau_{\mathrm{rec}})  - \frac{3}{2} \psi(k,\tau_{*}) \equiv  2 \psi(k,\tau_{\mathrm{rec}}) + {\mathcal R}_{*}(k)  - \frac{R_{\gamma}}{4} \Omega_{\mathrm{B}}(k),
 \label{LONG2}
 \end{equation}
 where it has been used, according to Eq. (\ref{BP4}), that in the large-scale limit
\begin{equation}
\delta_{\gamma}(k,\tau_{\mathrm{rec}}) = 4 \psi(k,\tau_{\mathrm{rec}}) + \delta_{\gamma}(k, \tau_{*}) - 4 \psi(k,\tau_{*}) + {\mathcal O}(k^2 \tau^2).
\label{LONG2a}
\end{equation}
Equation (\ref{LONG2}) then follows from Eq. (\ref{LONG2a}) by 
recalling that the initial conditions at $\tau_{*}$ are given, in the case 
of the magnetized adiabatic mode, as:
\begin{equation}
\delta_{\gamma}(k,\tau_{*}) = - 2 \phi_{*}(k) - R_{\gamma} \Omega_{\mathrm{B}}(k), 
\qquad {\mathcal R}_{*}(k) = - \psi_{*}(k) - \frac{\phi_{*}(k)}{2}.
\label{LONG2b}
\end{equation}
Since the phase appearing in the integrand of Eq. (\ref{HT15}) is estimated by positing $x(\tau) \simeq y_{\mathrm{rec}}$ the cancellation between the ordinary and the integrated SW contributions is maximal, and, in this sense, such an approximation can be improved:
\begin{eqnarray}
&& \Delta^{(\mathrm{SW})}_{\mathrm{I}}(k,\mu,\tau_{0}) + 
\Delta^{(\mathrm{ISW})}_{\mathrm{I}}(k,\mu,\tau_{0}) = 
\nonumber\\
&&\biggl\{ [2 \psi(k,\tau_{0}) - 2 \psi(k, \tau_{\mathrm{rec}})] + \biggl[2 \psi(k,\tau_{\mathrm{rec}}) + 
{\mathcal R}_{*}(k) - \frac{ R_{\gamma} \Omega_{\mathrm{B}}(k)}{4}
\biggr]\biggr\} e^{- i \mu y_{\mathrm{rec}}}, 
\nonumber\\
&& = \biggl[ {\mathcal R}_{*}(k) + 2 \psi(k,\tau_{0}) - \frac{ R_{\gamma} \Omega_{\mathrm{B}}(k)}{4}\biggr]e^{- i \mu y_{\mathrm{rec}}}.
\label{LONG3}
\end{eqnarray}
The second (intermediate) equality appearing in Eq. (\ref{LONG3}) has been 
included just to show the explicit cancellation at $\tau_{\mathrm{rec}}$.
In a pure CDM model (i.e. no dark energy, no magnetic fields and $\Omega_{\mathrm{M}0} =1$), the quantity in squared brackets in Eq. (\ref{LONG3}) can be written as:
\begin{equation}
 {\mathcal R}_{*}(k) + 2 \psi(k,\tau_{0}) - \frac{ R_{\gamma} \Omega_{\mathrm{B}}(k)}{4}=    
 -\frac{{\mathcal R}_{*}(k)}{5} +  \frac{ R_{\gamma} \Omega_{\mathrm{B}}(k)}{4}.
\label{LONG5}
\end{equation}
The estimate provided by Eq. (\ref{LONG5}) can be improved
by taking into account the observation of Eq. (\ref{APP2}). In the latter 
case it can be shown that the SW contribution turns out to be 
\begin{eqnarray}
\Delta^{(\mathrm{SW})}_{\mathrm{I}}(k,\mu,\tau_{0}) &=& \biggl[- \frac{{\mathcal R}_{*}(k)}{5} 
{\mathcal S}{\mathcal W}_{{\mathcal R}}(\alpha_{\mathrm{rec}}) + \frac{R_{\gamma} \Omega_{\mathrm{B}}(k)}{20} \,{\mathcal S}{\mathcal W}_{\mathrm{B}}(\alpha_{\mathrm{rec}})\biggr] e^{- i \mu y_{\mathrm{rec}}},
\label{LONG5a}\\
{\mathcal S}{\mathcal W}_{{\mathcal R}}(\alpha) &=& 1 + \frac{4}{3 \alpha} - \frac{16}{3 \alpha^2}
+ \frac{16( \sqrt{\alpha +1} -1)}{3 \alpha^3},
\label{SWR}\\
{\mathcal S}{\mathcal W}_{\mathrm{B}}(\alpha) &=& 1 - \frac{12}{\alpha} + \frac{48}{\alpha^2} + 
\frac{32(  1 - \sqrt{\alpha +1})}{\alpha^3},
\label{SWB}
\end{eqnarray}
where $\alpha_{\mathrm{rec}}$ has been introduced in Eq. (\ref{APP2}).
The result of Eq. (\ref{LONG5a}) corresponds to the approximation 
scheme labeled by [{\it b}] in the above list of items.
The evolution of $\psi$ is obtained by solving explicitly the corresponding 
evolution equations for ${\mathcal R}$ and by recalling the connection 
between ${\mathcal R}$ and $\psi$ (see Eq. (\ref{gauge4})). The functions 
reported in Eqs. (\ref{SWR}) and (\ref{SWB}) have been derived in the appendix \ref{APPA}
by using, consistently, the synchronous gauge description. It is interesting that the synchronous 
description leads to an explicit derivation of the SW and ISW contributions which is 
technically different from the one of the conformally Newtonian gauge. 
By taking the limit $\alpha_{\Lambda} \to \infty$ (for $\alpha$ fixed) in Eq. (\ref{LONG7}), 
the standard CDM result is recovered. In the pure CDM case the long-wavelength limit implies 
\begin{equation}
\frac{\psi'}{{\mathcal H}} + \psi + \frac{{\mathcal H}^2 - {\mathcal H}'}{{\mathcal H}^2} \psi = - {\mathcal R}(k,\tau),
\label{LONG9a}
\end{equation}
which can be easily solved analytically
\begin{equation}
\psi(k,\alpha) = - {\mathcal R}_{*}(k) {\mathcal I}_{1}(\alpha) + 
R_{\gamma} \Omega_{\mathrm{B}}(k) {\mathcal I}_{2}(\alpha) 
\label{LONG10}
\end{equation}
where 
\begin{eqnarray}
&&{\mathcal I}_{1}(\alpha) = \frac{\alpha  [\alpha (9\alpha +2) - 8] + 
16 [\sqrt{\alpha +1}-1]}{15 \alpha^3}, 
\nonumber\\
&& {\mathcal I}_{2}(\alpha) = \frac{3\{\alpha [\alpha (\alpha -2) + 8] + 16 [ 1 - \sqrt{\alpha +1}]\}}{20 \alpha^3}
\label{LONG11}
\end{eqnarray}
In the case of the $\Lambda$CDM model $|\psi(k,\tau)|$ grows linearly as a function of the conformal 
time coordinate as a consequence of the presence of the dark-energy phase. To obtain a quantitative 
estimate Eq. (\ref{LONG9a}) can be used but in the case when $\alpha_{\Lambda}$ and it is given 
as in Eqs. (\ref{LONG7}) and (\ref{LONG8}). Similarly, 
by summing up Eqs. (\ref{DE1}) and (\ref{DE2}) we shall have 
\begin{equation}
\psi'' + 2 ({\mathcal H}' - {\mathcal H}^2) \psi = 4\pi G a^2 \biggl[ (w_{\mathrm{de}} +1)\delta_{\mathrm{s}} \rho_{\mathrm{de}} + \frac{4}{3} \delta_{\mathrm{s}} \rho_{\mathrm{B}}\biggr].
\label{DE4a}
\end{equation}
In the case where $w_{\mathrm{de}} = -1$ (which is the $\Lambda$CDM case),  Eq. (\ref{DE4a}) reduces to 
\begin{equation}
\psi'' = \frac{16 \pi G a^2}{3} R_{\gamma} \rho_{\mathrm{R}} \Omega_{\mathrm{B}}(k).
\label{DE5}
\end{equation}
In the limit $\Omega_{\mathrm{B}}(k) \to 0$ we will have that $\psi(k,\tau) = c_1(k) + c_{2}(k) \tau$. The latter 
expression holds asymptotically. Direct numerical integration
must be eventually employed and the result can be expressed as:
\begin{equation}
\psi(k,\tau_{0}) = -0.4507 \, {\mathcal R}_{*}(k) + 0.1125\, R_{\gamma} \Omega_{\mathrm{B}}(k).
\label{DE6}
\end{equation}
Equation (\ref{DE6}) must be compared with the value of $\psi(k,\tau)$ at
$\tau_{\mathrm{rec}}$ (corresponding to $z_{{\mathrm{rec}}} = 1090.51$)
\begin{equation}
\psi(k,\tau_{\mathrm{rec}}) = -0.5882 \, {\mathcal R}_{*}(k) + 0.1467\, R_{\gamma} \Omega_{\mathrm{B}}(k).
\label{DE7}
\end{equation}
The result of Eq. (\ref{DE7}) is also compatible  with the estimates obtainable, 
from Eq. (\ref{LONG7}), in the limit $\alpha_{\Lambda}\to \infty$:
\begin{equation}
\psi(k,\tau_{\mathrm{rec}}) = -0.6 \, {\mathcal R}_{*}(k) + 0.15\, R_{\gamma} \Omega_{\mathrm{B}}(k).
\label{DE8}
\end{equation}
From  Eqs. (\ref{DE6}) and (\ref{DE7}) we can therefore say that 
\begin{equation}
\Delta \psi(k) = \psi(k,\tau_{0}) - \psi(k,\tau_{\mathrm{rec}}) = 0.1375 \,{\mathcal R}_{*} - 0.0342 \,R_{\gamma} \Omega_{\mathrm{B}}(k).
\label{DE9}
\end{equation}
Concerning the result of  Eq. (\ref{DE9})  few comments are in order:
\begin{itemize}
\item{} in the absence of large-scale magnetic fields $\Delta \psi(k) \simeq 0.137 \,{\mathcal R}_{*}(k)$; the latter result depends upon the cosmological parameters and the quoted 
value refers to Eq. (\ref{LONG8}); 
\item{} the comparative growth of the magnetized contribution is smaller 
than in the curvature case, i.e.   
$\Delta \psi(k) \simeq -0.0342 \, R_{\gamma} \Omega_{\mathrm{B}}(k)$;
the sign difference stems directly from the form of the Hamiltonian constraint (see Eqs. (\ref{ham1}) and (\ref{LSS1}))
as well as from the solution of Eq. (\ref{LSS2}) in the large-scale limit.
\end{itemize}
It is relevant to point out that the value 
$\Delta\psi(k) \simeq 0.137 \,{\mathcal R}_{*}(k)$ is compatible with the results obtained in \cite{hugor} (i.e. $0.14\, {\mathcal R}_{*}(k)$) under the same approximations discussed here but in the absence of large-scale magnetic fields. 

\subsection{Transfer functions and semi-analytical estimates} 
In the $w$CDM case the numerical integration is unavoidable even in the absence 
of large-scale magnetic fields. 
The semi-analytical treatment can be carried on, up to some 
point, and the obtained results are a useful complement of the 
numerical discussion (see section \ref{sec5}). More precisely the transfer 
function can be computed directly in terms of the evolution 
of the appropriate background quantities whose specific form will 
have to be studied numerically.
Using Eq. (\ref{INT2}) the line of sight solution (\ref{LONG1}) can be written as 
\begin{eqnarray}
\Delta_{\mathrm{I}\ell}(k,\tau_{0}) &=& - 2 \int_{\tau_{\mathrm{rec}}}^{\tau_{0}} 
\biggl[ {\mathcal R}_{*}(k) \frac{\partial T_{{\mathcal R}}}{\partial \tau} - 
R_{\gamma} \Omega_{\mathrm{B}}(k) \frac{\partial T_{\mathrm{B}}}{\partial \tau}
\biggr] j_{\ell}[x(\tau)] \, d\tau
\nonumber\\
&+& \biggr[ - \frac{{\mathcal R}_{*}(k)}{5} {\mathcal S}{\mathcal W}_{{\mathcal R}}(\alpha_{\mathrm{rec}}) + \frac{R_{\gamma} \Omega_{\mathrm{B}}}{20} \,{\mathcal S}{\mathcal W}_{\mathrm{B}}(\alpha_{\mathrm{rec}})\biggr] j_{\ell}[x(\tau_{\mathrm{rec}})],
\label{INT1}
\end{eqnarray}
where, we recall that $x(\tau) = k (\tau_{0} - \tau)$ and $x(\tau_{\mathrm{rec}})$ coincides, by definition 
with $y_{\mathrm{rec}}$; 
${\mathcal S}{\mathcal W}_{{\mathcal R}}(\alpha_{\mathrm{rec}})$ and 
${\mathcal S}{\mathcal W}_{\mathrm{B}}(\alpha_{\mathrm{rec}})$ have been 
already introduced in Eq. (\ref{LONG5a}). The functions $T_{\mathcal R}(\tau)$ 
and $T_{\mathrm{B}}(\tau)$ are given by 
\begin{eqnarray}
T_{{\mathcal R}}(\tau) &=& {\mathcal R}_{*}(k) \biggl[1 - \frac{{\mathcal H}(\tau)}{a^2(\tau)} \int_{0}^{\tau} a^2(x) \, d\,x\biggr],
\label{INT1a}\\ 
T_{\mathrm{B}}(\tau)  &=& \int_{0}^{\tau} {\mathcal B}(\tau,x) \biggl[ 
\frac{a^2(x)}{{\mathcal H}(x)} - \frac{a^2(\tau)}{{\mathcal H}(\tau)} \biggr] dx 
+ \int_{0}^{\tau} a^2(x) d\, x \int_{0}^{x} {\mathcal B}(\tau, y) \, d y; 
\label{INT1b}
\end{eqnarray}
for two generic arguments $x$ and $y$, the function ${\mathcal B}(x,y)$ is 
defined as 
\begin{equation}
{\mathcal B}(x,y)  = \frac{{\mathcal H}(x) {\mathcal H}(y)}{a^2(x)} \frac{ [3 c_{\mathrm{st}}^2(y) -1]}{3 [w_{\mathrm{t}}(y) +1]} \frac{\rho_{\mathrm{R}}(y)}{\rho_{\mathrm{t}}(y)};
\label{INT1c}
\end{equation}
$c_{\mathrm{st}}^2(y)$ is the sound speed of the total plasma already introduced 
in Eq. (\ref{LONG7}) and here appearing as a function of a generic integration variable $y$.
Equations (\ref{INT1a}) and (\ref{INT1b}) can be derived, after some algebra, from 
Eqs. (\ref{gauge4}), (\ref{LSS1}), (\ref{LSS2}) and (\ref{LSS3}). 
The derivation of Eqs. (\ref{INT1a})--(\ref{INT1b}) is specifically discussed in appendix \ref{APPB}. 

The results of Eqs. (\ref{INT1}), (\ref{INT1a}) and (\ref{INT1b}) imply 
that the angular power spectrum for the temperature autocorrelations consists of three distinct terms parametrizing, 
respectively, the ordinary Sachs-Wolfe constribution, the ISW term and the cross-correlation 
between the SW and the ISW terms:
\begin{equation}
C^{(\mathrm{TT})}_{\ell}  = C_{\ell}^{(\mathrm{SW})} + C_{\ell}^{(\mathrm{ISW})} + C_{\ell}^{(\mathrm{cross})}. 
\label{EX5}
\end{equation}
The ordinary SW contribution turns out to be, within the present approximations 
\begin{eqnarray}
&& C_{\ell}^{(\mathrm{SW})} = \frac{4 \pi}{25} {\mathcal S}{\mathcal W}_{{\mathcal R}}^2(\alpha_{\mathrm{rec}}) \int_{0}^{\infty} \frac{dk}{k} {\mathcal P}_{{\mathcal R}}(k) j^2_{\ell}(k\tau_{0}) 
 \nonumber\\
&&+ \frac{\pi}{100}  {\mathcal S}{\mathcal W}_{\mathrm{B}}^2(\alpha_{\mathrm{rec}}) \int_{0}^{\infty} \frac{dk}{k} {\mathcal P}_{\Omega}(k) j^2_{\ell}(k\tau_{0})
\nonumber\\
&&+ \frac{2\pi}{25} R_{\gamma} \,\cos{\beta}  
{\mathcal S}{\mathcal W}_{\mathrm{B}}(\alpha_{\mathrm{rec}}){\mathcal S}{\mathcal W}_{{\mathcal R}}(\alpha_{\mathrm{rec}})
\int_{0}^{\infty} \frac{d k}{k} \sqrt{{\mathcal P}_{{\mathcal R}}(k)} \sqrt{{\mathcal P}_{\Omega}(k)} j_{\ell}^2(k\tau_0),
\label{SW1}
\end{eqnarray}
where $\beta$ is the correlation angle\footnote{The general idea behind \cite{maxa1,maxa2} (see 
also \cite{max1,max2,max3}) has been to include large-scale magnetic fields at all the stages of the Einstein-Boltzmann hierarchy and, in particular, at the level  of the initial conditions. This means that the solutions such as the magnetized adiabatic mode and the various non-adiabatic modes are regular (in a technical sense) and well defined when the magnetized contribution is correctly supplemented by the curvature contribution. The correlation between the components is automatic already at the level of the initial conditions.  At earlier times the correlation is also suggested, incidentally,  by models where large-scale magnetic fields are generated during an inflationary stage \cite{max5}.}.
In Eq. (\ref{SW1}) we recalled that 
$\tau_{\mathrm{rec}} \ll \tau_{0}$, implying that $k (\tau_{0} - \tau_{\mathrm{rec}}) \simeq k \tau_{0}$. 
The relevant power spectra and the spherical Bessel functions can be written as\footnote{The scales $k_{\mathrm{p}}$ 
 and $k_{\mathrm{L}}$ are the two pivot scales at which the amplitudes of ${\mathcal P}_{{\mathcal R}}(k)$ 
 and of ${\mathcal P}_{\Omega}(k)$ are assigned. The pivotal values of these 
 quantities will be reminded in a moment but can be also found in Eqs. (3) and (5) of Ref. \cite{max1} (see also \cite{max2,max3} for further details).}
 \begin{equation}
 {\mathcal P}_{{\mathcal R}} = {\mathcal A}_{{\mathcal R}} \biggl(\frac{k}{k_{\mathrm{p}}}\biggr)^{n_{\mathrm{s}} -1}, \qquad {\mathcal P}_{\Omega}(k) 
 = {\mathcal E}_{\mathrm{B}} \biggl(\frac{k}{k_{\mathrm{L}}}\biggr)^{2(n_{\mathrm{B}}-1)},\qquad j_{\ell}(z) = \sqrt{\frac{\pi}{2 z}} J_{\ell + 1/2}(z),
 \label{SW1a}
 \end{equation}
 the integrals over the Bessel functions can be analytically performed since, as it is 
 well known  \cite{grad}
\begin{equation}
 \int_{0}^{\infty} J_{\nu}( \alpha y) \, J_{\mu}(\alpha y) y^{-\lambda} \, dy= 
 \frac{\alpha^{\lambda-1} \Gamma(\lambda)\Gamma\biggl(\frac{\nu + \mu -\lambda +1}{2}\biggr)}{2^{\lambda} \Gamma\biggl(\frac{- \nu + \mu + \lambda+1}{2}\biggr) \Gamma\biggl(\frac{\nu + \mu+ \lambda+1}{2}\biggr) \Gamma\bigg(\frac{\nu - \mu + \lambda +1}{2}\biggr)}.
 \label{EX11}
 \end{equation}
The ISW contribution is more complicated and, as previously remarked, it cannot 
be computed in fully analytic terms since each term contains two time derivatives of the 
transfer functions:
\begin{eqnarray}
&& C_{\ell}^{(\mathrm{ISW})} = 16 \pi \int_{0}^{\infty} \frac{d k}{k} {\mathcal P}_{\mathcal R}(k) \int_{0}^{\tau_{0}} \frac{d T_{{\mathcal R}}}{d\tau_{1}}\, j_{\ell}(k \Delta\tau_{1}) d\tau_{1} \int_{0}^{\tau_{0}}
\frac{d T_{{\mathcal R}}}{d\tau_{2}} j_{\ell}(k \Delta\tau_{2})
\nonumber\\
&& + 16 \pi R_{\gamma}^2\int_{0}^{\infty} \frac{d k}{k} {\mathcal P}_{\Omega}(k) \int_{0}^{\tau_{0}} \frac{d T_{\mathrm{B}}}{d\tau_{1}}\, j_{\ell}(k \Delta\tau_{1}) d\tau_{1} \int_{0}^{\tau_{0}}
\frac{d T_{\mathrm{B}}}{d\tau_{2}} j_{\ell}(k \Delta\tau_{2})
\nonumber\\
&&+ 32 \pi R_{\gamma} \cos{\beta} \int_{0}^{\infty} \frac{d k}{k} \sqrt{{\mathcal P}_{\Omega}(k)} \sqrt{{\mathcal P}_{{\mathcal R}}(k)} \int_{0}^{\tau_{0}} \frac{d T_{\mathrm{B}}}{d\tau_{1}}\, j_{\ell}(k \Delta\tau_{1}) d\tau_{1} \int_{0}^{\tau_{0}}
\frac{d T_{{\mathcal R}}}{d\tau_{2}} j_{\ell}(k \Delta\tau_{2}),
\label{EX7}
\end{eqnarray}
where $\Delta\tau_{1} = (\tau_{0} - \tau_{1})$ and $\Delta\tau_{2} = (\tau_{0} - \tau_{2})$ and 
$\tau_{1}$ and $\tau_{2}$ are integration variables. 
The last contribution listed in Eq. (\ref{EX5}) is the cross term whose explicit 
expression turns out to be 
\begin{eqnarray}
&& C_{\ell}^{(\mathrm{cross})} = \frac{16 \pi}{5} {\mathcal S}{\mathcal W}_{{\mathcal R}}^2(\alpha_{\mathrm{rec}})\int_{0}^{\infty} \frac{d k}{k}  {\mathcal P}_{\mathcal R}(k)
\int_{0}^{\tau_{0}} \biggl(\frac{dT_{{\mathcal R}}}{d\tau}\biggr) j_{\ell}(k \tau_{0})
 j_{\ell}(k \Delta\tau) d\tau
\nonumber\\
&&+  \frac{4 \pi}{5} R_{\gamma}^2 {\mathcal S}{\mathcal W}_{\mathrm{B}}^2(\alpha_{\mathrm{rec}})\int_{0}^{\infty} \frac{d k}{k}  {\mathcal P}_{\Omega}(k)
\int_{0}^{\tau_{0}} \biggl(\frac{dT_{\mathrm{B}}}{d\tau}\biggr) j_{\ell}(k \tau_{0})
 j_{\ell}(k \Delta\tau) d\tau
\nonumber\\
&&+ \frac{16\pi}{5} R_{\gamma} \cos{\beta} {\mathcal S}{\mathcal W}_{{\mathcal R}}(\alpha_{\mathrm{rec}})\int_{0}^{\infty} \frac{d k}{k}  \sqrt{{\mathcal P}_{\Omega}(k)} 
\sqrt{{\mathcal P}_{{\mathcal R}}(k)} 
\int_{0}^{\tau_{0}} \biggl(\frac{dT_{\mathrm{B}}}{d\tau}\biggr) j_{\ell}(k \tau_{0})
 j_{\ell}(k \Delta\tau) d\tau
\nonumber\\
&&+ \frac{16\pi}{5} R_{\gamma} \cos{\beta} {\mathcal S}{\mathcal W}_{\mathrm{B}}(\alpha_{\mathrm{rec}})\int_{0}^{\infty} \frac{d k}{k}  \sqrt{{\mathcal P}_{\Omega}(k)} 
\sqrt{{\mathcal P}_{{\mathcal R}}(k)} 
\int_{0}^{\tau_{0}} \frac{dT_{{\mathcal R}}}{d\tau} j_{\ell}(k \tau_{0})
 j_{\ell}(k \Delta\tau) d\tau,
\label{EX8}
\end{eqnarray}
where $k \Delta \tau = k(\tau_{0} -\tau)$.

Even if the  time evolution of $T_{{\mathcal R}}(\tau)$ and $T_{\mathrm{B}}(\tau)$ must be 
obtained numerically, Eqs. (\ref{EX7}) and (\ref{EX8}) can be further 
simplified by anticipating the integration over $k$. Since the both ${\mathcal P}_{{\mathcal R}}(k)$ and ${\mathcal P}_{\Omega}(k)$ have a power dependence upon $k$, the expressions appearing in Eq. (\ref{EX7}) can be made 
just more explicit by taking into account that the integral 
(see Eq. (\ref{LEG4}))
\begin{equation}
\int_{0}^{\infty} k^{-\lambda}\, J_{\nu}(k a) \,J_{\nu}(k b) \, d k 
\end{equation}
is proportional to a confluent hypergeometric function \cite{grad,abr1}. Using 
the quadratic transformations \cite{abr1}, the confluent hypergeometric 
functions can be related to Legendre functions. This analysis 
is reported in appendix \ref{APPC} (see Eqs. (\ref{LEG4}) and (\ref{LEG14})).
The expressions of appendix \ref{APPC} generalize the results of \cite{KS1,KS2} holding
 in the Harrison-Zeldovich limit. 

It is appropriate to remind about the conventions 
leading to the power spectra mentioned in Eq. (\ref{SW1a}). This topic has been 
thoroughly discussed (see \cite{max1,max2,max3} and references therein) and will just be swiftly mentioned 
here.  The power spectra 
of the magnetic fields are assigned exactly in the same way as the spectra of curvature perturbations:
it would be rather weird to have a convention for the power spectrum 
of curvature perturbations and a totally different one for the magnetic power spectrum. 
In the $w$CDM paradigm (as in the standard $\Lambda$CDM case)
the temperature and polarization inhomogeneities are solely sourced by  
the fluctuations of the spatial curvature ${\mathcal R}$ whose Fourier modes obey
\begin{equation}
\langle {\mathcal R}(\vec{k}) {\mathcal R}(\vec{p}) \rangle = \frac{2\pi^2}{k^3} {\mathcal P}_{{\mathcal R}}(k)\delta^{(3)}(\vec{k} + \vec{p}), \qquad 
{\mathcal P}_{{\mathcal R}}(k)  = {\mathcal A}_{\mathcal R} \biggl(\frac{k}{k_{\mathrm{p}}}\biggr)^{n_{\mathrm{s}}-1};
\label{PS1}
\end{equation}
in the $\Lambda$CDM case and in the light of the WMAP 5yr data alone \cite{WMAP51,WMAP52,WMAP53} 
 ${\mathcal A}_{{\mathcal R}} = (2.41 \pm 0.11) \times 10^{-9}$; as already mentioned in section 
 \ref{sec1} the pivot scale  $k_{\mathrm{p}}$ 
 is  $0.002\,\mathrm{Mpc}^{-1}$.
In full analogy with Eq. (\ref{PS1}) the ensemble average of the Fourier modes of the magnetic field are 
given by
\begin{equation}
\langle B_{i}(\vec{k}) \, B_{j}(\vec{p}) \rangle = \frac{2\pi^2 }{k^3} P_{ij}(k) {\mathcal P}_{\mathrm{B}}(k) \delta^{(3)}(\vec{k} + \vec{p}), \qquad {\mathcal P}_{{\mathrm{B}}}(k) = {\mathcal A}_{\mathrm{B}} \biggl(\frac{k}{k_{\mathrm{L}}}\biggr)^{n_{\mathrm{B}}-1}, 
\label{PS2}
\end{equation}
where $P_{ij}(k) = (k^2 \delta_{ij} - k_{i} k_{j})/k^2$; 
${\mathcal A}_{\mathrm{B}}$ the spectral amplitude of the magnetic field at the pivot scale $k_{\mathrm{L}} = \mathrm{Mpc}^{-1}$ \cite{max1,max2}.  In the case when  $n_{\mathrm{B}} > 1$ (i.e. blue magnetic field spectra),  ${\mathcal A}_{\mathrm{B}} = 
(2\pi)^{n_{\mathrm{B}} -1} \, B_{\mathrm{L}}^2 /\Gamma[(n_{\mathrm{B}} -1)/2]$; if $n_{\mathrm{B}} < 1$ 
(i.e. red magnetic field spectra), ${\mathcal A}_{\mathrm{B}} =[ (1 -n_{\mathrm{B}})/2] (k_{\mathrm{A}}/k_{\mathrm{L}})^{(1 - n_{\mathrm{B}})}B_{\mathrm{L}}^2$ where $k_{\mathrm{A}}$ is the infra-red cut-off of the spectrum. In the case of white spectra (i.e. $n_{\mathrm{B}} =1$) the two-point 
function is logarithmically divergent in real space and this is fully analog to what happens in Eq. (\ref{PS1}) when $n_{\mathrm{s}} =1$, i.e. the Harrison-Zeldovich (scale-invariant) spectrum. By selecting $k_{\mathrm{L}}^{-1}$ of the order of the Mpc 
scale the comoving field $B_{\mathrm{L}}$ represents the (frozen-in) magnetic field intensity at the onset 
of the gravitational collapse of the protogalaxy \cite{max3}. On top of the power spectrum of the magnetic energy density 
it is necessary to compute and regularize also the power spectrum of the anisotropic stress and explicit discussions 
can be found in \cite{max2,max3}.  

\renewcommand{\theequation}{4.\arabic{equation}}
\setcounter{equation}{0}
\section{Magnetized initial conditions with dark energy}
\label{sec4}
If the dark energy component is fluctuating there are, in principle various ways in which initial conditions of the Einstein-Boltzmann hierarchy can be set. In general terms the pre-decoupling plasma contains 
five\footnote{This  statement holds under the assumption that 
electrons and protons are tightly coupled by Coulomb scattering. 
If we count separately the electrons and the ions the total number of 
components increases to $6$.} physically different components (see, e.g. section \ref{sec2}).  In the $\Lambda$CDM case  the dark energy 
does not fluctuate and, therefore, the way initial conditions are set does 
not differ from the CDM scenario insofar as the entropy fluctuations 
of the whole plasma vanish for large scales i.e. for $k/{\mathcal H} \ll 1$. The latter condition 
can be expressed by requiring that ${\mathcal S}_{\mathrm{ij}} =0$ where i and j denote 
a generic pair of fluids in the mixture and where 
\begin{eqnarray}
&&{\mathcal S}_{\mathrm{ij}} = 3 ( \zeta_{\mathrm{i}} - \zeta_{\mathrm{j}}) 
\label{entro1}\\
&& \zeta_{\mathrm{i}}  = - \psi + \frac{\delta^{(\mathrm{cn})}_{\mathrm{i}}}{3(w_{\mathrm{i}} +1)}  = \xi + \frac{\delta^{(\mathrm{S})}_{\mathrm{i}}}{3(w_{\mathrm{i}} +1)},
\label{entro2}
\end{eqnarray}
are, by definition, the entropy fluctuations of the system. 
The second equality in Eq. (\ref{entro2}) follows by transforming 
the fluctuation variables from the conformally Newtonian to the 
synchronous gauge. In what follows the condition of Eq. (\ref{entro1}) 
forbids non-adiabatic fluctuations in the dark energy sector as well as in the fluid sector. 

If the dark energy background is consistently included in the initial conditions together with large-scale magnetic fields the initial conditions 
of the Einstein-Boltzmann hierarchy must be supplemented, in the conformally Newtonian gauge, by the following pair of conditions 
\begin{eqnarray}
\delta_{\mathrm{de}}(k,\tau) &=& 
- \frac{3}{2} (w_{\mathrm{de}} + 1) \phi_{*}(k) - 
\frac{3}{4} (w_{\mathrm{de}} +1) R_{\gamma} \Omega_{\mathrm{B}}(k) + {\mathcal O}\biggl(\frac{k}{{\mathcal H}}\biggr),
\label{inde1}\\
\theta_{\mathrm{de}}(k,\tau) &=&
 \frac{(1 - w_{\mathrm{de}})}{(1+ w_{\mathrm{de}})(1 - 3 w_{\mathrm{de}})} 
 \phi_{*}(k) \frac{k^2}{{\mathcal H}} - \frac{w_{\mathrm{de}}}{(1 + w_{\mathrm{de}})(1 - 3 w_{\mathrm{de}})}  R_{\gamma} 
 \Omega_{\mathrm{B}}(k) \frac{k^2}{{\mathcal H}},
 \label{inde2}
 \end{eqnarray}
 where the second equation holds under the assumption 
 that $w_{\mathrm{de}} \neq -1$ and that $w_{\mathrm{de}}< 0$.
 For the actual numerical integration it turns out to be very useful 
 to reshuffle the dependence of the dark energy variables 
 in terms of two generalized potentials defined, in Fourier space, as 
 \begin{eqnarray}
\theta_{\mathrm{de}}(k,\tau) &=& \frac{k^2}{a \sqrt{\rho_{\mathrm{de}}}} g(k,\tau)
\label{inde3}\\
\delta_{\mathrm{s}} \rho_{\mathrm{de}}(k,\tau) &=& 
\frac{\sqrt{\rho_{\mathrm{de}}}}{a} ( 1 + w_{\mathrm{de}}) \biggl[ f(k,\tau) 
- \frac{3}{2} {\mathcal H} ( 1 - w_{\mathrm{de}}) g(k,\tau)\biggr],
\label{inde4}\\
 \delta_{\mathrm{s}} p_{\mathrm{de}}(k,\tau) &=& 
\frac{\sqrt{\rho_{\mathrm{de}}}}{a} ( 1 + w_{\mathrm{de}}) \biggl[ f(k,\tau) 
+ \frac{3}{2} {\mathcal H} ( 1 - w_{\mathrm{de}}) g(k,\tau)\biggr].
\label{inde5}
\end{eqnarray}
Using Eqs. (\ref{inde3}), (\ref{inde4}) and (\ref{inde5}) into Eqs. (\ref{DE1}) and (\ref{DE2})
we do get a system for $f$ and $g$:
\begin{eqnarray}
g' &=& f + a \sqrt{\rho_{\mathrm{de}}} \phi, 
\label{inde6}\\
f' &=& - 2 {\mathcal H} f - \biggl\{ k^2 - \frac{3}{4} (1 -w)[ 2 {\mathcal H}' - ( 3 w_{\mathrm{de}} + 5) {\mathcal H}^2] \biggr\} g 
\nonumber\\
&+& 3 a \sqrt{\rho_{\mathrm{de}}}\biggl[ \psi' + \frac{{\mathcal H} ( 1 - w_{\mathrm{de}})}{2}
 \phi\biggr]\biggr\}.
 \label{inde7}
 \end{eqnarray}
The corresponding equations in the synchronous gauge can be obtained just by replacing, in Eqs. (\ref{inde6})
and (\ref{inde7}), 
\begin{equation}
f\to \tilde{f},\qquad g \to \tilde{g},\qquad \phi \to 0 , \qquad \psi' \to \frac{h'}{6}.
\label{inde8}
\end{equation}
The resulting equations are even simpler to integrate numerically:
\begin{eqnarray}
 \tilde{g}' &=& \tilde{f} , 
\label{inde9}\\
\tilde{f}' &=& - 2 {\mathcal H} \tilde{f} - \biggl\{ k^2 - \frac{3}{4} (1 -w)[ 2 {\mathcal H}' - ( 3 w_{\mathrm{de}} + 5) {\mathcal H}^2] \biggr\} \tilde{g} + a \sqrt{\rho_{\mathrm{de}}} \frac{h'}{2}.
 \label{inde10}
 \end{eqnarray}
 Equations (\ref{inde9}) and (\ref{inde10}) are the ones which are numerically integrated. The initial 
 conditions of the remaining components of the Einstein-Boltzmann hierarchy 
 do not change and are given as in \cite{max1,max2}. To be more specific Eq. (2.54) of Ref. \cite{max2} 
 contains the initial conditions used in the numerical discussion which will be reported in section \ref{sec5}.
\renewcommand{\theequation}{5.\arabic{equation}}
\setcounter{equation}{0}
\section{Numerical results}
\label{sec5}
The choice of parameters of the pivotal $\Lambda$CDM model in the light of the WMAP 5yr data alone is given by 
\cite{WMAP51,WMAP52,WMAP53}
\begin{equation}
( \Omega_{\mathrm{b}}, \, \Omega_{\mathrm{c}}, \Omega_{\mathrm{de}},\, h_{0},\,n_{\mathrm{s}},\, \epsilon_{\mathrm{re}}) \equiv 
(0.0441,\, 0.214,\, 0.742,\,0.719,\, 0.963,\,0.087),
\label{Par1}
\end{equation}
where, consistently 
with the established notations (see, e.g., Eqs. (\ref{HTA}) and (\ref{LONG1})), $\epsilon_{\mathrm{re}}$ denotes the 
optical depth to reionization. The parameters of Eq. (\ref{Par1}) maximize the likelihood when the barotropic index of the 
dark energy is fixed to $-1$. 

\begin{figure}[!ht]
\centering
\includegraphics[height=6cm]{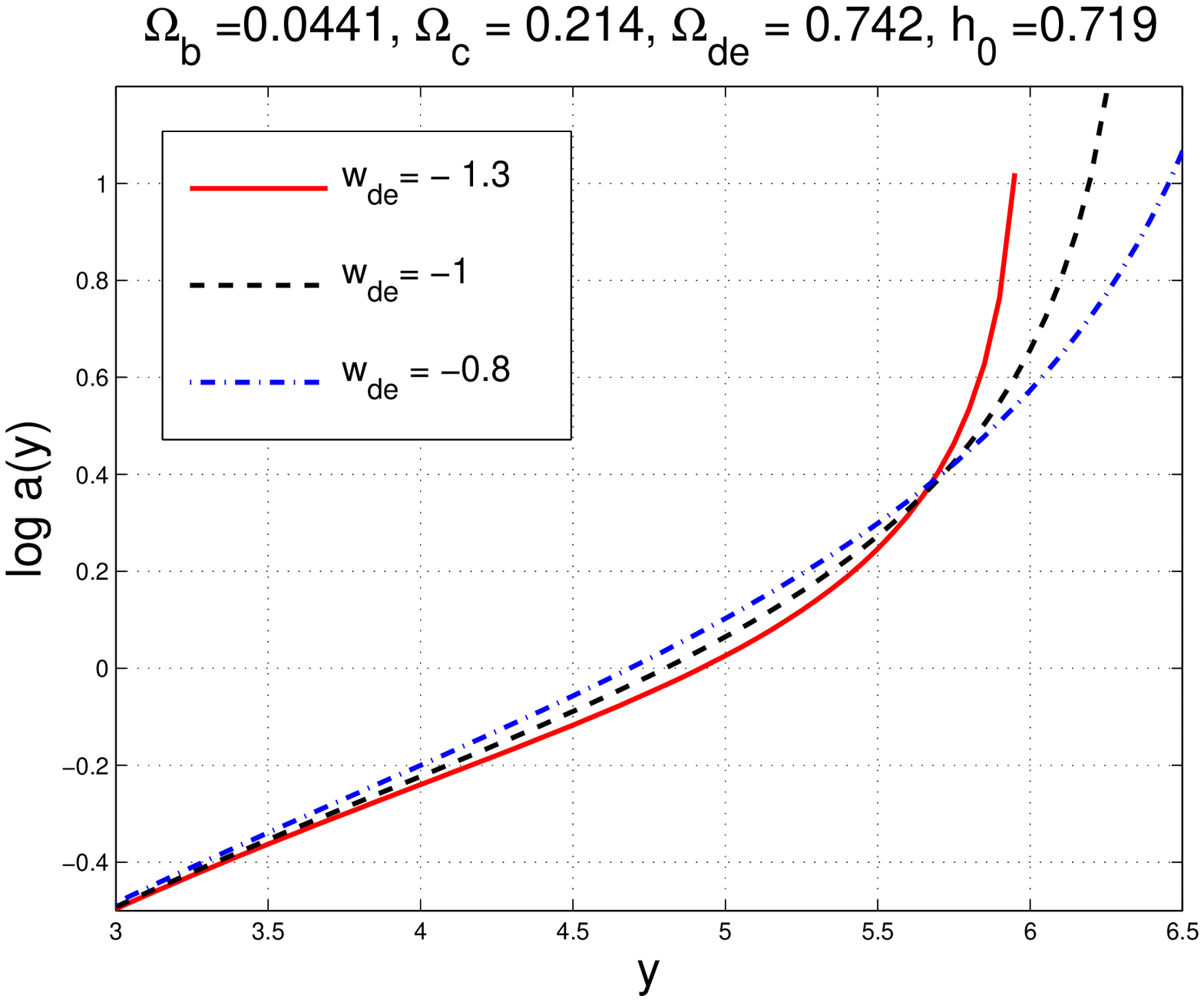}
\includegraphics[height=6cm]{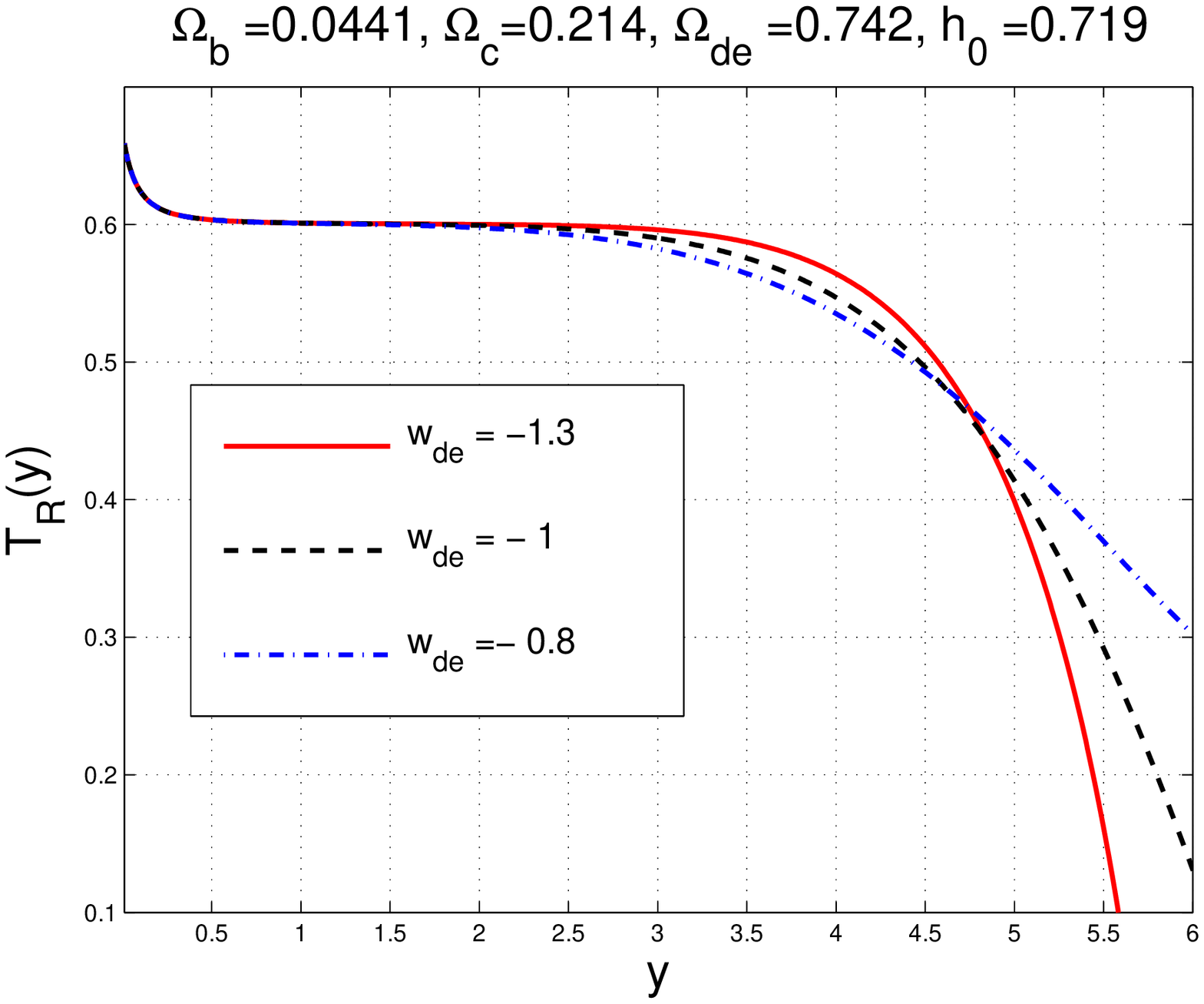}
\caption[a]{The evolutions of the scale factor and of $T_{{\mathcal R}}(y)$ are illustrated for different values of the barotropic index of dark energy $w_{\mathrm{de}}$.}
\label{FIGone}      
\end{figure}

\begin{figure}[!ht]
\centering
\includegraphics[height=6cm]{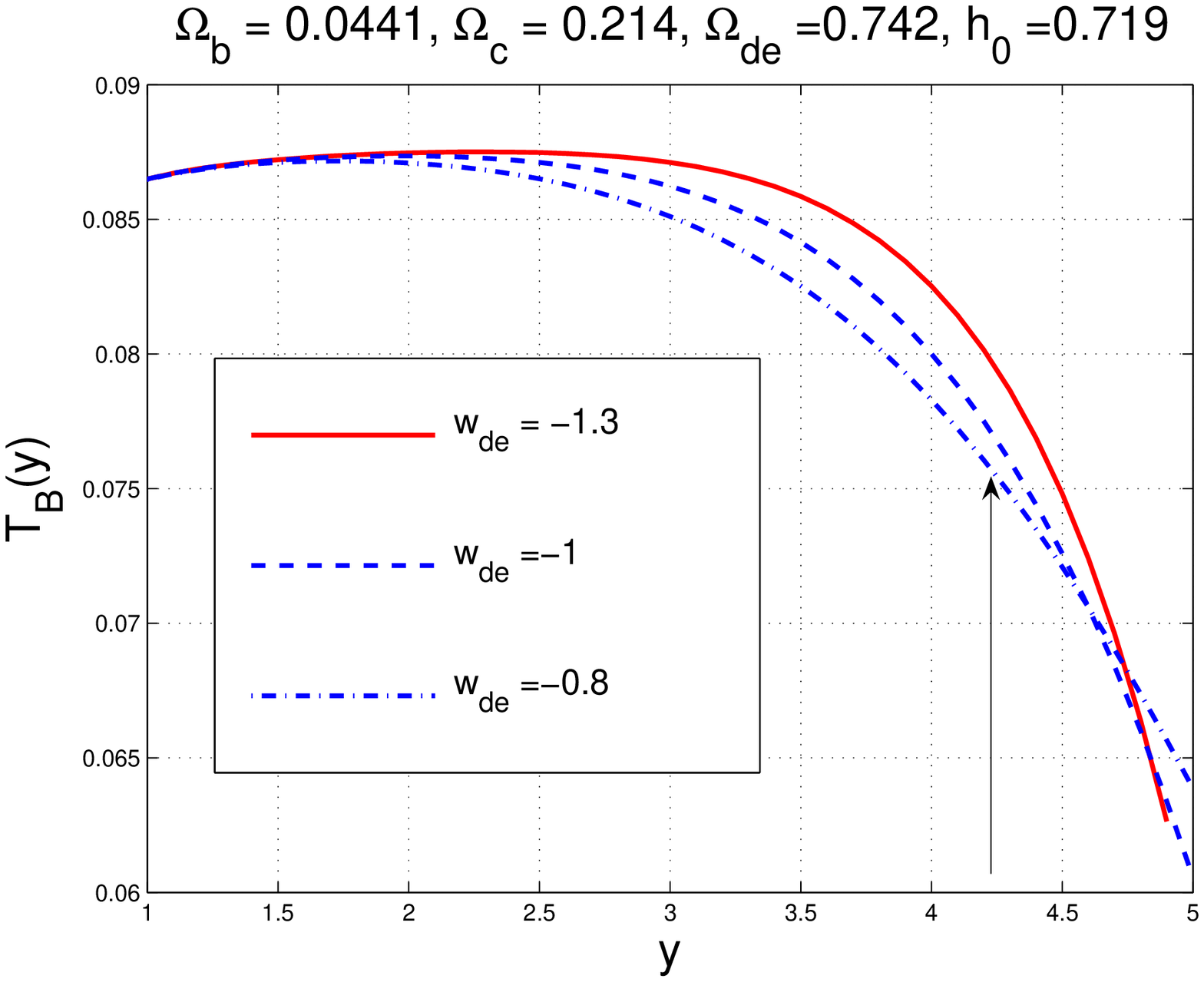}
\includegraphics[height=6cm]{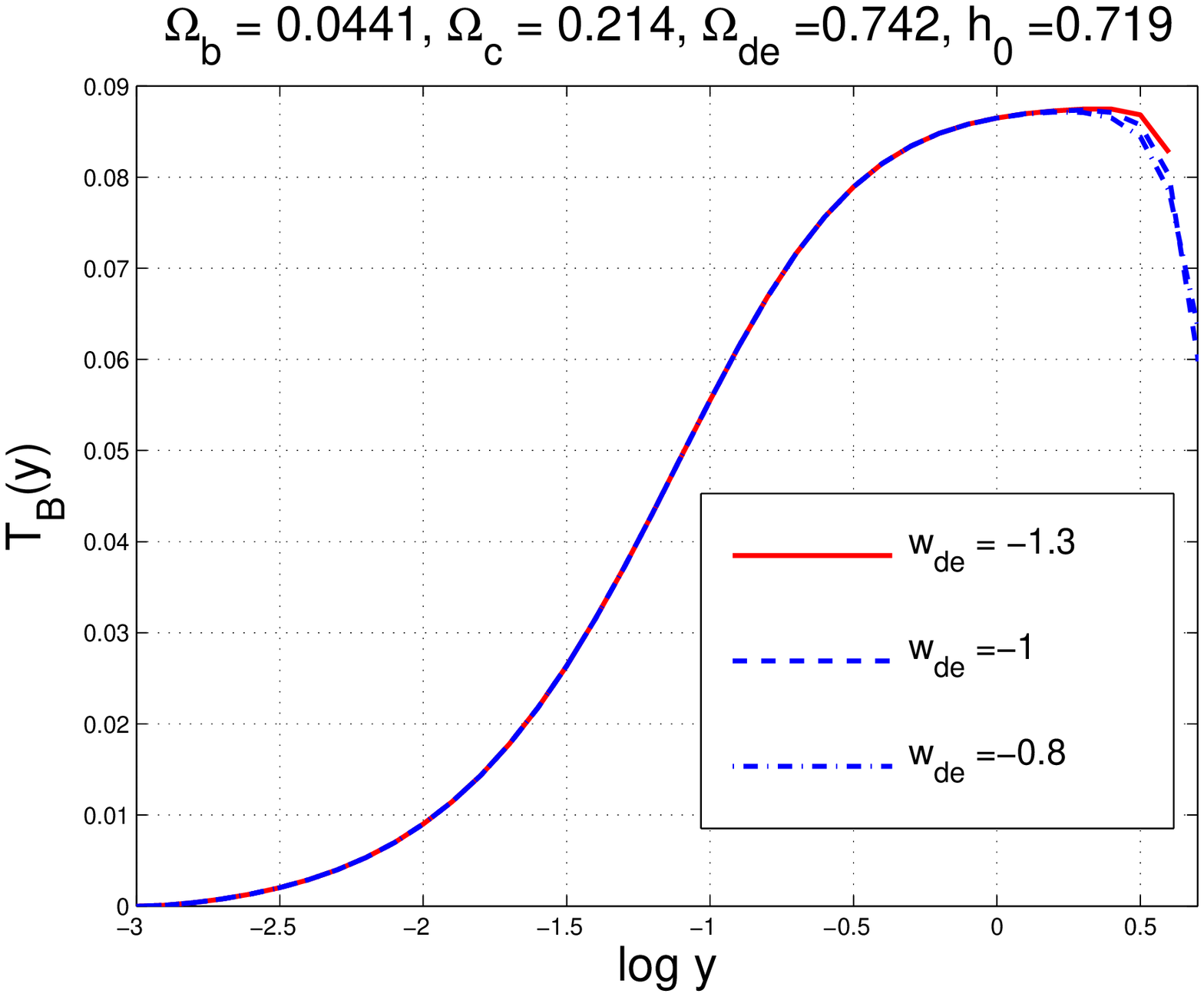}
\caption[a]{In the plots at the left and at the right the evolution of $T_{\mathrm{B}}(y)$ is illustrated, respectively, on a linear and on a 
logarithmic time scale. The case labeled with $w_{\mathrm{de}} =-1$ 
corresponds to the best fit of the WMAP 5yr data alone in the light of the 
$\Lambda$CDM scenario. }
\label{FIGtwo}      
\end{figure}

The first step of the numerical analysis is to compute $T_{\mathcal R}(\tau)$ and $T_{\mathrm{B}}(\tau)$ when the
parameters are given exactly by Eq. (\ref{Par1}) but the barotropic index of the dark energy 
assumes arbitrary values which can differ from the $\Lambda$CDM choice (i.e. $w_{\mathrm{de}} = -1$). 
The functions $T_{\mathcal R}(\tau)$ and $T_{\mathrm{B}}(\tau)$ have been 
introduced in Eq. (\ref{INT1}) and control the two relevant large-scale contributions to the ISW effect.  
According to Eqs. (\ref{INT1a}) and (\ref{INT1b}),  $T_{\mathcal R}(\tau)$ and $T_{\mathrm{B}}(\tau)$ can be 
assessed once the (numerical) evolution of the background geometry has been obtained. Such a technique is rather 
useful for the analytic estimate of the asymptotic limits and it otherwise demands the full (numerical) solution 
of the background geometry; the results of  Figs. \ref{FIGone} and \ref{FIGtwo} $T_{\mathcal R}(\tau)$ and $T_{\mathrm{B}}(\tau)$ are obtained by direct numerical integration. It has been explicitly verified that, indeed, Eqs. (\ref{INT1a}) and (\ref{INT1b}) are in excellent agreement with the numeric result once the full numerical expression of the scale factor is known.
In Fig. \ref{FIGone} (plot at the left) on the vertical axis 
the common logarithm of the scale factor is illustrated while on the horizontal axis the convenient time variable is given by 
$y = 3.33 \times 10^{-4} (\tau/\mathrm{Mpc})$  where $\tau$, as usual, is the conformal time coordinate.
Using $y$ as integration variable, Eq. (\ref{FL1}) and the related  initial conditions become very simple
\footnote{To avoid potential confusions, it is appropriate to remark that the integration variable of Eqs. (\ref{Par2}) and  (\ref{Par3}) has nothing to do with the variables $x$ and $y$ introduced in section \ref{sec3}. In Eqs. (\ref{Par2}) and (\ref{Par3}) 
the variable $y$ is simply given by $h_{0} y = \tau H_{0}$ where $H_{0}$ is the present value 
of the Hubble constant and $h_{0}$ its (dimensionless) indetermination. Conversely, in section \ref{sec3} $x$ and $y$ 
have been used as (dimensionfull)  integration variables coinciding exactly with the conformal time coordinate.} 
\begin{eqnarray}
\frac{d a}{d y} &=& \sqrt{ \omega_{\mathrm{R}} + \omega_{\mathrm{M}} a 
+ \omega_{\mathrm{de}} a^{1 - 3 w_{\mathrm{de}}}},
\label{Par2}\\
a(y_{\mathrm{i}}) &=& \frac{\omega_{\mathrm{M}}}{4} y_{\mathrm{i}}^2 
+ \sqrt{\omega_{\mathrm{R}}} y_{\mathrm{i}}, 
\label{Par3}
\end{eqnarray}
where $\omega_{\mathrm{R}} = h_{0}^2 (\Omega_{\gamma} + \Omega_{\nu})$, 
$\omega_{\mathrm{M}} = h_{0}^2 (\Omega_{\mathrm{b}} + \Omega_{\mathrm{c}})$ and 
$\omega_{\mathrm{de}} = h_{0}^2 \Omega_{\mathrm{de}}$.
The physical range of the parameters and the present normalization of the scale factor imply that $a_{0}=1$. To illustrate the trends induced by different values of $w_{\mathrm{de}}$ it is useful to plot the quantities of Fig. \ref{FIGone} also a bit outside their physical range  (i.e. for $a> a_{0}$).  The initial condition of Eq. (\ref{Par3}) are imposed by making use of  an exact solution 
of Eq. (\ref{Par2}) (as well as of the other Friedmann-Lema\^itre equations (\ref{FL1})) in the limit $\rho_{\mathrm{de}} \to 0$.
The form of the initial conditions of Eq. (\ref{Par3}) implies that $y_{\mathrm{i}}$ is well before matter-radiation equality.

\begin{figure}[!ht]
\centering
\includegraphics[height=6cm]{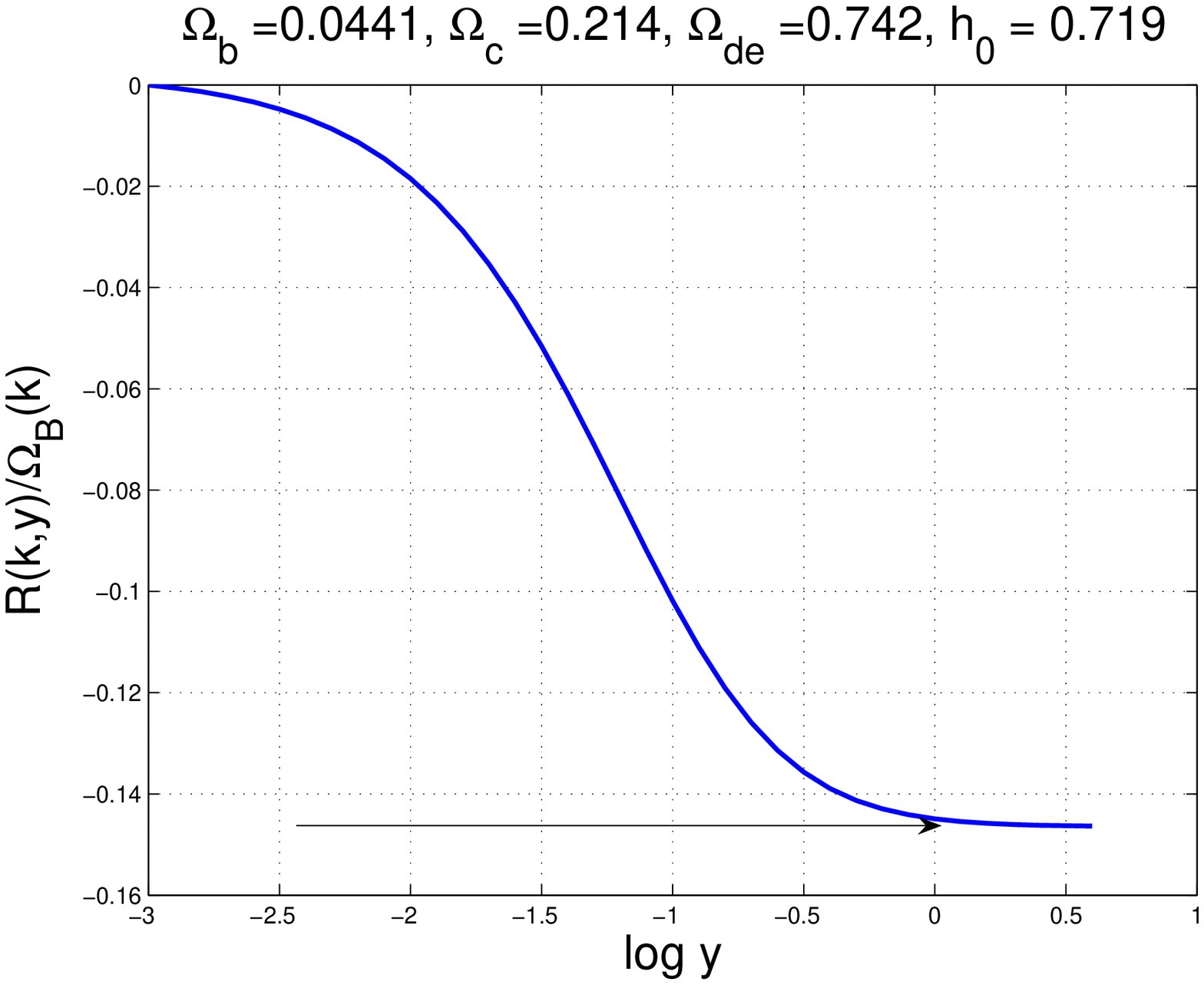}
\includegraphics[height=6cm]{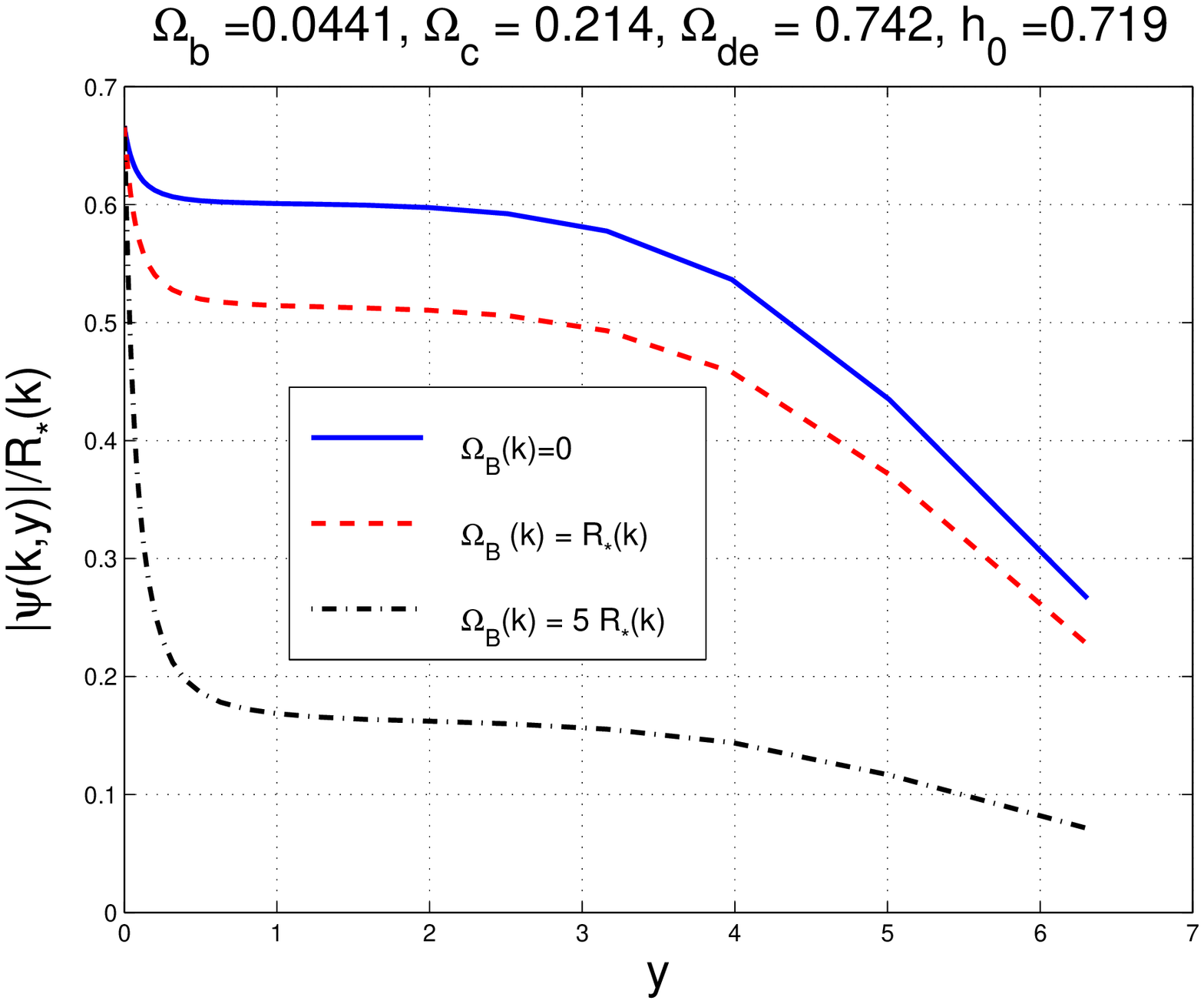}
\caption[a]{The evolution of ${\mathcal R}(k,y)$ in the case ${\mathcal R}_{*}(k) \to 0$ (plot at the left). The evolution of $\psi(k,y)/{\mathcal R}_{*}(k)$ 
for different values of the magnetic energy density.}
\label{FIGthree}      
\end{figure}

Always in Fig. \ref{FIGone} (plot at the right) the profile of $T_{{\mathcal R}}(y)$ is illustrated. 
As expected, around  $y_{\mathrm{i}}$ (i.e. deep in the radiation dominated epoch) $T_{{\mathcal R}}(y_{\mathrm{i}}) = 2/3$. 
Then, as $y$ increases, $T_{\mathcal R}(y)$ reaches a flat plateau for intermediate times around decoupling where 
$T_{{\mathcal R}}(y) \simeq 3/5$ (see Fig. \ref{FIGone}, plot at the right). Absent any late dominance of dark energy, $T_{{\mathcal R}}(y) \to 3/5$ exactly. The  (late) deviation from the CDM asymptote (i.e. $3/5$) depends upon the specific value of $w_{\mathrm{de}}$ and determines the ISW
contribution.
In Fig. \ref{FIGtwo} the function $T_{\mathrm{B}}(y)$ is illustrated 
both on a linear time scale (plot at the left) and using a logarithmic 
time scale (plot at the right) which better represents the early evolution 
before matter radiation equality and across decoupling. 
The arrow appearing in the 
left plot of Fig. \ref{FIGtwo} marks, approximately, the present 
time, i.e. when $a(y_{0}) \simeq a_{0} =1$. In short the shape 
of $T_{\mathrm{B}}(y)$ can be understood as follows:
\begin{itemize}
\item{} unlike the 
case of $T_{\mathcal R}(y)$ (which reaches a constant asymptote 
deep in the radiation-dominated epoch, i.e. for  $y\ll 1$), $T_{\mathrm{B}}(y) \to 0$ in the limit $y\to 0$;
\item{} in the CDM paradigm $T_{\mathrm{B}}(y)$ would reach the asymptotic value $ 3 R_{\gamma}/20 \simeq 0.089$;
\item{} both in the $\Lambda$CDM  and in the {\it w}CDM 
cases the would be asymptotes turn into an intermediate plateau 
(see Fig. \ref{FIGtwo}, in particular plot at the right);
\item{} the intermediate plateau is accurately estimated by $3 R_{\gamma}/20$ while the presence of a dark energy background is responsible  for the deviation of $T_{\mathrm{B}}(y)$ from the CDM asymptote.
\end{itemize}
The vanishing of $T_{\mathrm{B}}(y)$  for $y\to 0$  
is a consequence of the evolution of ${\mathcal R}(k,y)$ and, in particular, of Eqs. (\ref{LSS2}) and (\ref{LSS3}). 
The same observation implies that whenever ${\mathcal R}_{*}(k)\to 0$, ${\mathcal R}(k,y) \to 0$ for $y\to 0$:
such an occurrence is verified  (and illustrated) in the left plot of Fig. \ref{FIGthree} where ${\mathcal R}(k,y)$
is shown to vanish in the limit   ${\mathcal R}_{*}(k)\to 0$.

It can be speculated that the the contribution 
coming from $\Omega_{\mathrm{B}}(k)$ and ${\mathcal R}_{*}(k)$ might interfere destructively leading to an overall suppression of the large-scale contribution and, in particular, of the quadrupole. In Fig. \ref{FIGthree} (plot at the right) this possibility seems 
excluded unless $\Omega_{\mathrm{B}}(k) \simeq {\mathcal R}_{*}(k)$. 
But such an extreme value would jeopardize the agreement of the theory with the acoustic region. 

\begin{figure}[!ht]
\centering
\includegraphics[height=6cm]{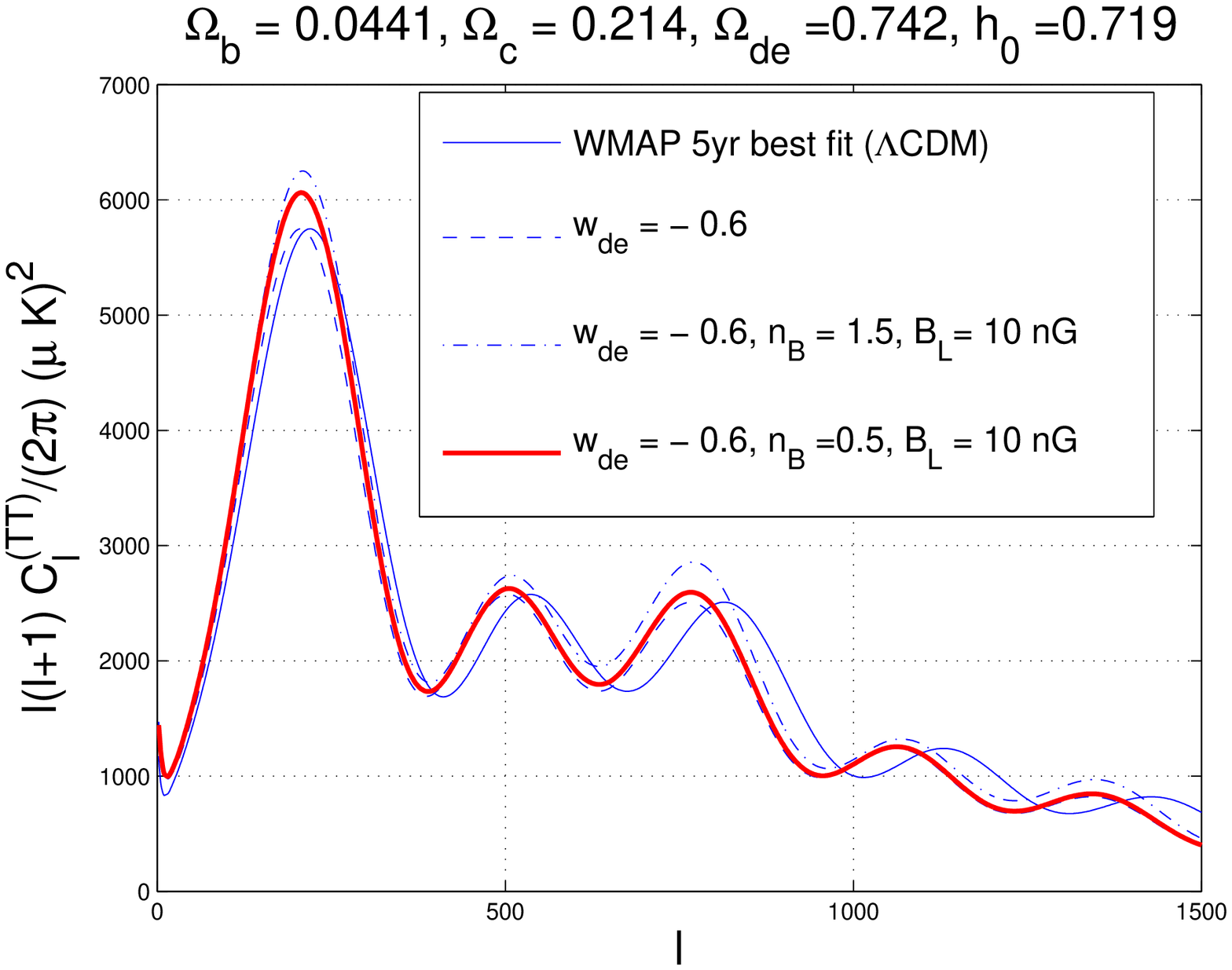}
\includegraphics[height=6cm]{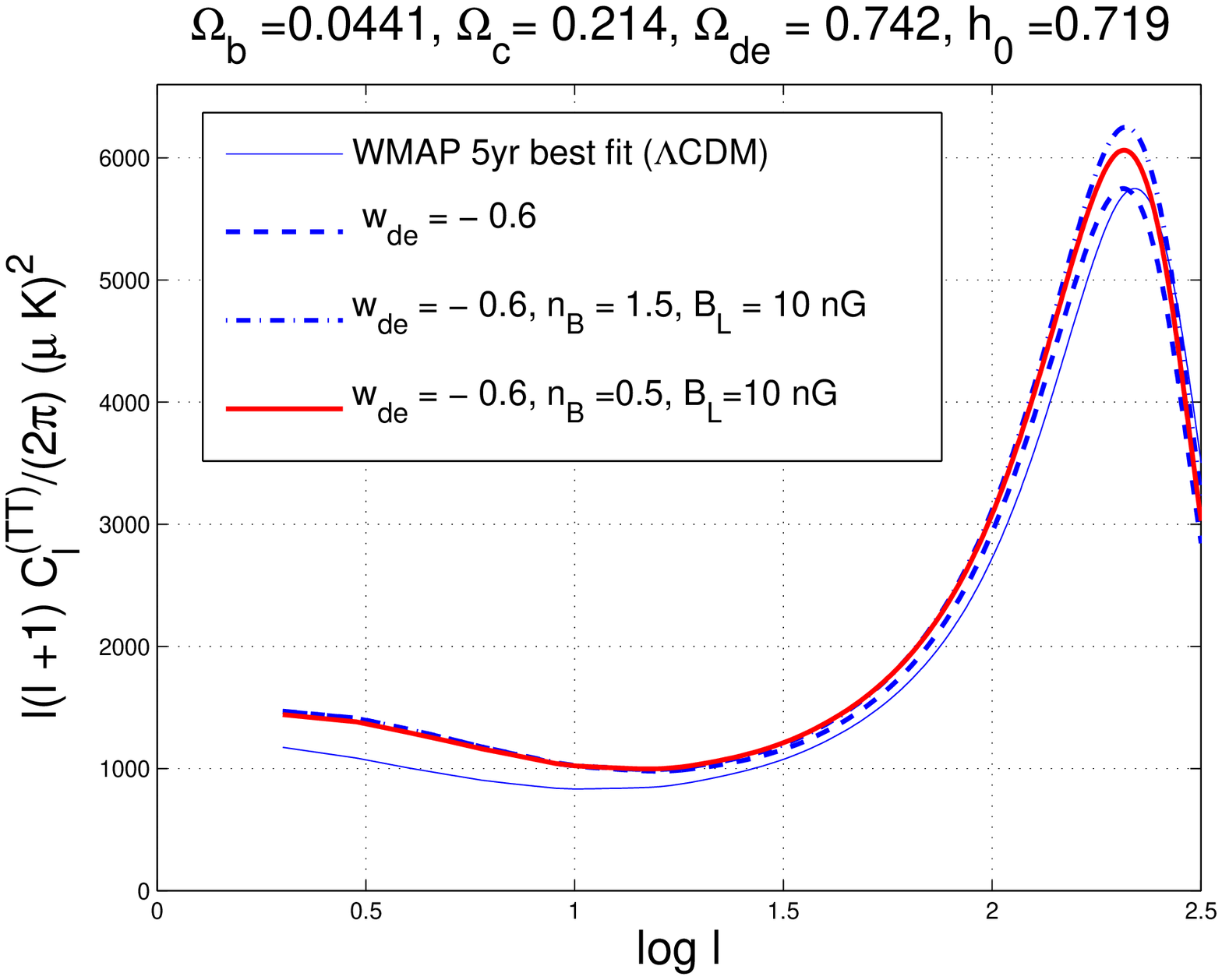}
\includegraphics[height=6cm]{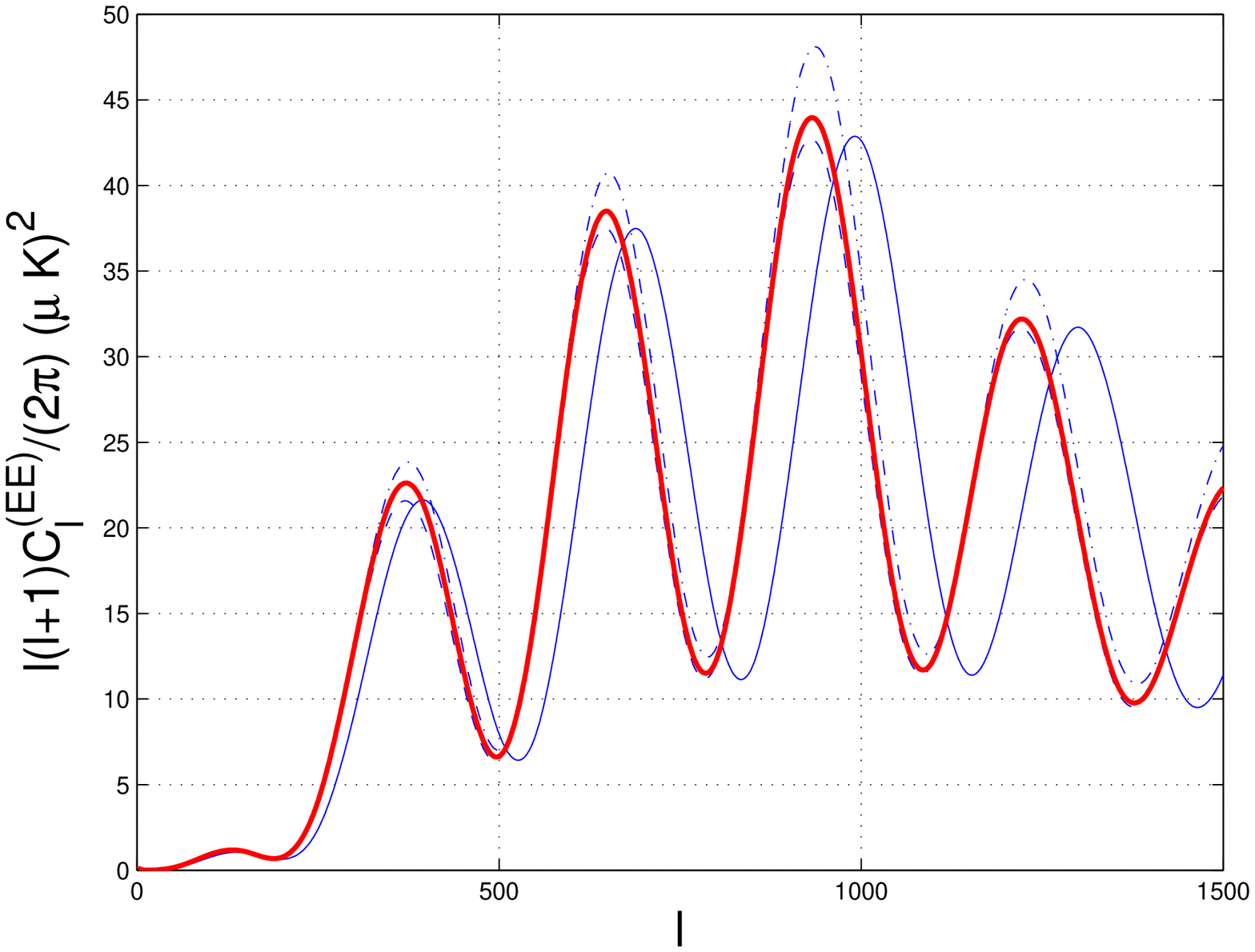}
\includegraphics[height=6cm]{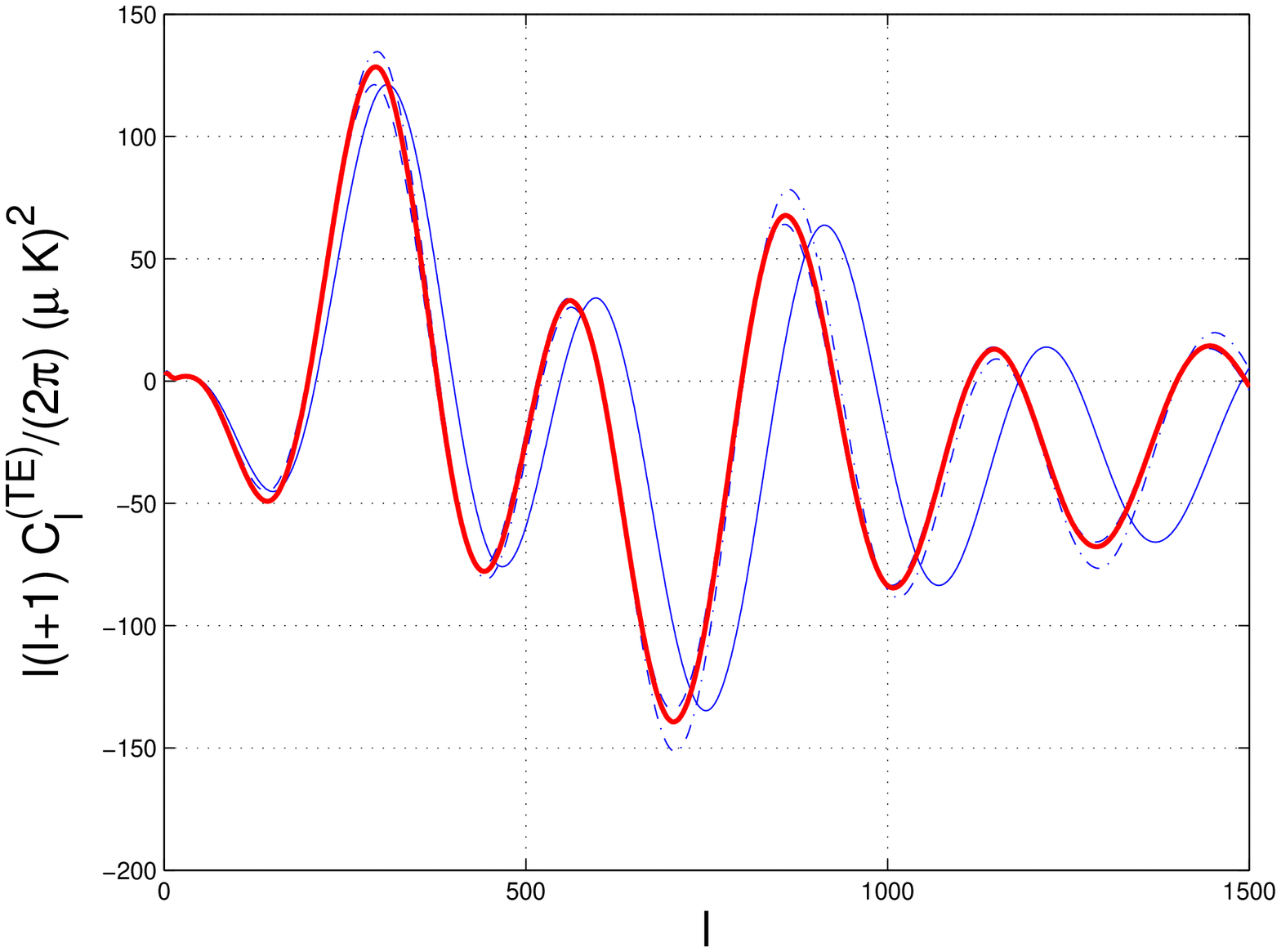}
\caption[a]{The angular power spectra of the CMB observables in the light 
of the {\it w}CDM scenario when large-scale magnetic fields are included. In all the plots 
the barotropic index of the (fluctuating) dark energy background has been 
fixed to $w_{\mathrm{de}} =-0.6$.}
\label{FIGfour}      
\end{figure}

To scrutinize this point in a model-independent perspective, 
consider the condition $\Omega_{\mathrm{B}}(k) \simeq {\mathcal R}_{*}(k)$ in 
loose terms and postulate that the amplitude of curvature perturbations matches approximately 
the amplitude of the power spectrum of the magnetic energy density. Neglecting, for simplicity, the dependence upon 
the spectral indices we get 
a typical amplitude for the comoving amplitude of the magnetic field intensity 
which is of the order of $B_{\mathrm{L}} \simeq 22.68 \, \mathrm{nG}$ where 
it has been assumed that ${\mathcal A}_{{\mathcal R}} = 2.4\times 10^{-9}$, $k_{\mathrm{L}} = 
1/\mathrm{Mpc}$ and $k_{\mathrm{p}} = 0.002 \, \mathrm{Mpc}^{-1}$.
The putative value of about $23$ nG is indeed quite large and it would totally disrupt the structure of the acoustic oscillations. 
This aspect can be appreciated from Fig. \ref{FIGfour} where a magnetic field of $10$ nG already jeopardizes the observed features of the temperature autocorrelations \footnote{The threshold field of about $23$ nG is obtained 
by requiring that ${\mathcal P}_{\Omega}(k) \simeq {\mathcal P}_{{\mathcal R}}(k)$. The latter condition implies, from Eq. (\ref{SW1a}), that ${\mathcal E}_{\mathrm{B}} \simeq {\mathcal A}_{{\mathcal R}}$. Since ${\mathcal E}_{\mathrm{B}} \propto 
\Omega_{\mathrm{B L}}^2$ and $\Omega_{\mathrm{BL}} = 9.536  \times 10^{-8} (B_{\mathrm{L}}/\mathrm{nG})^2$, we do get 
$B_{\mathrm{L}} \sim 22.6 \mathrm{nG}$. This argument holds rigorously in the scale-invariant limit and it can be easily 
generalized to the case $n_{\mathrm{B}} \neq 1$.}. 

In the minimal scenario \cite{max1,max6}, the parameter space of the magnetized CMB anisotropies can be 
described in terms of the magnetic spectral index $n_{\mathrm{B}}$ and the regularized magnetic field 
amplitude $B_{\mathrm{L}}$.  These two quantities and their relations with $\Omega_{\mathrm{B}}$ and $\sigma_{\mathrm{B}}$ (introduced in section \ref{sec2}) have been thoroughly discussed in previous papers \cite{max1,max2,max3}. 
The basic advantages of such a parametrization have been illustrated at the end of section \ref{sec3} (see, in  particular, Eq. (\ref{PS2}) and Eq. (\ref{PS1}) for the analog conventions employed in the definition of   the power spectra of curvature perturbations).
In Fig. \ref{FIGfour} the parameters are fixed to the best fit of the WMAP 5yr data alone 
in the light of the $\Lambda$CDM paradigm (corresponding to $w_{\mathrm{de}} = -1$). 
Always in Fig. \ref{FIGfour} the barotropic index of the dark energy is increased 
from $-1$ to $-0.6$ while the magnetized background is switched on. The 
magnetic field intensity has been chosen to be rather large (i.e. ${\mathcal O}(10\, \mathrm{nG})$).
Examples of blue (i.e. $n_{\mathrm{B}} =1.5$) 
and  red (i.e. $n_{\mathrm{B}} =0.5$) spectral indices are illustrated. From
the TT correlations \footnote{The TT power spectra denote, as usual, the autocorrelations of the temperature. 
The TE power spectra are the cross-correlations between temperature and poalrization. The EE power spectra 
denote the polarization autocorrelations. Within the conventions employed in the present paper, 
the definitions of the TT, TE and EE power spectra in terms of the solutions 
of the heat transfer equations can be found in \cite{max2} (see also \cite{max3} and references therein). 
The B-mode polarization induced by Faraday rotation \cite{far1,far2} will not be 
specifically addressed in this context. It has been actually shown in \cite{far1} (second paper of the list) that the present data on 
the B-mode autocorrelations only allow for upper bounds on the Faraday-induced B-mode. An exception 
to this statement are the QUAD data \cite{quad1,quad2,quad3} but, at the moment, its not clear to what extent the results are contaminated by systematics \cite{quad2}.} in semilogarithmic coordinates (top-right plot) as well as from the TE and EE correlations the general interplay of the dark energy and magnetized backgrounds is more clear and can be summarized as follows:
\begin{itemize}
\item{} in the large-scale domain (i.e. $\ell < 50$) both the dark energy and the magnetized contribution augment the power 
in the TT angular power spectra;
\item{} possible interference effects leading to an overall suppression of the lower multipoles are excluded
if the acoustic oscillations are to be reproduced correctly; this statement holds in the absence of any 
non-adiabatic contribution (i.e. $\delta p_{\mathrm{nad}} =0$ in Eq. (\ref{LSS2})) but it might change 
if $\delta p_{\mathrm{nad}}\neq 0$;
\item{} the interplay between the magnetized and the dark energy backgrounds leads to a combined 
distortion of the peaks in the TT, TE and EE angular power spectra; since the level of distortion 
increases with the multipole the higher peaks look also shifted in comparison with 
the patterns exhibited by the underlying $\Lambda$CDM model with the same 
parameters and in the absence of any magnetic field.
\end{itemize}
The value $w_{\mathrm{de}} =-0.6$ chosen in Fig. \ref{FIGfour} is rather extreme insofar as it is excluded by 
current fits to the WMAP 5yr data (either alone or in combination with other data sets). It is however useful 
highlight some general trend induced by the dark energy fluctuations. 

Figures \ref{FIGone}, \ref{FIGtwo} and \ref{FIGthree}  do also contain examples where $w_{\mathrm{de}} < -1$.  When analyzing the cosmological data sets in the light of a fluctuating  dark energy background 
it can happen that the central value of the barotropic index maximizing the 
likelihood for a specific data set gets smaller than $-1$. In the latter case future singularities are expected (see for instance \cite{rip1}).  In what follows the cases $w_{\mathrm{de}} <-1$ 
will be considered for completeness. Indeed the central values of the barotropic index is still debatable and does  depend upon the data sets which are combined in the analysis. For the purposes of the present paper, the situation 
is summarized in three forthcoming paragraphs.

By analyzing the WMAP 5yr data alone in the light of the $w$CDM model we do get $w_{\mathrm{de}} = -1.06_{-0.42}^{0.41}$ while the remaining parameters are determined to be
\begin{eqnarray}
&&( \Omega_{\mathrm{b}}, \, \Omega_{\mathrm{c}}, \Omega_{\mathrm{de}},\, h_{0},\,n_{\mathrm{s}},\, \epsilon_{\mathrm{re}}) \equiv 
\nonumber\\
&&(0.046^{0.018}_{-0.018},0.221_{-0.082}^{0.087}, 0.733^{0.10}_{-0.11},0.74^{0.15}_{-0.14},0.963^{0.016}_{-0.016},0.086^{0.017}_{-0.016});
\label{Par4}
\end{eqnarray}
in the case of the parameters of Eq. (\ref{Par4}) the corresponding value of the 
normalization of the curvature perturbations is still ${\mathcal A}_{{\mathcal R}} = 2.41 \times 10^{-9}$ 
at the pivot scale of $k_{\mathrm{p}} =0.002\, \mathrm{Mpc}^{-1}$; the value of ${\mathcal A}_{{\mathcal R}}$ 
does not change in comparison with the corresponding value determined in connection with the parameters 
of Eq. (\ref{Par1}).

By combining the WMAP 5yr data with the data stemming from the analysis of the baryon acoustic oscillations (BAO in what follows) \cite{bao1,bao2,bao3} the preferred range of values of $w_{\mathrm{de}}$  becomes $-1.15_{-0.22}^{0.21}$; the remaining 
parameters turn out to be \cite{WMAP51,WMAP52,WMAP53}: 
\begin{eqnarray}
&&( \Omega_{\mathrm{b}}, \, \Omega_{\mathrm{c}}, \Omega_{\mathrm{de}},\, h_{0},\,n_{\mathrm{s}},\, \epsilon_{\mathrm{re}}) \equiv 
\nonumber\\
&&(0.0419^{0.0055}_{-0.0056}, 0.213^{0.021}_{-0.021},\, 0.745^{0.026}_{-0.026},0.739^{0.047}_{-0.048}, 0.958_{-0.015}^{0.015},\,0.083^{0.016}_{-0.016});
\label{Par5}
\end{eqnarray}
the value of ${\mathcal A}_{{\mathcal R}}$ determined in connection with the data 
of Eq. (\ref{Par5}) increases in comparison with the corresponding values holding 
in the case of Eqs. (\ref{Par1}) and (\ref{Par4}) and it is given by 
${\mathcal A}_{{\mathcal R}} = 2.48 \times 10^{-9}$.

Finally, by combining the WMAP 5yr data with the BAO and with all the supernovae \cite{sn1,sn2,sn3}
the central value of $w_{\mathrm{de}}$ maximizing the likelihood gets above $-1$ and it is 
$w_{\mathrm{de}} = -0.972_{-0.060}^{0.061}$; the remaining 
parameters are given, in this case, by \cite{WMAP51,WMAP52,WMAP53}
\begin{eqnarray}
&&( \Omega_{\mathrm{b}}, \, \Omega_{\mathrm{c}}, \Omega_{\mathrm{de}},\, h_{0},\,n_{\mathrm{s}},\, \epsilon_{\mathrm{re}}) \equiv 
\nonumber\\
&& (0.0467_{-0.0018}^{0.0018}, 0.231_{- 0.014}^{0.014}, 0.722_{-0.015}^{0.015},0.697_{-0.014}^{0.014}, 
0.962_{-0.014}^{0.014},0.085_{-0.016}^{0.016});
\label{Par6}
\end{eqnarray}
again the value of ${\mathcal A}_{{\mathcal R}}$ determined in connection with the data 
of Eq. (\ref{Par5}) diminishes in comparison with the corresponding values holding in the case of Eqs. (\ref{Par5}) and it is ${\mathcal A}_{{\mathcal R}} = 2.43 \times 10^{-9}$.

\begin{figure}[!ht]
\centering
\includegraphics[height=6cm]{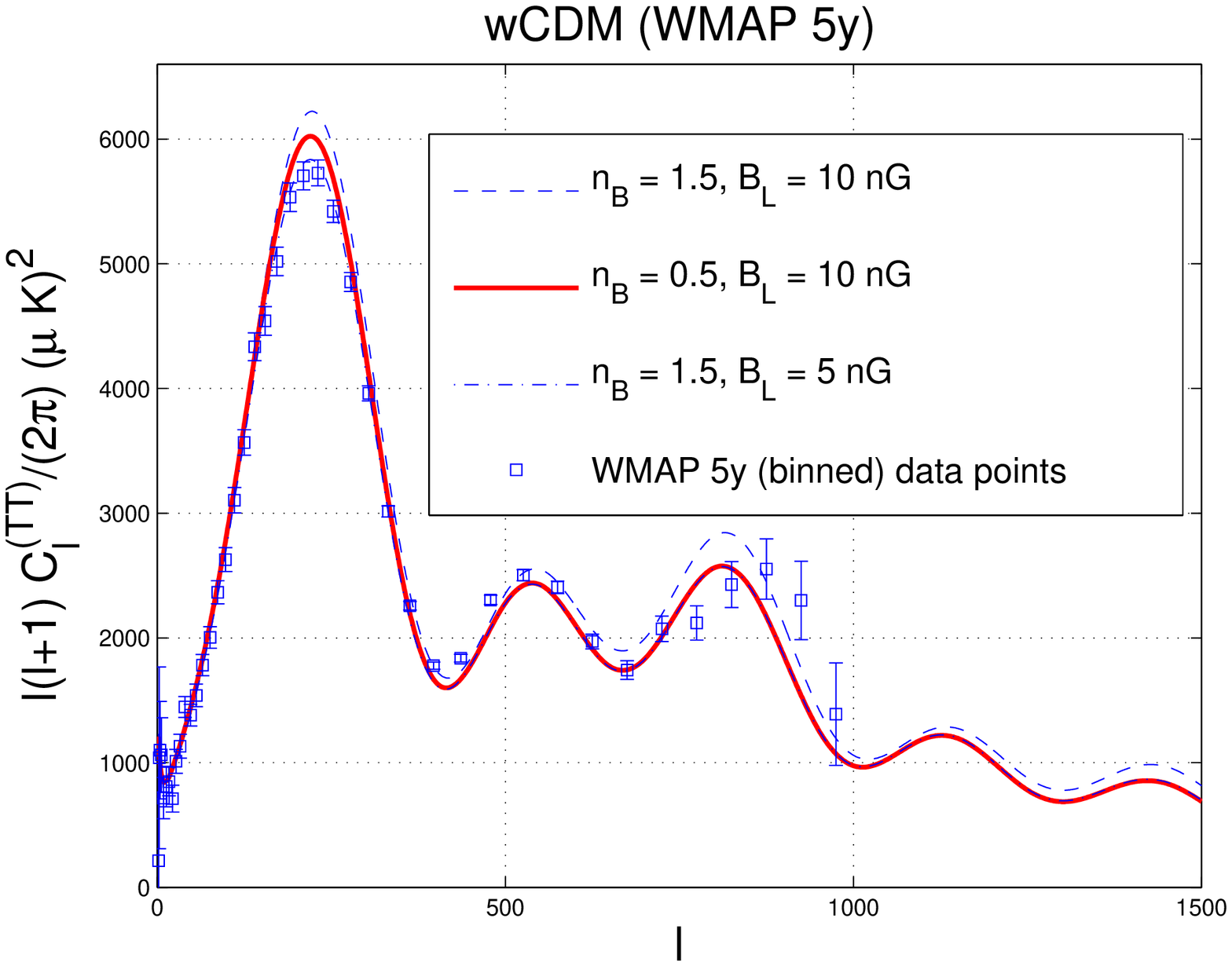}
\includegraphics[height=6cm]{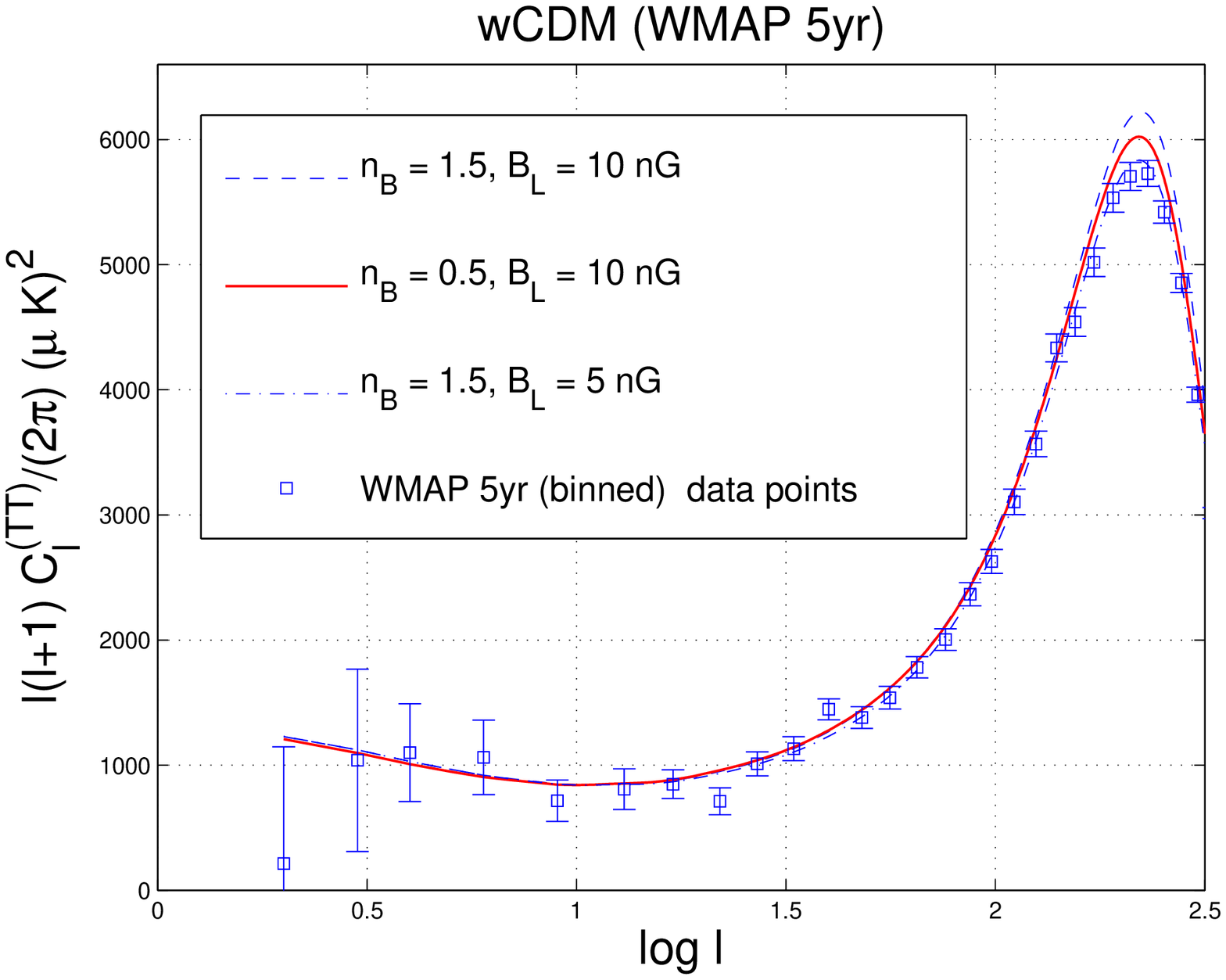}
\includegraphics[height=6cm]{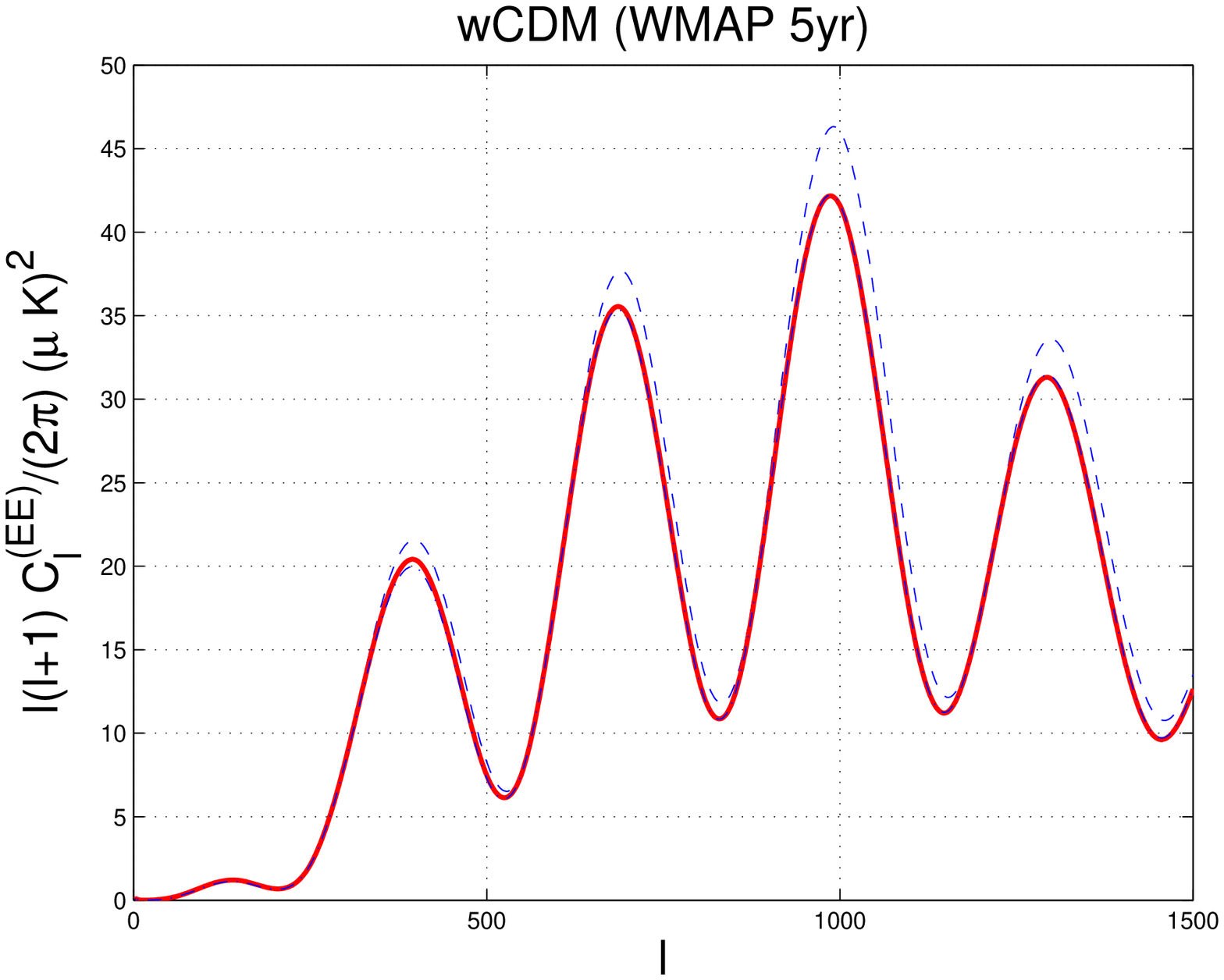}
\includegraphics[height=6cm]{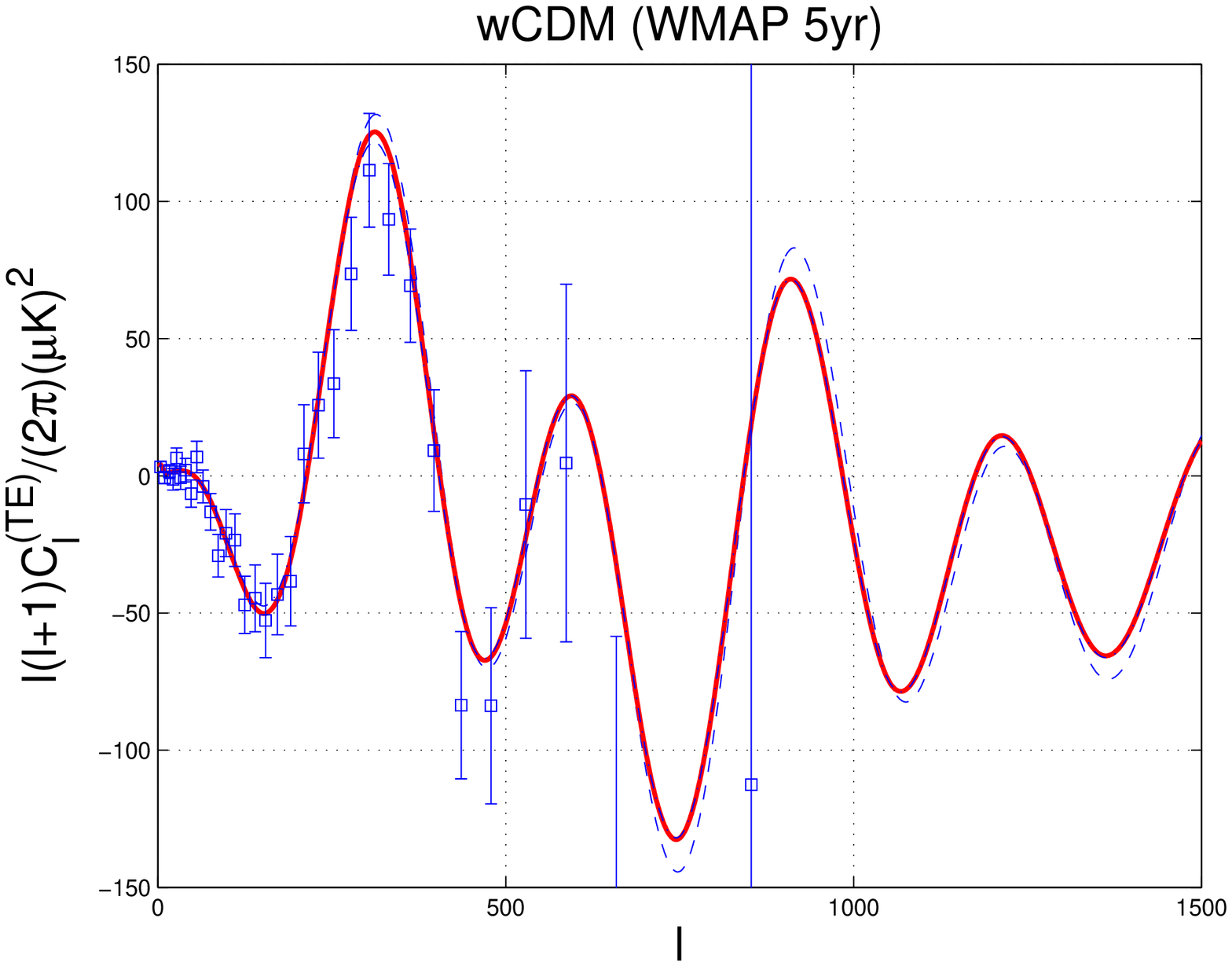}
\caption[a]{Large-scale magnetic fields are included together with dark energy fluctuations in the case of the {\it w}CDM 
scenario (see Eq. (\ref{Par4}) for the list of the parameters). In the TT and TE angular power spectra the binned points of the WMAP 5yr data are also reported. In all plots the fluctuating dark energy contribution has been included and, consistently with Eq. (\ref{Par4}), $w_{\mathrm{de}} = -1.06$.}
\label{FIGfive}      
\end{figure}

\begin{figure}[!ht]
\centering
\includegraphics[height=6cm]{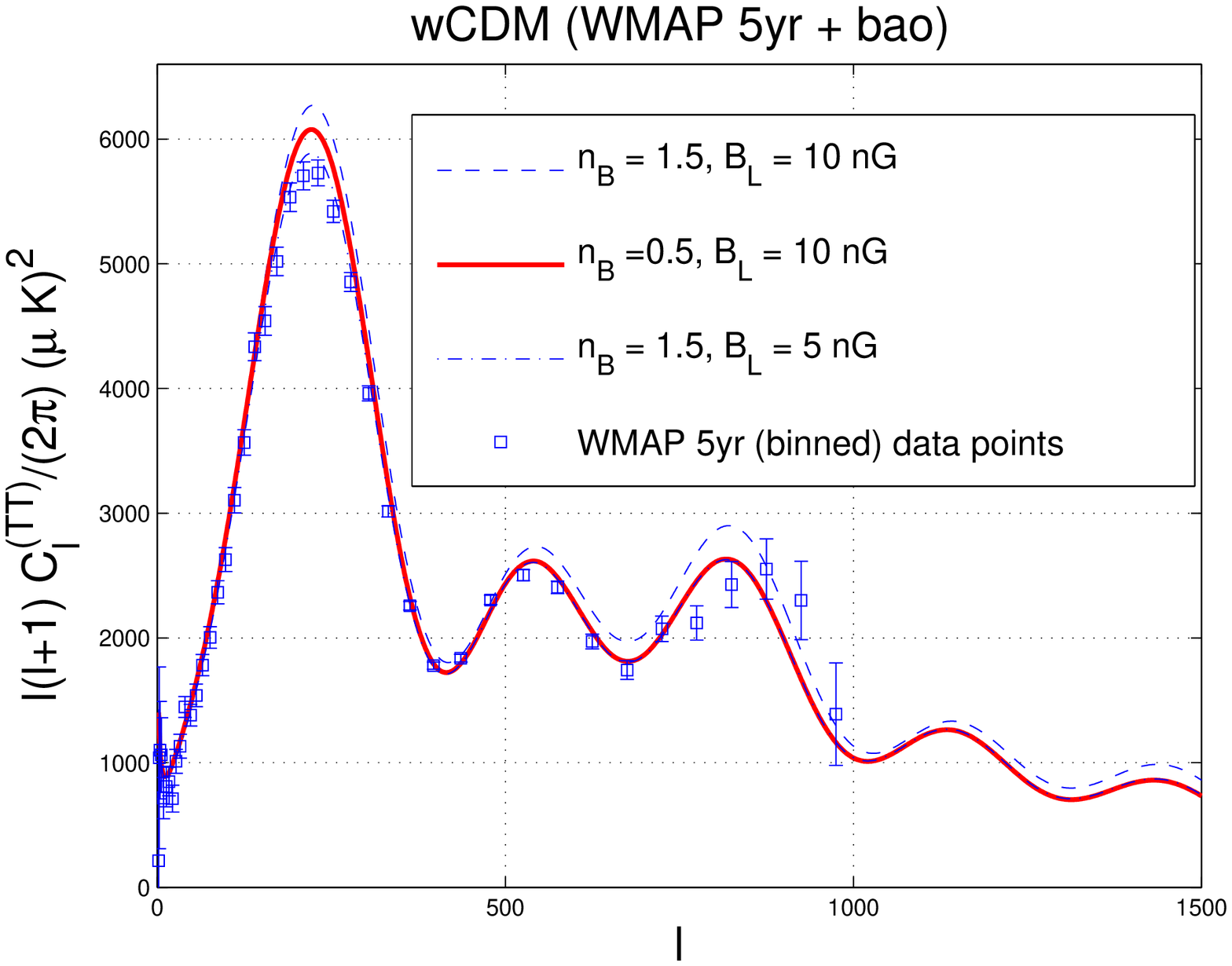}
\includegraphics[height=6cm]{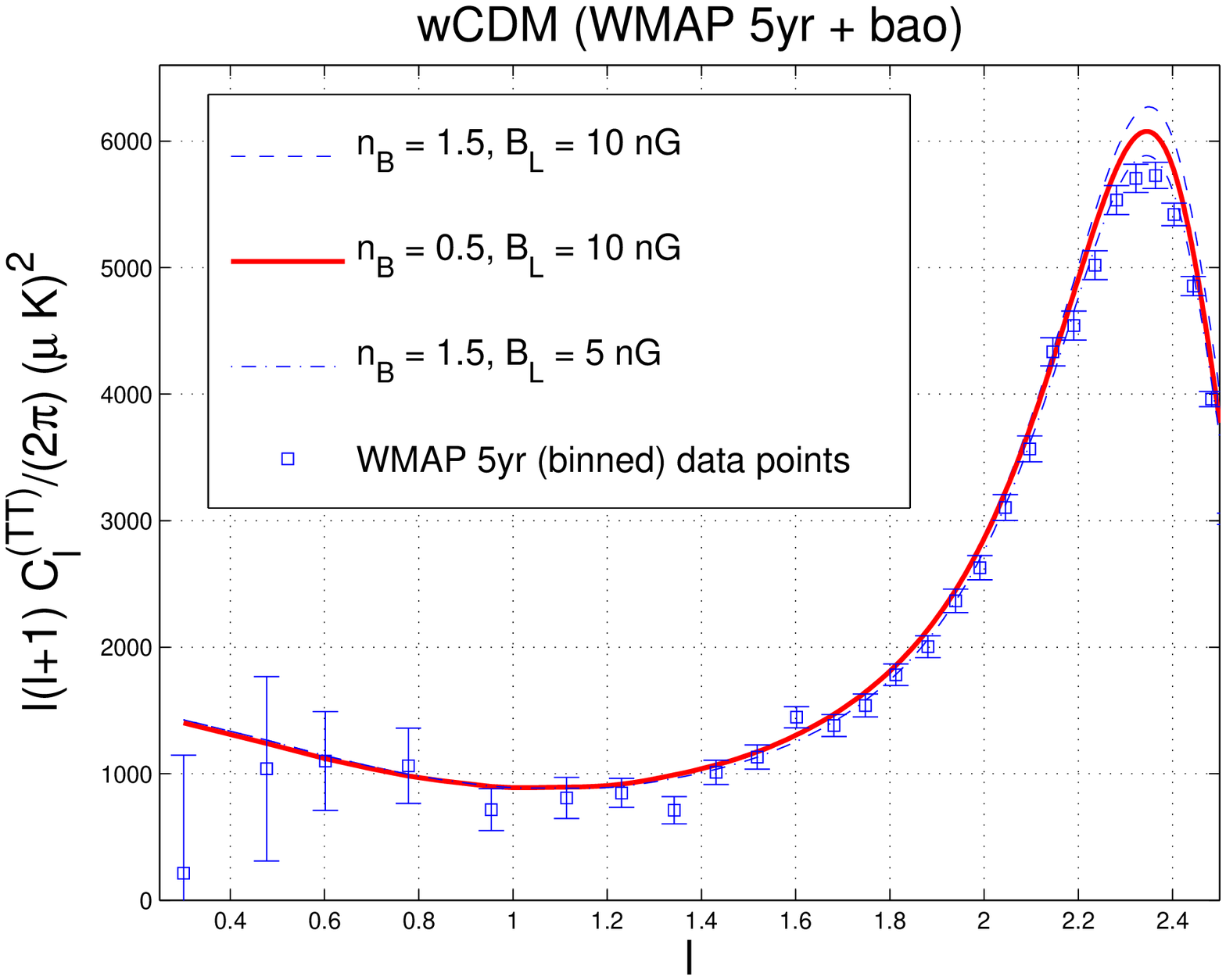}
\includegraphics[height=6cm]{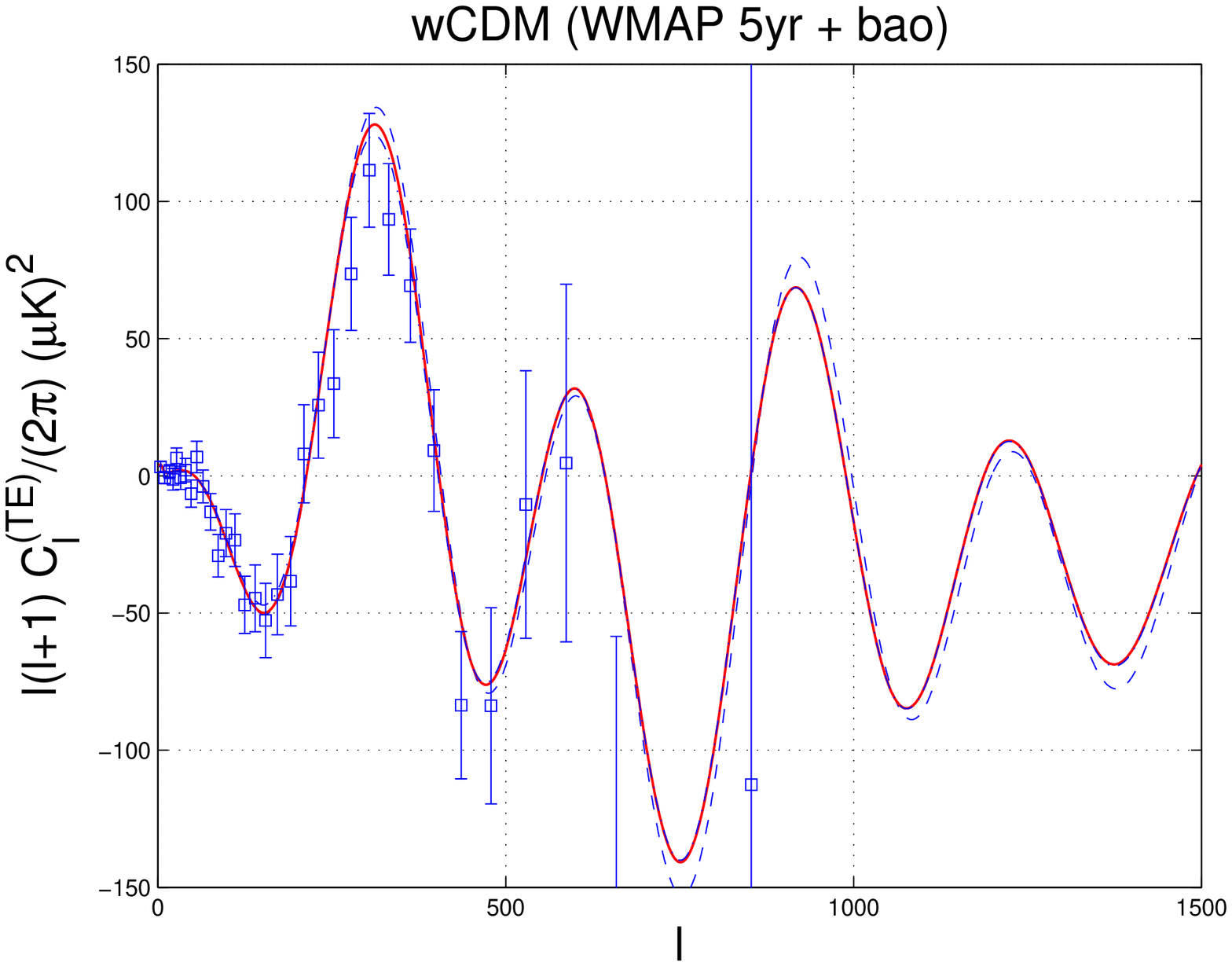}
\includegraphics[height=6cm]{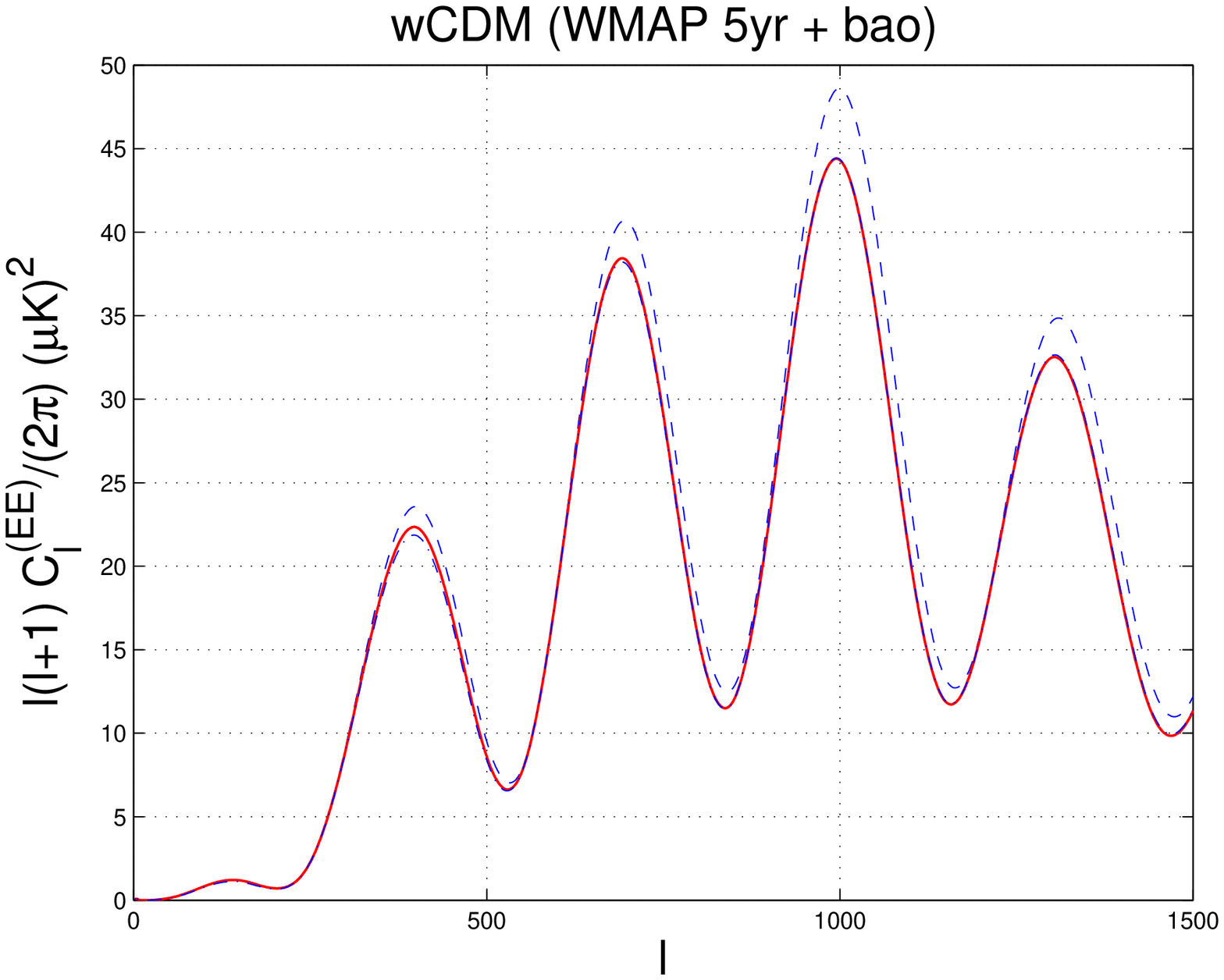}
\caption[a]{Large-scale magnetic fields are included for the  {\it w}CDM best fit when 
the WMAP 5yr data are combined with the ones stemming from the baryon acoustic oscillations. As in Fig. \ref{FIGfive} the fluctuations in the dark 
energy background have been taken into account; $w_{{\mathrm{de}}} = -1.15$ (as it must be from Eq. (\ref{Par5})).}
\label{FIGsix}      
\end{figure}

\begin{figure}[!ht]
\centering
\includegraphics[height=6cm]{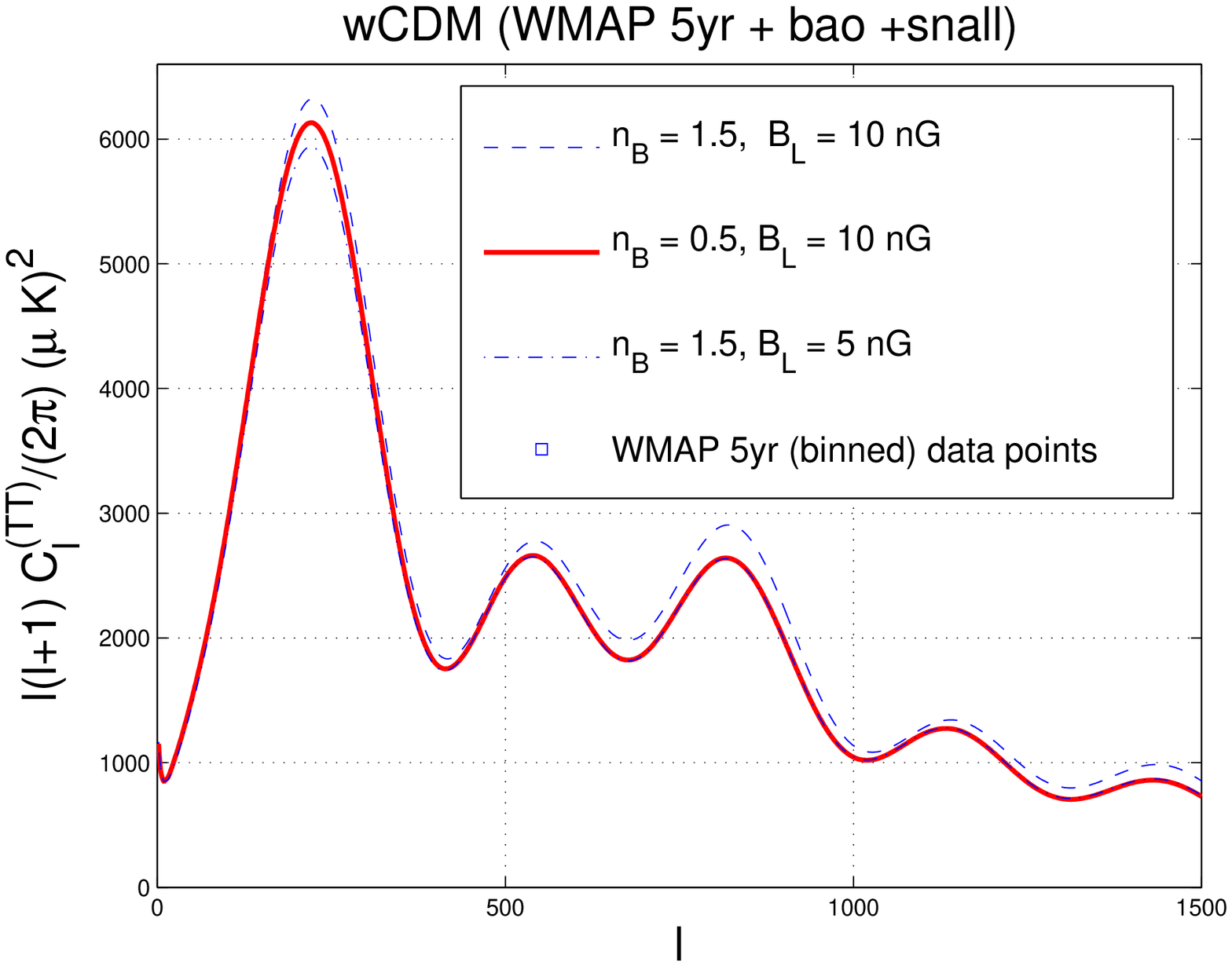}
\includegraphics[height=6cm]{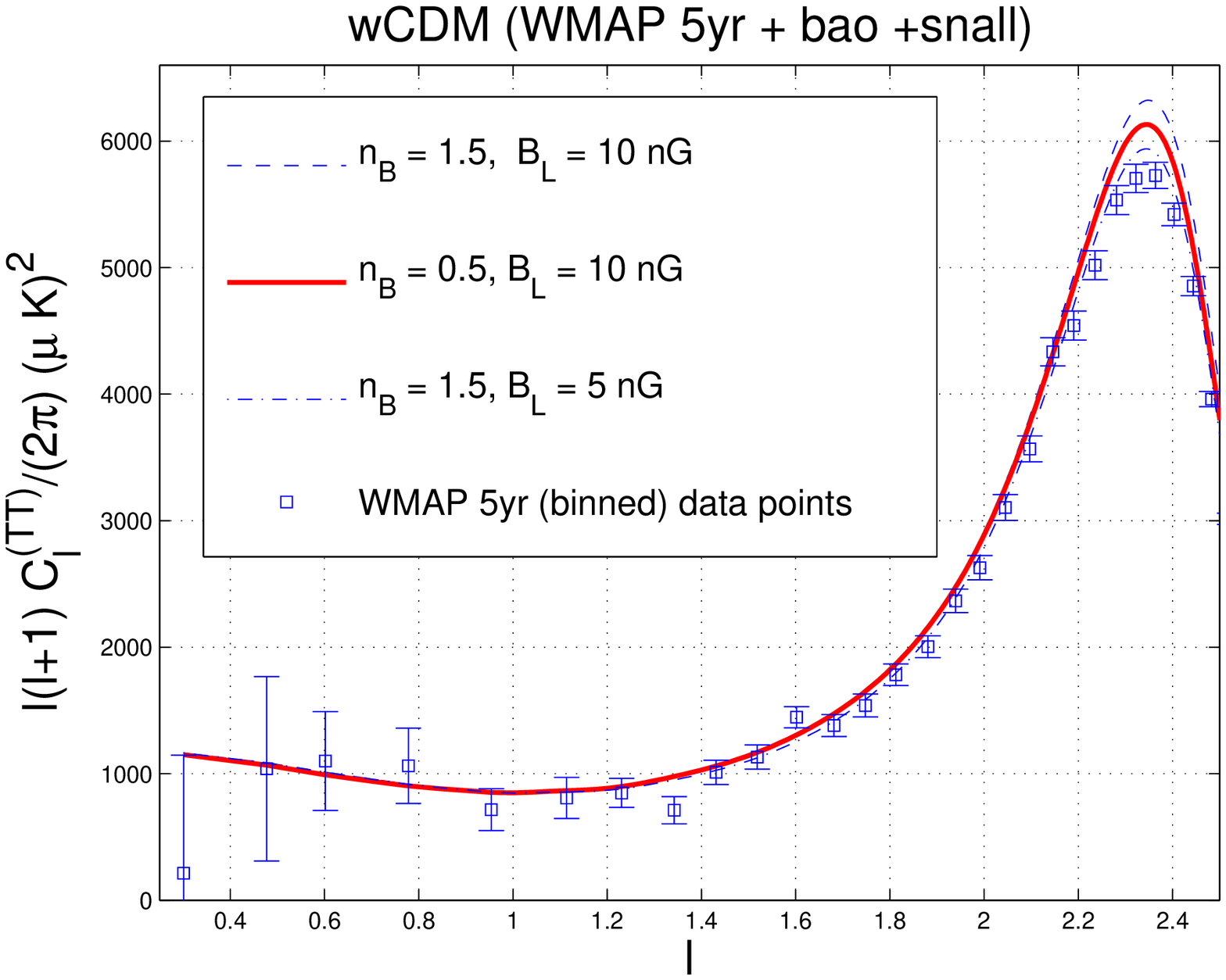}
\includegraphics[height=6cm]{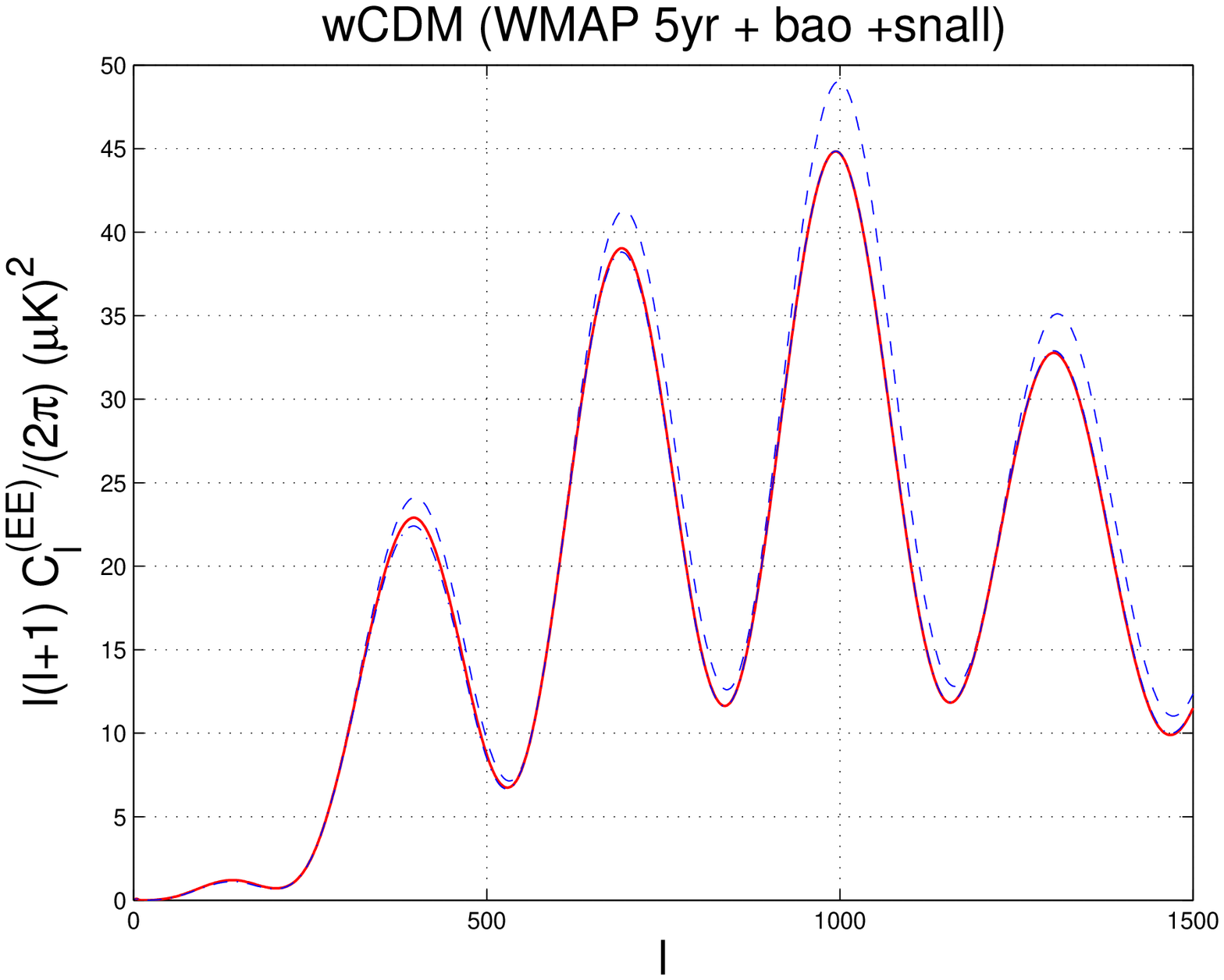}
\includegraphics[height=6cm]{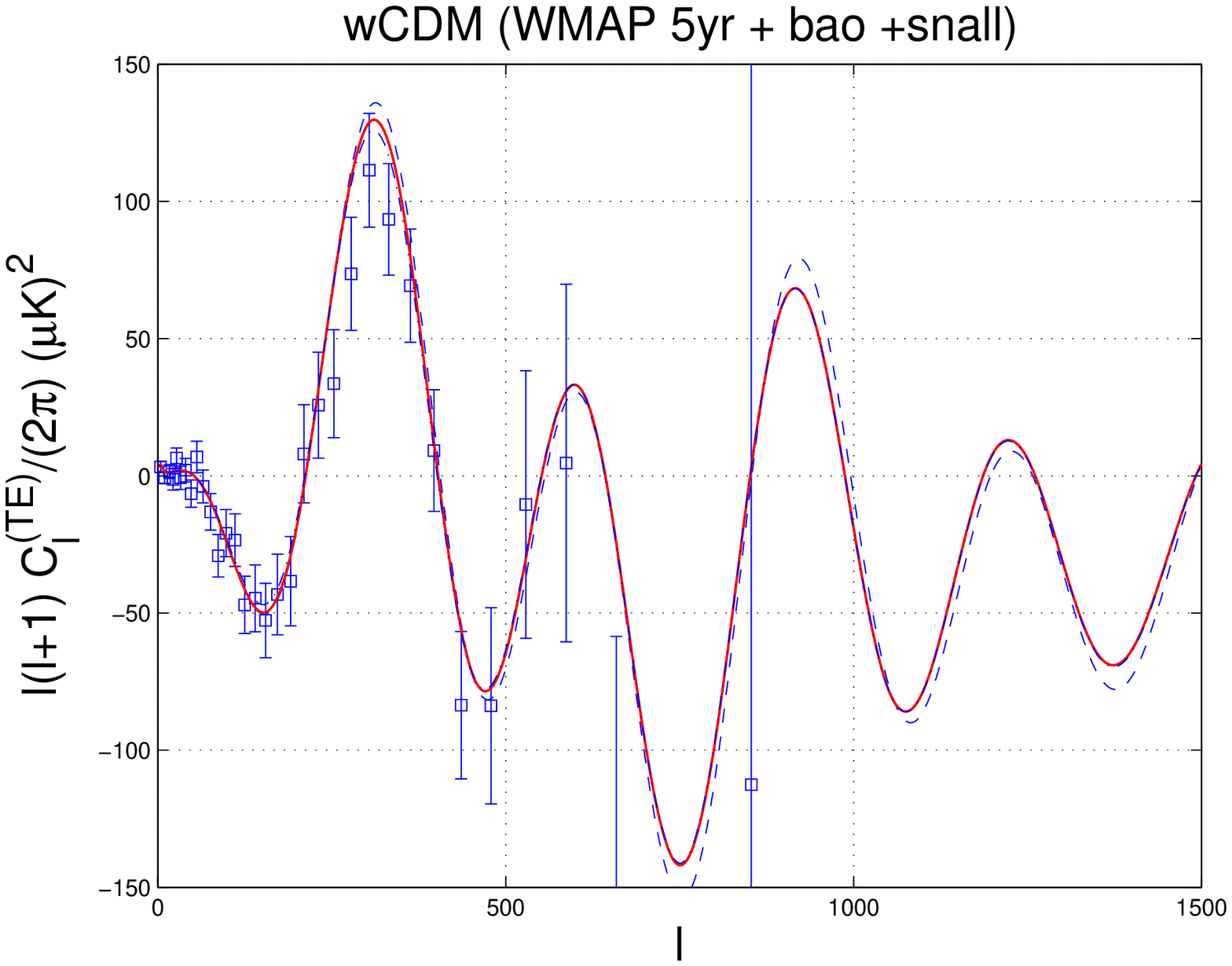}
\caption[a]{Large-scale magnetic fields are included for the  {\it w}CDM best fit when the WMAP 5yr data are combined with the  baryon acoustic oscillations and with all supernova data. As in Figs. \ref{FIGfive} and \ref{FIGsix} the fluctuations in the dark 
energy background have been taken into account; $w_{{\mathrm{de}}} = -0.972$ (in agreement with Eq. (\ref{Par6}))}
\label{FIGseven}      
\end{figure}

The data reported in Eqs. (\ref{Par4}), (\ref{Par5}) and (\ref{Par6}) 
will be referred  to as, respectively,  WMAP 5yr alone, WMAP 5yr + bao, and WMAP 5yr + bao + snall.
Bearing in mind the latter shorthand notations it is now appropriate to scrutinize how the simultaneous presence of large-scale magnetic fields and of the dark energy fluctuations affects the temperature and the polarization observables. This aspect is illustrated in three of the forthcoming figures, i.e. in Figs. \ref{FIGfive}, \ref{FIGsix} and \ref{FIGseven} where the magnetized background is studied together with the dark energy background for the reference models discussed in Eqs. (\ref{Par4}), (\ref{Par5}) and (\ref{Par6}).
In Fig. \ref{FIGfive} the parameters of the {\it w}CDM model have been chosen as in Eq. (\ref{Par4}) with $w_{\mathrm{de}} = -1.06$ corresponding to the best fit in the light of the WMAP 5yr data points. It is 
superficially clear that the string of parameters of Eq. (\ref{Par1})  does not 
differ, qualitatively, from the ones of Eq. (\ref{Par4}).  As expected, the inclusion of the magnetized contribution distorts the higher peaks. Three representative values of the parameter space of the magnetized {\it w}CDM scenario are illustrated:
\begin{itemize}
\item{({\it i})} the curves with $n_{\mathrm{B}} =1.5$ and $B_{\mathrm{L}} =10$ nG are representative of the region with high magnetic field intensity and 
blue spectral index (i.e. $n_{\mathrm{B}} >1$);
\item{({\it ii})} the curves with $n_{\mathrm{B}} =0.5$ and $B_{\mathrm{L}} =10$ nG are representative of the region with high magnetic field intensity and 
red spectral index (i.e. $n_{\mathrm{B}} <1$);
\item{({\it iii})} the curves with $n_{\mathrm{B}} = 1.5$ and $B_{\mathrm{L}} =5$ nG are representative of the region with intermediate magnetic field intensities and blue spectral index.
\end{itemize}

The regions of the parameter space corresponding to $(i)$ and $(ii)$ are roughly incompatible 
with the observed angular power spectra as it can be easily argued 
by superimposing the curves to the binned data of the TT and TE correlations \cite{WMAP51,WMAP52,WMAP53}. Furthermore, the values of the parameters of regions $(i)$ 
and $(ii)$  lie in a portion of the parameter space which has 
been excluded by estimating the parameters of the magnetized background in the light of the magnetized extension of the standard 
$\Lambda$CDM scenario. Does the inclusion of a fluctuating dark energy background make the difference? The answer to this question will be the subject of the remaining considerations of this section.  Notice also that 
the choice of the parameters corresponding to $(iii)$ seems to be qualitatively compatible with the observations. One of the purposes 
of the following discussion will  be to make such a statement 
more quantitative.

As a last remark concerning Fig. \ref{FIGfive}, 
the value of the barotropic index of 
dark energy (i.e. $w_{\mathrm{de}} = -1.06$) is rather close to the $\Lambda$CDM determination and this occurrence 
excludes interference effects such as the ones observed in Fig. \ref{FIGfour} where $w_{\mathrm{de}}$ differs from $-1$ appreciably (i.e. $w_{\mathrm{de}}=-0.6$). 
In Fig. \ref{FIGsix} and \ref{FIGseven} the dark energy background and the other {\it w}CDM parameters are fixed, respectively, 
as in Eqs. (\ref{Par5}) and (\ref{Par6}). The considerations 
suggested by Fig. \ref{FIGfour} are confirmed, in some sense, also 
by Figs. \ref{FIGfive} and \ref{FIGsix}. Extreme values of 
the magnetic field intensity ${\mathcal O}(10\, \mathrm{nG})$ 
are incompatible with the observations both in the case of red and blue spectral indices. 

The joined two-dimensional marginalized contours 
for the various cosmological parameters identified already by the analyses of the WMAP 3yr data are ellipses with an approximate Gaussian dependence on the confidence level as they must be in the Gaussian approximation (see, e. g. \cite{WMAP3a}). The parameter  space of the magnetized CMB anisotropies can then be scanned in the presence of dark energy fluctuations 
by using the same technique discussed in \cite{max1,max2} and applied to the case of the  $\Lambda$CDM paradigm.
\begin{figure}[!ht]
\centering
\includegraphics[height=6cm]{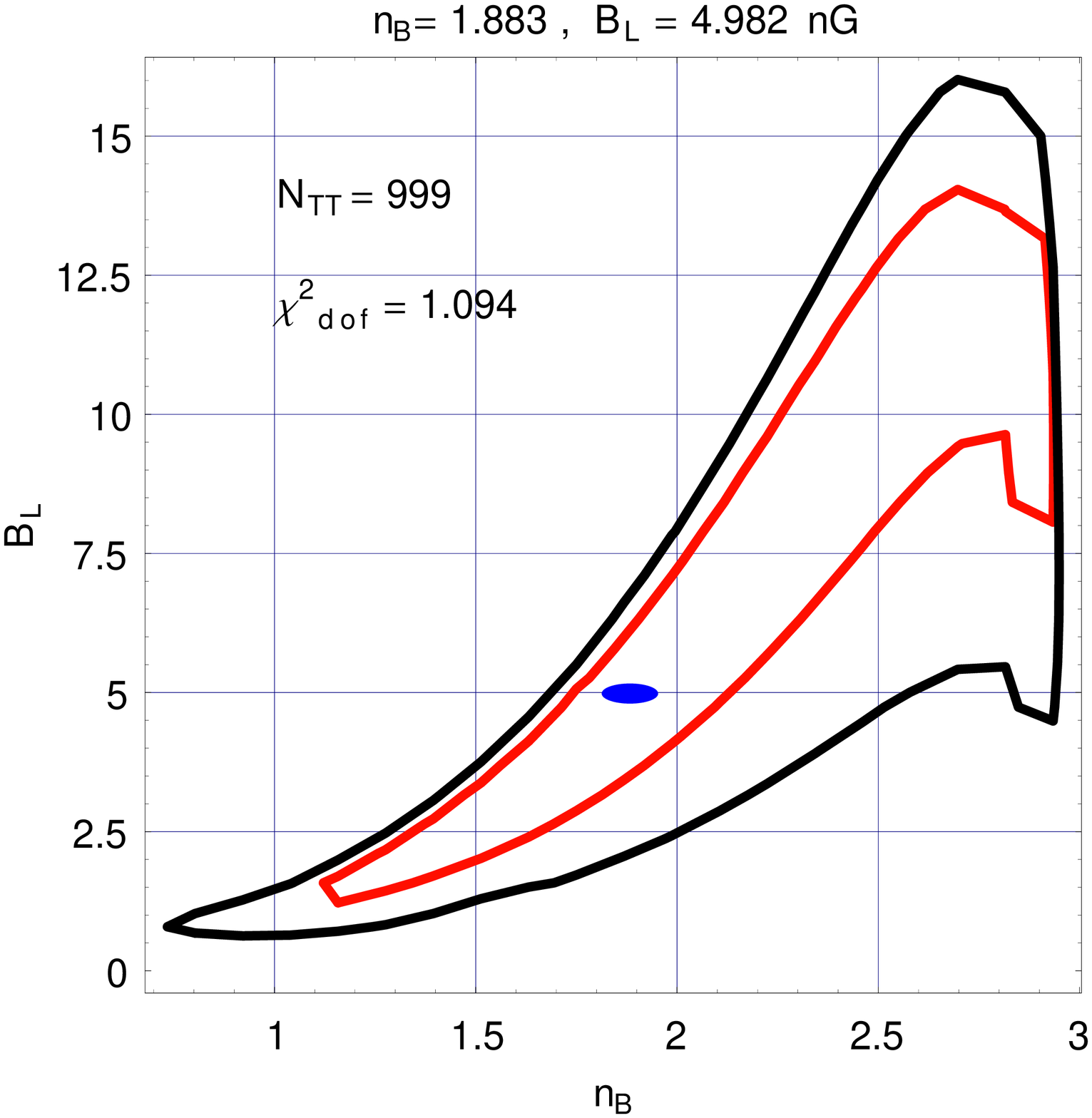}
\includegraphics[height=6.1cm]{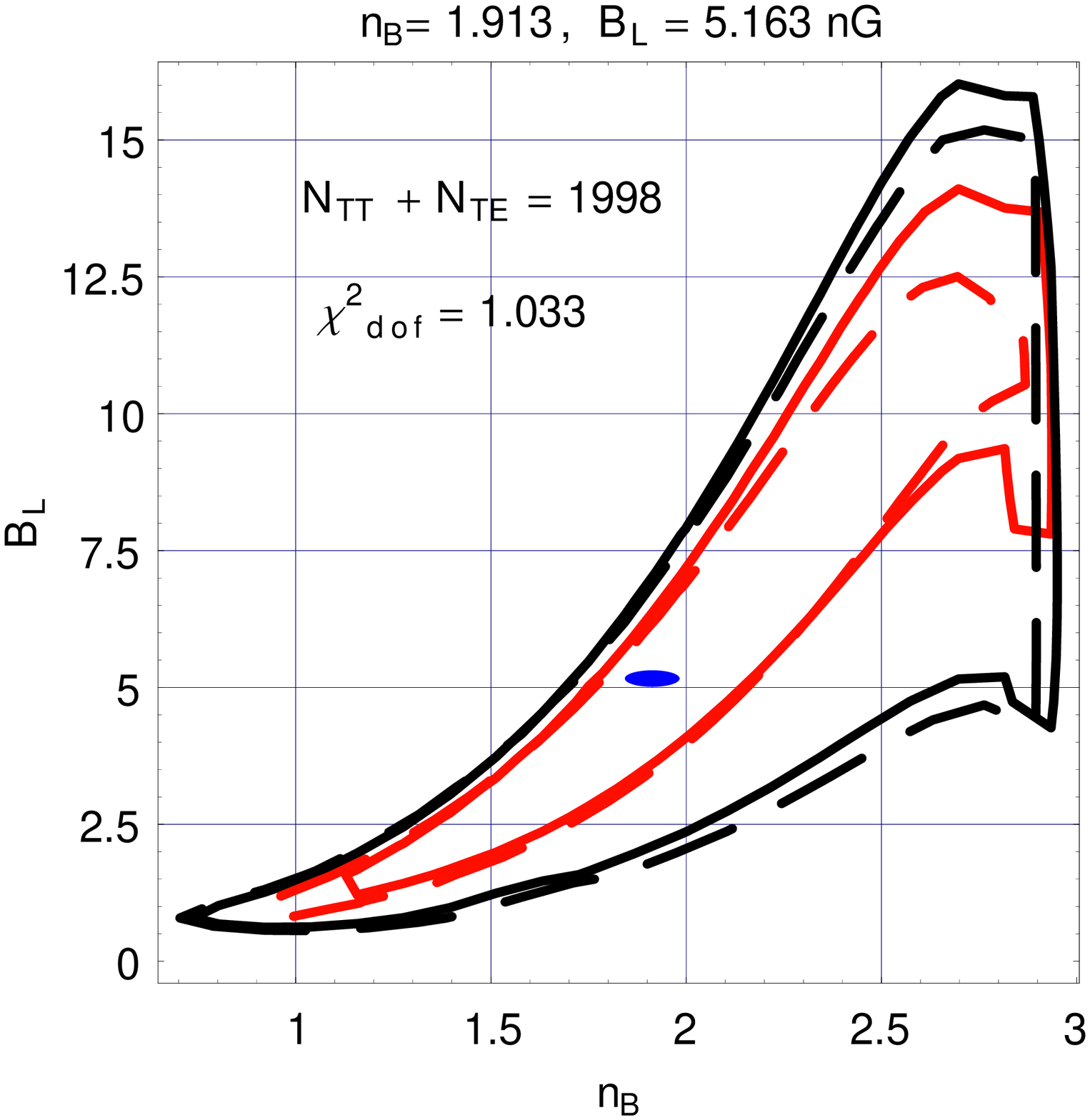}
\caption[a]{Likelihood contours in the plane $(n_{\mathrm{B}},\, B_{\mathrm{L}})$. In both plots the inner curve corresponds to $\Delta \chi^2 = 2.3$ (i.e. $68.3$\% or $1\, \sigma$ of likelihood content) while the outer curve  corresponds to $\Delta \chi^2 = 6.17$ (i.e. $95.4\,$\% or $2\, \sigma$ of likelihood content).}
\label{FIGeight}      
\end{figure}
In Fig. \ref{FIGeight}  the 
filled (ellipsoidal) spots in both plots mark the minimal value of the $\chi^2$ (i.e. $\chi^2_{\mathrm{min}}$) obtained in the 
context of the WMAP 5yr data alone (see, e.g., Eq (\ref{Par4})); the closed curves represent the
likelihood contours in the two parameters $n_{\mathrm{B}}$ and $B_{\mathrm{L}}$.
In both plots of Fig. \ref{FIGeight} the boundaries of the two regions 
contain $68.3\,$\% and $95.4\,$\% of likelihood as the 
values for which the $\chi^2$ has increased, respectively, by an amount $\Delta \chi^2 = 2.3$ and $\Delta \chi^2 = 6.17$. In the plot at the left of Fig. \ref{FIGone} the data points used for the analysis  correspond to the measured TT correlations contemplating $N_{\mathrm{TT}} =999$ experimental points from $\ell = 2$ to $\ell =100$. In the plot at the right the data points include, both, the TT and TE correlations 
and the total number of data points increases to $N_{\mathrm{TT}} + N_{\mathrm{TE}} =1998$.
The values of $n_{\mathrm{B}}$ and $B_{\mathrm{L}}$ minimizing the $\chi^2$ when 
only the TT correlations are considered  turn out to be $n_{\mathrm{B}}= 1.883$ and $B_{\mathrm{L}}= 4.982\, \mathrm{nG}$; in this case the 
reduced $\chi^2$ is $1.094$. When also the TE correlations are included in the 
analysis the reduced $\chi^2$, i.e. $\chi^2_{\mathrm{dof}}$,  diminishes from $1.094$ to $1.033$ and the 
values of $n_{\mathrm{B}}$ and $B_{\mathrm{L}}$ minimizing the $\chi^2$
become $n_{\mathrm{B}} =1.913$ and  $B_{\mathrm{L}} =5.163\, \mathrm{nG}$. The points of the parameter 
leading to $\chi^2_{\mathrm{min}}$ maximize the likelihood in the Gaussian approximation (see \cite{max1} and \cite{max2}).

In a frequentist approach, the boundaries of the confidence regions represent exclusion plots at  $68.3\,$\% and $95.4\,$\% confidence level. 
 In Fig. \ref{FIGeight} (plot at the right) the inner dashed curve corresponds to the 
$68.3\,$\% boundary while the outer dashed line corresponds to the  $95.4\,$\% boundary as determined in the context of the magnetized $\Lambda$CDM  scenario and with the very same set of data \cite{max1}. 
By comparing the dashed and full contour plots in Fig. \ref{FIGeight} (right plot) 
the shapes of the excluded regions are compatible in the two cases. The overlap seems even to increase by 
increasing the confidence range. At the same time the addition of a fluctuating dark energy background pins down systematically larger values of the magnetic field  parameters.  More specifically, by looking at the parameters corresponding 
to $\chi^2_{\mathrm{min}}$, we can say that
\begin{eqnarray}
&&(n_{\mathrm{B}},B_{\mathrm{L}})_{\Lambda\mathrm{CDM}} = ( 1.598,\,3.156 \mathrm{nG})\to (n_{\mathrm{B}} , B_{\mathrm{L}})_{\mathrm{{\it w}CDM}} =(1.883, \,4.982\, \mathrm{nG}), 
\label{NTT}\\
&&(n_{\mathrm{B}}, B_{\mathrm{L}})_{\Lambda\mathrm{CDM}} = ( 1.616,\,3.218 \mathrm{nG})\to (n_{\mathrm{B}},B_{\mathrm{L}})_{\mathrm{{\it w}CDM}} = (1.913, \,5.163\, \mathrm{nG}), 
\label{NTE}
\end{eqnarray}
where Eq. (\ref{NTT}) corresponds to $N_{\mathrm{TT}} =999$ (i.e. 
only the unbinned points of the TT correlations are used) while Eq. 
(\ref{NTE}) does correspond to $N_{\mathrm{TT}} +N_{\mathrm{TE}} =1998$ (i.e. the unbinned points of the TT and TE correlations are used).

Equations (\ref{NTT}) and (\ref{NTE}) show that the magnetic field parameters minimizing the $\chi^2$ (and maximizing the likelihood) do (slightly) increase when the dark energy component is allowed to fluctuate. Of course 
the slight numerical difference does not change the physical conclusions 
obtained in the context of the magnetized $\Lambda$CDM class 
of models: the preferred region is for blue spectral indices and moderate 
magnetic field intensities. 
\begin{figure}[!ht]
\centering
\includegraphics[height=6cm]{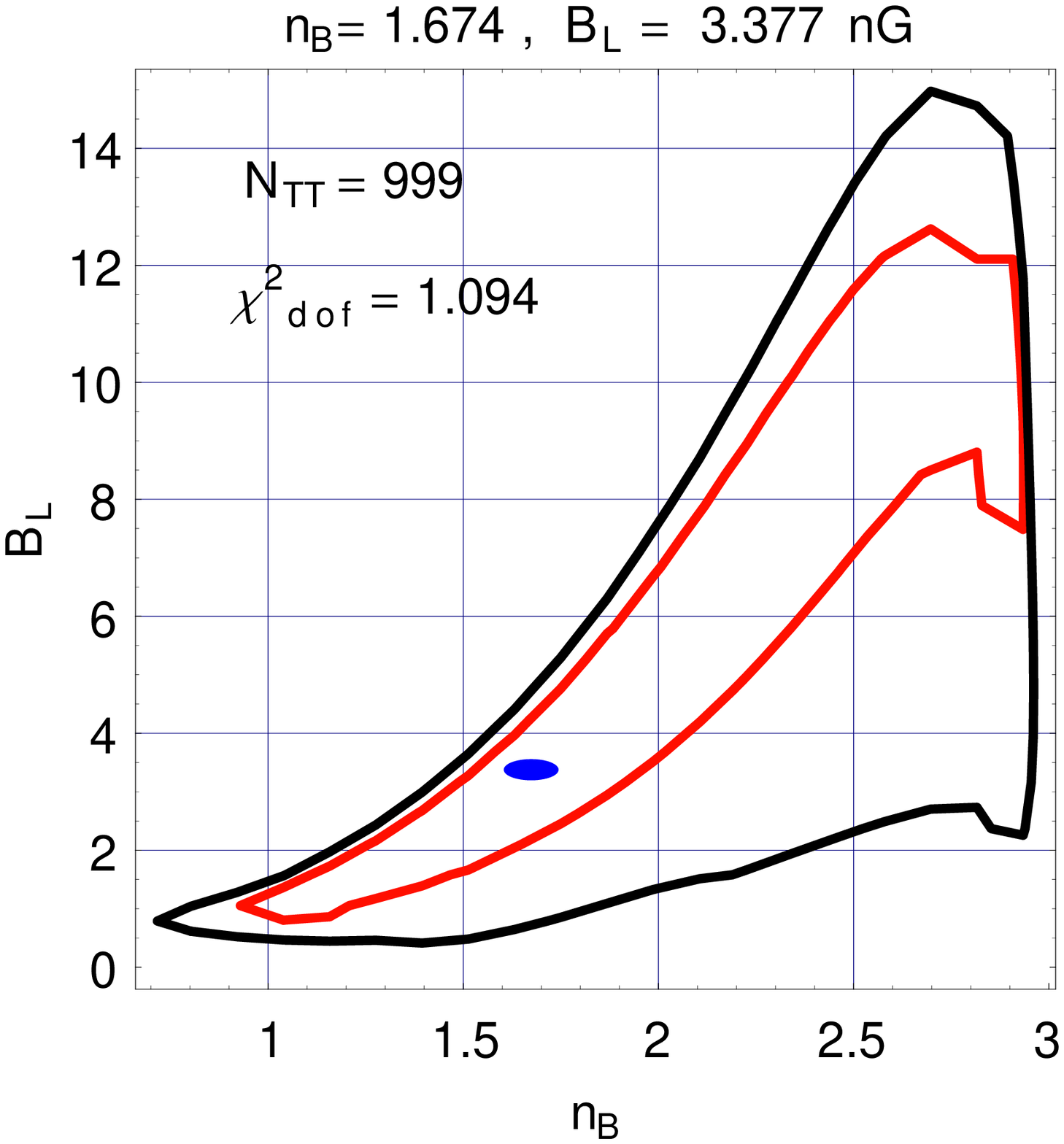}
\includegraphics[height=6.1cm]{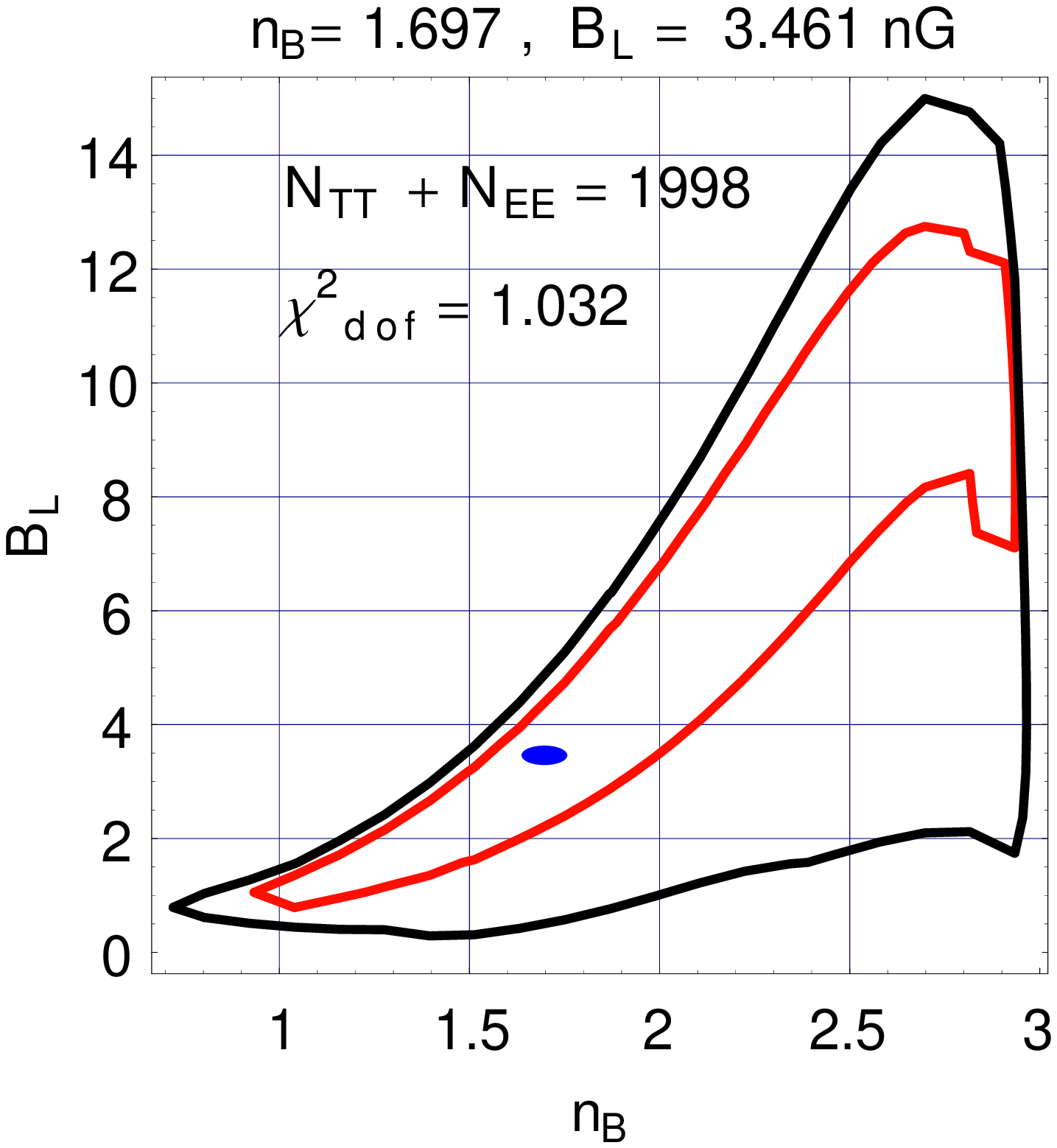}
\caption[a]{Same quantities as in Fig. \ref{FIGeight} but in the  case when 
{\it w}CDM scenario is analyzed in the light of the WMAP 5yr +bao data (see 
Eq. (\ref{Par5})).}
\label{FIGnine}      
\end{figure}
In Fig. \ref{FIGnine} the likelihood contours are obtained when the best 
fit parameters are determined from the {\it w}CDM parameters arising when the WMAP 5yr data are combined with the data stemming from the baryon acoustic oscillations (see Eq. (\ref{Par5})). In this case, again, the shapes 
of the likelihood contours are qualitatively a bit different but quantitatively 
compatible with the cases of Fig. \ref{FIGeight} and of Eq. \ref{FIGnine}.
\begin{figure}[!ht]
\centering
\includegraphics[height=6cm]{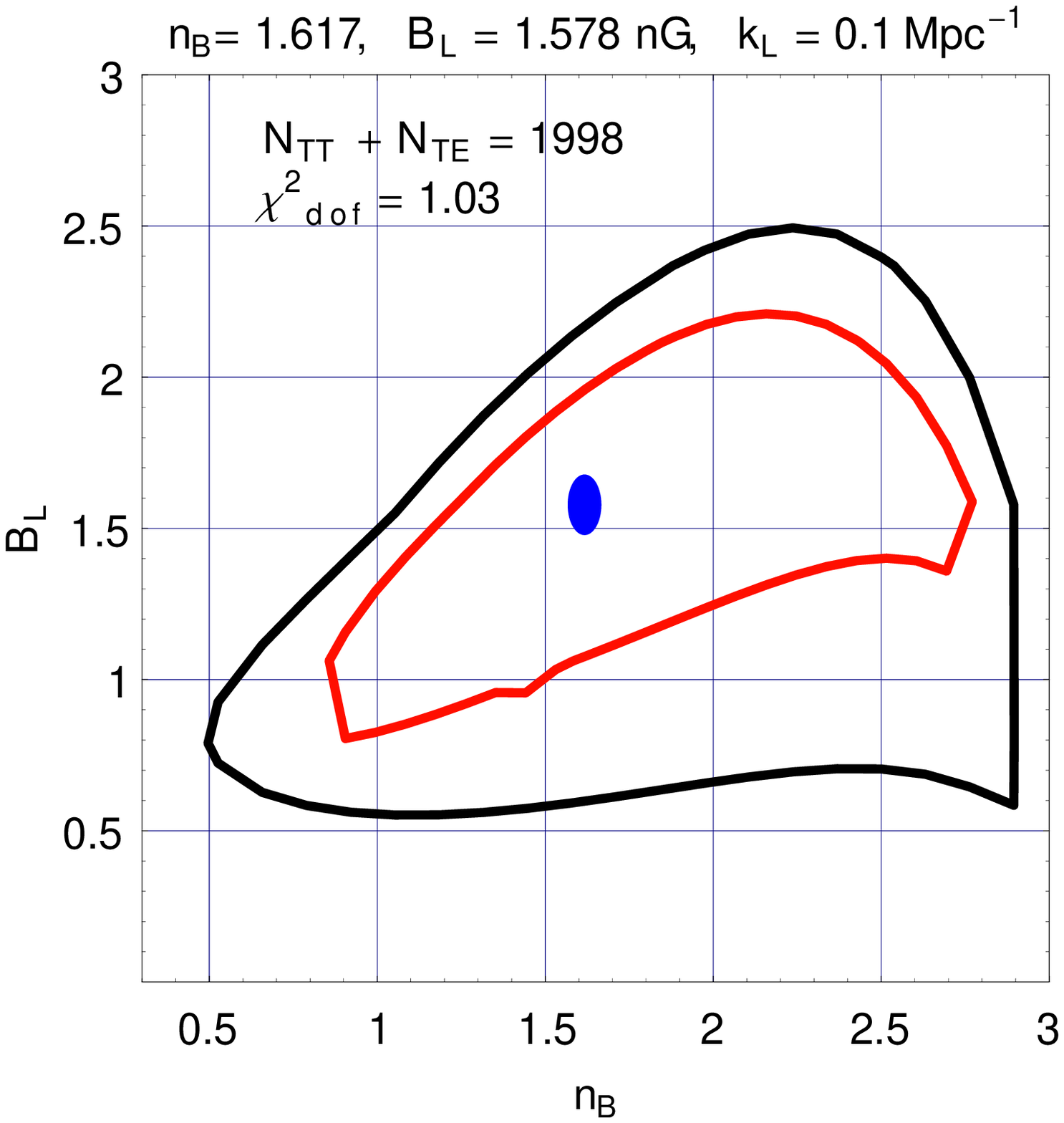}
\includegraphics[height=6cm]{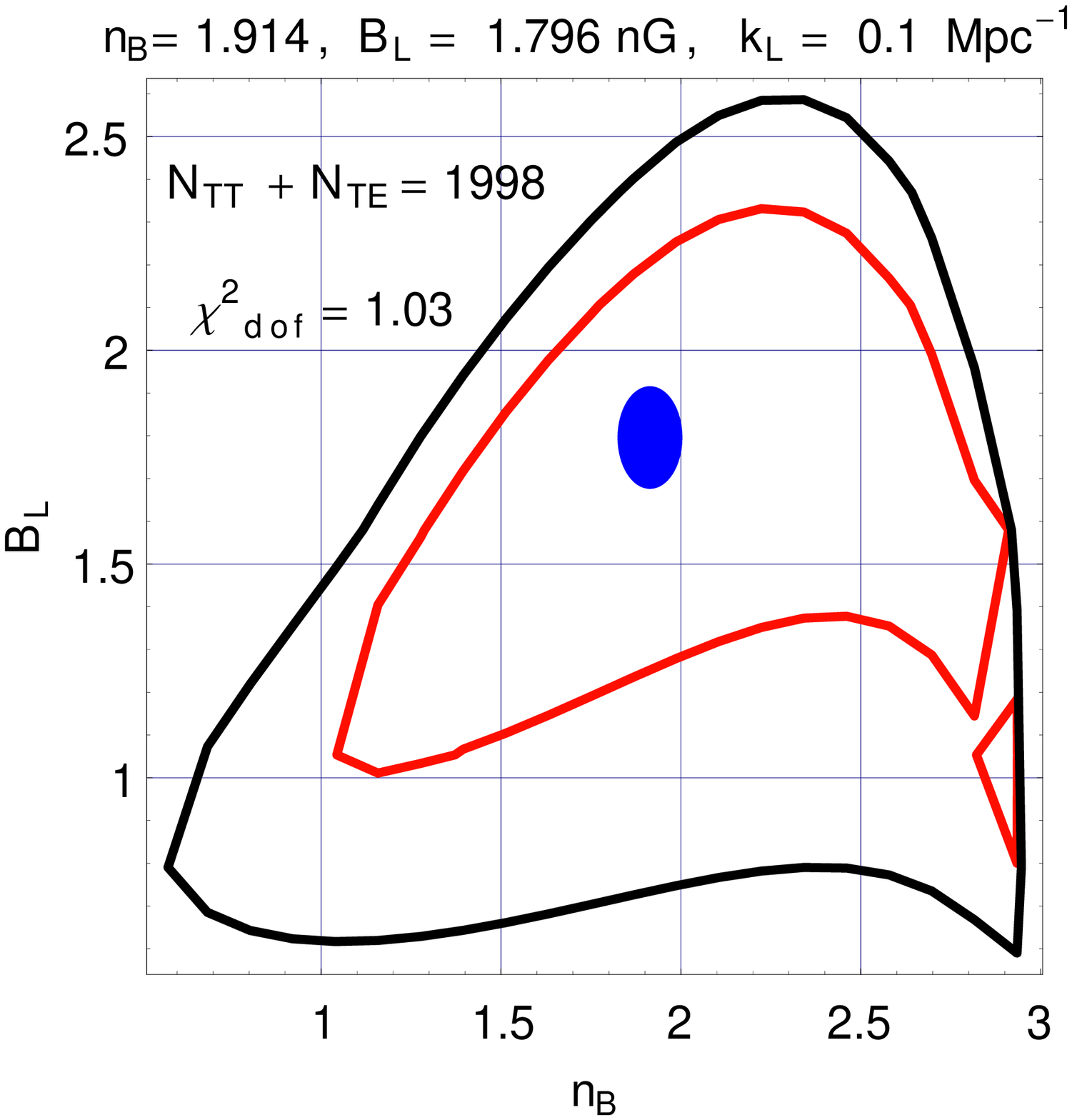}
\caption[a]{The effect of change in the magnetic pivot scale $k_{\mathrm{L}}$ is illustrated in the case 
of the WMAP 5yr data alone in the $\Lambda$CDM case (left plot) and in the {\it w}CDM case (plot at the right).}
\label{FIGten}      
\end{figure}
 As expected a decrease of $k_{\mathrm{L}}$ 
lowers the value of the regularized magnetic field intensity 
$B_{\mathrm{L}}$. This effect is illustrated in Fig. \ref{FIGten}. 
In the right plot, the likelihood contours are computed when $k_{\mathrm{L}} = 0.1 \, \mathrm{Mpc}^{-1}$; the 
parameters of the $w$CDM paradigm  are determined from Eq. (\ref{Par4}).  In the plot at the left 
the same analysis is performed for  the $\Lambda$CDM case when, again, the other parameters are 
determined from the WMAP 5yr data alone. 
The exclusion plots of Figs. \ref{FIGeight}, \ref{FIGnine} and \ref{FIGten} explain, 
a posteriori, why in the plots of Fig. \ref{FIGfive}, \ref{FIGsix} and \ref{FIGseven} the minimal 
amount of distortions was provided by the dot-dashed curves. 
The dot-dashed curves corresponded, in those figures, 
to the region $(iii)$ i.e. the region where the moderate magnetic field 
intensities corresponded to blue magnetic spectral indices. 

\renewcommand{\theequation}{6.\arabic{equation}}
\setcounter{equation}{0}
\section{Concluding remarks}
\label{sec6}
The analysis of magnetized CMB anisotropies has been extended, for the first time, to the case when the 
dark energy background fluctuates. This class of models, conventionally labeled 
as magnetized $w$CDM, generalizes the magnetized $\Lambda$CDM scenario. 
The ordinary and integrated Sachs-Wolfe effects have been analyzed in the light of the $w$CDM 
paradigm. In the minimal realization of the scenario (i.e. in the absence of non-adiabatic 
components in the initial conditions) the contribution of the magnetized background 
is not sufficient to compensate the effect induced on the adiabatic mode by the late-time dynamics 
of the dark energy background. Possible compensating effects arise for excessively 
large values of the magnetic field intensity; these values are already constrained by 
the structure of the acoustic oscillations. 

Since the modified initial conditions of  the Einstein-Boltzmann hierarchy allow for the simultaneous presence of dark energy perturbations and of large-scale magnetic fields, the morphology of the distortions arising in the 
magnetized $w$CDM scenario has been compared with what happens when the dark energy background is not dynamical.
The modifications induced by the fluctuating dark energy component on the shapes of the $1$- and $2$-$\sigma$ contours in the parameter space of the magnetized background have been computed. 
At $95$ \% confidence level,  the allowed spectral indices and magnetic field intensities turn out to be systematically larger than those determined in the framework of the magnetized $\Lambda$CDM paradigm.

\newpage

\begin{appendix}
\renewcommand{\theequation}{A.\arabic{equation}}
\setcounter{equation}{0}
\section{Synchronous frame}
\label{APPA}
It is understood that all the quantities appearing in this section of the appendix are evaluated in the 
synchronous frame. The evolution equation of the intensity brightness perturbations are given by
\begin{eqnarray}
&&\Delta_{\mathrm{I}}' + (i k \mu + \epsilon') \Delta_{\mathrm{I}} = - \xi' - \frac{\mu^2}{2}(h + 6 \xi)'  + \epsilon'\biggl[ \Delta_{\mathrm{I}0} + \mu v_{\mathrm{b}} + 
\frac{(1 - 3 \mu^2)S_{\mathrm{P}}(k,\tau)}{4} \biggr],
\label{HT1}\\
&& v_{\mathrm{b}}' + {\mathcal H} v_{\mathrm{b}} + \frac{\epsilon'}{R_{\mathrm{b}}} ( 3 i \Delta_{\mathrm{I}1} + v_{\mathrm{b}}) + i k \frac{\Omega_{\mathrm{B}} - 4 \sigma_{\mathrm{B}}}{4 R_{\mathrm{b}}}.
\label{HT2}
\end{eqnarray}
The formal solution of Eq. (\ref{HT1}) can be written, after integrating once  by parts, as:
\begin{eqnarray}
&&\Delta_{\mathrm{I}}(k,\mu,\tau_{0}) = \int_{0}^{\tau_{0}}\, e^{- \epsilon(\tau,\tau_{0})}\biggl[ - \xi'  - \frac{(h + 6 \xi)'''}{2 k^2}\biggr] e^{- i \mu x(\tau)}\, d\tau
\nonumber\\
&+&\int_{0}^{\tau_{0}} {\mathcal K}(\tau) e^{- i \mu x(\tau)} 
\biggl[  \Delta_{\mathrm{I}0} - \frac{(h + 6 \xi)''}{2 k^2} + \mu v_{\mathrm{b}} + 
\frac{i \mu}{2 k}(h + 6 \xi)' + \frac{(1 - 3 \mu^2)S_{\mathrm{P}}(k,\tau)}{4}  \biggr],
\label{HT3}
\end{eqnarray}
where, as in the bulk of the paper, $x(\tau) = k (\tau_{0} \tau)$. The visibility function is defined in terms of the   
optical depth are defined as in Eq. (\ref{LONG1}). The integration by parts can be further used in Eq. (\ref{HT1}):
\begin{eqnarray}
\Delta_{\mathrm{I}}(k,\mu,\tau_{0}) &=& \int_{0}^{\tau_{0}} \frac{e^{-i  \mu x(\tau)}}{4 k^2} \tilde{S}_{\mathrm{P}}(k,\tau_{0}),
\label{HT4a}\\
 \tilde{S}_{\mathrm{P}}(k,\tau_{0}) &=& {\mathcal K}(\tau) [ 4 k^2 \Delta_{\mathrm{I}0} - (h + 6 \xi)'' + 4 i k v_{\mathrm{b}}'  + k^2 S_{\mathrm{P}} + 3 S_{\mathrm{P}}'']
 \nonumber\\
 &+& {\mathcal K}' [ 6 S_{\mathrm{P}}' + 4 i k v_{\mathrm{b}}  - 2 ( h + 6 \xi)'] + 3 
S_{\mathrm{P}} {\mathcal K}'' 
\nonumber\\
&& - e^{- \epsilon(\tau,\tau_{0})} [ 4 k^2 \xi'  + 2 ( h + 6 \xi)'],
\label{HT4b}
\end{eqnarray}
but in spite of the simplifications obtained in Eqs. (\ref{HT4a}) and (\ref{HT4b}),  Eq. 
(\ref{HT3}) is still the most direct for separating the ordinary and the integrated Sachs-Wolfe contribution
in the presence of a sudden drop in the opacity at $\tau_{\mathrm{rec}}$. In this case Eqs. (\ref{HT3}) implies 
\begin{eqnarray}
&& \Delta_{\mathrm{I}}(k,\mu,\tau_{0}) 
= \Delta^{(\mathrm{SW})}_{\mathrm{I}}(k,\mu,\tau_{0}) + \Delta^{(\mathrm{ISW})}_{\mathrm{I}}(k,\mu,\tau_{0}),
\label{HT5}\\
&& \Delta^{(\mathrm{ISW})}_{\mathrm{I}}(k,\mu,\tau_{0})= \int_{\tau_{\mathrm{rec}}}^{\tau_{0}} \biggl[ - \xi' - \frac{(h + 6 \xi)'''}{2 k^2}\biggr] e^{- i \mu x(\tau)}\, d\tau,
\label{HT6}\\
&& \Delta^{(\mathrm{SW})}_{\mathrm{I}}(k,\mu,\tau_{0})=  \biggl[\Delta_{\mathrm{I}0}   - \frac{(h + 6 \xi)''}{2 k^2}\biggr]_{\tau_{\mathrm{rec}}} e^{- i \mu y_{\mathrm{rec}}}.
\label{HT7}
\end{eqnarray}
Recalling that, by definition,  $y_{\mathrm{rec}} = x(\tau_{\mathrm{rec}}) = k (\tau_{0} - \tau_{\mathrm{rec}})$ the  SW contribution can be easily rearranged as
\begin{equation}
\Delta^{(\mathrm{SW})}_{\mathrm{I}}(k,\mu,\tau_{0}) = 
\biggl[\Delta_{\mathrm{I}0}(k,\tau)  - \xi(k,\tau) + {\mathcal H}\frac{(h + 6 \xi)'}{k^2}\biggr]_{\tau_{\mathrm{rec}}} e^{- i \mu y_{\mathrm{rec}}},
\label{HT8}
\end{equation}
where the evolution equation 
\begin{equation}
(h + 6\xi)'' + 2 {\mathcal H} (h + 6 \xi)' = 2 k^2 \xi + {\mathcal O}(k^2\tau^2)
\label{HT8a}
\end{equation}
has been used explicitly. In Eq. (\ref{HT8a}) the various anisotropic stresses have been 
neglected; Eq. (\ref{HT8a}) 
is the synchronous version of Eq. (\ref{ij2}), as it can be easily verified by using the procedure swiftly 
illustrated in section \ref{sec1}. Recalling that $\Delta_{\mathrm{I}0} = 4 \delta_{\gamma}$, the solution of synchronous gauge version of Eq. (\ref{BP4}) implies 
\begin{equation}
\Delta_{\mathrm{I}0} = - \frac{R_{\gamma} \Omega_{\mathrm{B}}}{4} + \frac{h}{6} + {\mathcal O}(k^2\tau^2).
\label{HT8b}
\end{equation}
To obtain the explicit form of Eq. (\ref{HT8}) we need the evolution of $\xi(k,\tau)$ and $h(k,\tau)$ across equality.
By using Eq. (\ref{gauge5}) and (\ref{synch1}) we obtain
\begin{eqnarray}
&&\xi(k,\tau) = {\mathcal R}_{*}(k) - \frac{R_{\gamma} \Omega_{\mathrm{B}}(k)}{8} {\mathcal G}_{1}(\alpha),
\label{HT8c}\\
&& h(k,\tau) = \frac{3}{4} R_{\gamma} \Omega_{\mathrm{B}}(k) {\mathcal G}_{1}(\alpha) + k^2 \tau_{1}^2 {\mathcal R}_{*}(k) 
{\mathcal G}_{2}(\alpha) - \frac{3}{4} R_{\gamma} \Omega_{\mathrm{B}}(k)  k^2 \tau_{1}^2 {\mathcal G}_{3}(\alpha),
\label{HT8d}\\
&& \frac{(h + 6 \xi)'}{2 k^2} = \frac{{\mathcal R}_{*}(k)}{15} {\mathcal G}_{4}(\alpha) + \frac{R_{\gamma} \Omega_{\mathrm{B}}}{20} {\mathcal G}_{5}(\alpha),
\label{HT8e}
\end{eqnarray}
where $\alpha = a/a_{\mathrm{eq}} = [(\tau/\tau_1)^2 + 2 (\tau/\tau_{1})]$ solves Eq. (\ref{FL1}) across equality and where 
the five functions ${\mathcal G}_{i}(\alpha)$ are given by:
\begin{eqnarray}
&& {\mathcal G}_{1}(\alpha) = 2\frac{[ \alpha (\alpha - 4) + 8 (\sqrt{\alpha +1} -1)]}{\alpha^2},
\label{Gone}\\
&& {\mathcal G}_{2}(\alpha) = \frac{[ \alpha^3 - 2 \alpha^2 + 8\alpha + 16 ( 1 - \sqrt{\alpha + 1})]}{{5 \alpha^2} },
\label{Gtwo}\\
&& {\mathcal G}_{3}(\alpha) = \frac{\{ 32 (\sqrt{\alpha + 1} -1) + 
\alpha [ \alpha ( 3 \alpha - 26)  - 16] + 60 \sqrt{\alpha + 1}[\alpha - \ln{(\alpha + 1)]}\}}{{45 \alpha^2} },
\label{Gthree}\\
&& {\mathcal G}_{4}(\alpha) = \frac{\sqrt{1 + \alpha}[ 3\alpha^2 - 4 \alpha +8] -8}{\alpha^2}, 
\label{Gfour}\\
&& {\mathcal G}_{5}(\alpha)= \frac{\sqrt{\alpha+1}[ 24 - \alpha^2 + 8 \alpha] - 4 ( 5 \alpha + 6)}{\alpha^2}.
\label{Gfive}
\end{eqnarray}
Using Eqs. (\ref{Gone})--(\ref{Gfive}) the ordinary SW contribution can be written as 
\begin{equation}
\Delta^{(\mathrm{SW})}_{\mathrm{I}}(k,\mu,\tau_{0})=  \biggl[ - \frac{{\mathcal R}_{*}(k)}{5} 
{\mathcal S}{\mathcal W}_{{\mathcal R}}(\alpha_{\mathrm{rec}}) + \frac{R_{\gamma} \Omega_{\mathrm{B}}(k)}{20} \,{\mathcal S}{\mathcal W}_{\mathrm{B}}(\alpha_{\mathrm{rec}}) + {\mathcal O}(k^2 \tau^2)]  e^{- i \mu y_{\mathrm{rec}}}
\label{Gsix}
\end{equation}
where ${\mathcal S}{\mathcal W}_{{\mathcal R}}(\alpha_{\mathrm{rec}})$ and ${\mathcal S}{\mathcal W}_{\mathrm{B}}(\alpha_{\mathrm{rec}})$ are the two functions appearing, respectively, in Eqs. (\ref{SWR}) and (\ref{SWB}).
\renewcommand{\theequation}{B.\arabic{equation}}
\setcounter{equation}{0}
\section{Explicit derivation of $T_{\mathcal R}(\tau)$ and $T_{\mathrm{B}}(\tau)$}
\label{APPB}
Consider Eq. (\ref{gauge4}) in terms of the conformally Newtonian 
variables. The first equality can be written, for immediate convenience, as 
\begin{equation}
{\mathcal R} = - \psi + \frac{H^2}{\dot{H}} \biggl[ \psi + \frac{\dot{\psi}}{H}\biggr],\qquad \frac{\partial}{\partial t} \biggl(\frac{a \psi}{H}\biggr) = a {\mathcal R} \frac{\dot{H}}{H^2},
\label{INT4}
\end{equation}
where we only passed from the conformal time coordinate $\tau$ to the 
cosmic time coordinate $t$ and we recalled that $dt \, = a(\tau) d\tau$; it is also 
practical to use the well known identities $\dot{H} =( {\mathcal H}' - {\mathcal H}^2)/a^2$ and $H = {\mathcal H}/a$.
From Eq. (\ref{INT4}) by  integrating once with respect to $t$ 
the solution for $\psi(k,t)$ then becomes
\begin{equation}
\psi(k,t) = - {\mathcal R}(k,t) + \frac{H}{a} \int_{0}^{t} \frac{a \dot{{\mathcal R}}}{H}\, dt' 
+\frac{H}{a} \int_{0}^{t} {\mathcal R}(k,t') a(t') \, dt',
\label{INT11}
\end{equation}
where we integrated by parts and used that 
\begin{equation}
\frac{\partial}{\partial t} \biggl( \frac{a {\mathcal R}}{H}\biggr) = 
- \frac{\dot{H}}{H^2} a {\mathcal R} + \frac{\dot{a}}{H} {\mathcal R} + \frac{a \dot{{\mathcal R}}}{H}.
\label{INT10}
\end{equation}
From Eqs. (\ref{LSS1}), (\ref{LSS2}) and (\ref{LSS3}), neglecting the 
electric and Ohmic contributions, $\dot{{\mathcal R}}(k,t)$ is given by:
\begin{equation}
\dot{{\mathcal R}} = - \frac{H\, \delta p_{\mathrm{nad}}}{p_{\mathrm{t}} + \rho_{\mathrm{t}}} 
+ \biggl( c_{\mathrm{st}}^2 - \frac{1}{3}\biggr) \frac{H \delta_{\mathrm{s}} \rho_{\mathrm{B}}}{p_{\mathrm{t}} + \rho_{\mathrm{t}}},
\label{INT12}
\end{equation}
whose formal solution can be written as:
\begin{equation}
{\mathcal R}(k,t') = {\mathcal R}_{*}(k)  - \int_{0}^{t'} \frac{H(t') \delta p_{\mathrm{nad}}(k,t')}{p_{\mathrm{t}} + \rho_{\mathrm{t}}} dt'' +
\int_{0}^{t'}\frac{H(t'') \delta_{\mathrm{s}} \rho_{\mathrm{B}}(k,t'')}{p_{\mathrm{t}} + \rho_{\mathrm{t}}} 
\biggl(c_{\mathrm{st}}^2 - \frac{1}{3}\biggr) dt''.
\label{INT13}
\end{equation}
If we now insert Eq. (\ref{INT13}) into Eq. (\ref{INT11}) and go  back to the conformal time coordinate we can write 
\begin{equation}
\psi(k,\tau) = -{\mathcal R}_{*}(k) T_{\mathcal R}(\tau) + 
R_{\gamma} \Omega_{\mathrm{B}}(k) T_{\mathrm{B}}(\tau)
\label{INT13a}
\end{equation}
where $T_{\mathcal R}(\tau)$ and $T_{\mathrm{B}}(\tau)$ are given by 
Eqs. (\ref{INT1a}) and (\ref{INT1b}).
For sake of simplicity and for comparison with other treatments consider, for instance, 
the case of the $\Lambda$CDM scenario  where 
\begin{equation}
a(t) = a_{1} \biggl[\sinh{\biggl(\frac{3}{2} H_{0} t\biggr)}\biggr]^{2/3}.
\label{INT8}
\end{equation}
In the cosmic time coordinate $T_{{\mathcal R}}(t)$ can be written as 
\begin{equation}
T_{{\mathcal R}}(t) = 1 - \frac{H(t)}{a(t)}\int_{0}^{t} a(t') d t' 
\label{INT8a}
\end{equation}
It is now easy to show that, for $t\to 0$, $a(t) \simeq t^{2/3}$ and $T_{{\mathcal R}}(t) \simeq - (3/5)$.
For $t\to \infty$ $T_{{\mathcal R}}(t) \simeq e^{- H_{0} t}$. 
Unfortunately, in the most general situation of the $w$CDM scenario, the semi-anaytic estimates are not that simple and the numerical evaluation of 
$T_{\mathcal R}(\tau)$ and $T_{\mathrm{B}}(\tau)$ is often 
mandatory (see, for instance, the initial part of section \ref{sec5}).
\renewcommand{\theequation}{C.\arabic{equation}}
\setcounter{equation}{0}
\section{Generalization of certain integral representations}
\label{APPC}
In the present study there is often the need of generalizing certain 
integral representations of the power spectrum of the ISW and of its 
correlation with the SW effect. For instance, in the simplest 
case of adiabatic fluctuations we shall have that the autocorrelation if the ISW 
and its cross correlation with the SW term boils down to the following 
pair of expressions:
\begin{eqnarray}
&& C_{\ell}^{(\mathrm{ISW})} = \int_{\tau_{{\mathrm{rec}}}}^{\tau_{0}} \frac{d T_{{\mathcal R}}}{d\tau_{1}} \, d\tau_{1} \int_{\tau_{{\mathrm{rec}}}}^{\tau_{0}}  \frac{d T_{{\mathcal R}}}{d\tau_{2}} \, d\tau_{2} f_{\ell}(\tau_{1}, \tau_{2}, n_{\mathrm{s}}),
\label{LEG1}\\
&&C_{\ell}^{(\mathrm{cross})} = \frac{1}{5}\int_{\tau_{{\mathrm{rec}}}}^{\tau_{0}} \frac{d T_{{\mathcal R}}}{d\tau} \, d\tau \int_{\tau_{{\mathrm{rec}}}}^{\tau_{0}}  f_{\ell}(\tau,\tau_{{\mathrm{rec}}},n_{\mathrm{s}}),
\label{LEG2}
\end{eqnarray}
where 
\begin{equation}
 f_{\ell}(a, b, n_{\mathrm{s}}) = \frac{8\pi^2 {\mathcal A}_{{\mathcal R}} \, k_{\mathrm{p}}^{ 1 - n_{\mathrm{s}}}}{\sqrt{a \, b}}
 \int_{0}^{\infty} k^{n_{\mathrm{s}} -3} \, J_{\ell + 1/2}(k a)\, J_{\ell + 1/2}(k b).
\label{LEG3}
\end{equation}
The expressions of Eqs. (\ref{LEG1}) and (\ref{LEG2}) are just illustrative since the same problem arises in the case 
of the transfer functions of the magnetic component. To evaluate expressions like the ones of Eqs. (\ref{LEG1}) and (\ref{LEG2}) 
one might choose to perform first the (numerical) integrations over the conformal time coordinate and then the 
integrations over the modulus of the wavevector $k$. The idea is to reverse this order and integrate first over $k$.
In doing so we are led, in general terms, to integrals like  
\begin{eqnarray}
&& \int_{0}^{\infty} k^{-\lambda} J_{\nu}(k a) J_{\nu}(k b) d k =
\nonumber\\
&& \frac{a^{\nu} \Gamma\biggl(\nu + \frac{1 - \lambda}{2}\biggr)}{ 2^{\lambda} b^{\nu - \lambda + 1 } \Gamma\biggl(\frac{\lambda + 1}{2}\biggr) \Gamma(\nu + 1)}
F\biggl[ \nu + \frac{1 - \lambda}{2}, \frac{1 - \lambda}{2}; \nu + 1; \frac{a^2 }{b^2}\biggr].
\label{LEG4}
\end{eqnarray}
In terms of the illustrative examples of Eqs. (\ref{LEG1}) and (\ref{LEG2}), $\nu = (\ell +1/2)$ and $\lambda = 3 - n_{\mathrm{s}}$.
Equation (\ref{LEG4}) can be found, for instance, in Ref. \cite{grad} (see formula 6.574, page 675). 

It is more practical, for the purpose of approximate evaluations, to relate 
the hypergeometric functions to the associated Legendre functions. This was the strategy followed 
in the first attempt to evaluate analytically the ISW contribution in the 
$\Lambda$CDM model \cite{KS1}. The latter analysis has been performed is the case of (exactly) scale-invariant 
contribution. In spite of the fact that the explicit dependence 
upon the spectral index is unimportant for orders of magnitude 
estimates, the habit of computing analytically the ISW contribution 
for the exactly scale-invariant case persists \cite{elga}. In what 
follows the integral representation of the two-point function 
for the ISW will be directly related to the associated Legendre 
functions but  for a generic spectral index.

Consider an hypergeometric 
function parametrized as $F(\alpha,\, \beta;\, \gamma;\, z)$. 
It can be shown that if two of the numbers $(\gamma -1)$, $(\alpha - \beta)$ and 
$(\alpha + \beta -\gamma)$ are equal (or if one of them is equal to $1/2$) then 
there exist a quadratic transformation which allows to connect the hypergeometric 
function to a Legendre function whose argument will be suitable for 
(approximate) semi-analytic integrations.   Indeed in the case of Eq. (\ref{LEG4}) 
$\alpha = \nu + (1-\lambda)/2$, $\beta= (1 - \lambda)/2$ and $\gamma = \nu +1$;
this observation implies that $(\alpha - \beta)$ and $(\gamma -1)$ are both equal to $\nu$. Consequently, the general quadratic relation \cite{abr1}
\begin{equation}
F(\alpha,\, \beta;\, 2 \beta;\, z) = \biggl[ \frac{1+ \sqrt{1-z}}{2}\biggr]^{-2\alpha} 
F\biggl[\alpha,\, \alpha - \beta + \frac{1}{2};\, \beta + \frac{1}{2};\, \biggl(\frac{1 - \sqrt{1 - z}}{1 + \sqrt{1 - z}}\biggr)^2\biggr],
\label{LEG5}
\end{equation}
implies, in the case of Eq. (\ref{LEG4}) 
\begin{equation}
F\biggl[ \nu + \frac{1 - \lambda}{2}, \frac{1 - \lambda}{2}; \nu + 1; \frac{a^2 }{b^2}\biggr]
= \biggl(\frac{b}{b + a}\biggr)^{2 \nu + 1 - \lambda}
F\biggl[ \nu + \frac{1 - \lambda}{2},\,\nu + \frac{1}{2} ; 2\nu + 1; \frac{4 a b}{(a + b)^2}\biggr].
\label{LEG6}
\end{equation}
It also follows that that\footnote{We correct here a typo appearing in the formula 6.576 (number 2) page 576 of Ref. \cite{grad}.
In the numerator of the formula we read $2^{\nu}$ which should be instead $b^{\nu}$. Equation (\ref{LEG7}) 
represents the correct version of the quoted formula where $b^{\nu}$ (and not $2^{\nu}$) correctly appears in the numerator. }
\begin{eqnarray}
&& \int_{0}^{\infty} k^{-\lambda} J_{\nu}(k a) J_{\nu}(k b) d k =
\nonumber\\
&& \frac{a^{\nu} \, b^{\nu}}{ 2^{\lambda} ( a + b)^{2 \nu + 1 - \lambda}} \frac{\Gamma\biggl(\nu + \frac{1 - \lambda}{2}\biggr)}{\Gamma\biggl(\frac{1 + \lambda}{2}\biggr) \Gamma(\nu + 1)}F\biggl[ \nu + \frac{1 - \lambda}{2},\,\nu + \frac{1}{2} ; 2\nu + 1; \frac{4 a b}{(a + b)^2}\biggr].
\label{LEG7}
\end{eqnarray}
The hypergeometric function appearing in Eq. (\ref{LEG7}) can be further transformed 
according to the general relation (see \cite{abr1} , p. 560)
\begin{equation}
F(\alpha,\, \beta;\, 2\beta;\, z) = \biggl(1 - \frac{z}{2}\biggr)^{-\alpha} F\biggl[ \frac{\alpha}{2},\, \frac{\alpha + 1}{2};\, \beta + \frac{1}{2};\, \frac{z^2}{(2 - z)^2}\biggr].
\label{LEG8}
\end{equation}
In our case Eq. (\ref{LEG8}) implies 
\begin{eqnarray}
&& F\biggl[ \nu + \frac{1 - \lambda}{2},\, \nu + \frac{1}{2};\, 2 \nu + 1;\, \frac{4 a b}{(a + b)^2}\biggr] = 
\nonumber\\
&& \biggl[\frac{a^2 + b^2}{(a+ b)^2}\biggr]^{ - \nu - \frac{1}{2} + \frac{\lambda}{2}}
F\biggl[\frac{\nu}{2} + \frac{1- \lambda}{4},\, \frac{\nu}{2} - \frac{\lambda}{4} + \frac{3}{4}; 
\nu + 1; \frac{4 a^2 b^2 }{(a^2 + b^2)}\biggr].
\label{LEG9}
\end{eqnarray}
Using Eq. (\ref{LEG9}) we also have that:
\begin{eqnarray}
&&\int_{0}^{\infty} k^{-\lambda} \, J_{\nu}(k a) \, J_{\nu}(k b) d k = 
\frac{a^{\nu} \, b^{\nu}}{2^{\lambda} (a^2 + b^2)^{\nu + (1- \lambda)/2}} \frac{\Gamma\biggl(\nu + \frac{1- \lambda}{2}\biggr)}{\Gamma\biggl(\frac{\lambda+1}{2} \biggr) 
\Gamma(\nu +1)}
\nonumber\\
&\times& F\biggl[\frac{\nu}{2}- \frac{\lambda}{4} + \frac{3}{4},\, \frac{\nu}{2} + \frac{1- \lambda}{4}; \nu + 1; \frac{4 a^2 b^2 }{(a^2 + b^2)^2}\biggr];
\label{LEG10}
\end{eqnarray}
the hypergeometric function appearing in Eq. (\ref{LEG10}) can be finally 
related to the associated Legendre functions according to the general 
relation 
\begin{eqnarray}
&& F\biggl[ 1 + \frac{n + m}{2},\, \frac{1}{2} + \frac{m + n}{2};\, n + \frac{3}{2};\, \frac{1}{z^2}\biggr]=
\nonumber\\
&& e^{- i\,m \pi} \frac{z^{m + n +1}}{( z^2 -1)^{m/2}} \frac{2^{n + 1}}{\sqrt{\pi}} 
\frac{\Gamma\biggl(n + \frac{3}{2}\biggr)}{\Gamma(m + n +1)} Q_{n}^{m}(z),
\label{LEG11}
\end{eqnarray}
where $Q_{n}^{m}(z)$ are the associated Legendre functions of second order. 
Matching Eq. (\ref{LEG11}) with the expression appearing in Eq. (\ref{LEG10}) we shall have 
\begin{eqnarray}
&& F\biggl[\frac{\nu}{2}- \frac{\lambda}{4} + \frac{3}{4},\, \frac{\nu}{2} + \frac{1- \lambda}{4}; \nu + 1; \frac{4 a^2 b^2 }{(a^2 + b^2)^2}\biggr]=
\nonumber\\
&& \biggl( \frac{a^2 + b^2 }{2 a b}\biggr)^{\nu +1/2}  \biggl( \frac{a^2 - b^2 }{a^2 + b^2}\biggr)^{\lambda/2} \frac{\Gamma(\nu + 1)}{\Gamma\biggl(\nu + \frac{1+ \lambda}{2}\biggr)} \frac{2^{\nu+1/2}}{\sqrt{\pi}} e^{- i \pi \lambda/2} Q_{\nu -1/2}^{\lambda/2}\biggl(\frac{a^2 + b^2}{2 a b}\biggr).
\label{LEG12}
\end{eqnarray}
where we used the fact that 
\begin{equation}
Q_{n}^{- m}(y) = e^{- 2 i\pi m} \frac{\Gamma(n - m +1)}{\Gamma(n + m +1)} Q_{n}^{m}(y).
\label{LEG13}
\end{equation} 
Thus, in conclusion, we shall have that 
\begin{eqnarray}
&& \int_{0}^{\infty} k^{-\lambda} J_{\nu}(k a) J_{\nu}(k b) d k =
\nonumber\\
&& \frac{(a^2 -b^2)^{\lambda/2}}{2^{\lambda} \sqrt{a b}} \frac{\Gamma\biggl( \nu + \frac{1- \lambda}{2}\biggr)}{\Gamma\biggl(\frac{1 + \lambda}{2}\biggr) \Gamma\biggl(\nu 
+ \frac{1+ \lambda}{2} \biggr)} \frac{e^{- i \pi \lambda/2}}{\sqrt{\pi}} 
Q_{\nu -1/2}^{\lambda/2}\biggl(\frac{a^2 + b^2}{2 a b}\biggr).
\label{LEG14}
\end{eqnarray}
\end{appendix}
This expression generalizes, to the best of our knowledge, 
the analog expressions derived in  \cite{KS1} in the case of scale-invariant curvature perturbations
(corresponding, in the present parametrization, to $\lambda = 3 - n_{\mathrm{s}}$, i.e. $\lambda= 2$ for $n_{\mathrm{s}} =1$ 
and $\nu = \ell + 1/2$).  The very same expression can be used to integrate the magnetized contributions to the ISW. 
In the latter case, unlike the expressions of Eqs. (\ref{LEG1}) and (\ref{LEG2}), 
 the various terms will contain (at least) one $T_{\mathrm{B}}(\tau)$. For instance, in the case of the magnetized 
 contribution to the ISW there will be two derivatives of    $T_{\mathrm{B}}(\tau)$ and the new $\lambda$ will be given as 
 $\lambda = 4 - 2 n_{\mathrm{B}}$. All the different terms arising in Eqs. (\ref{EX7}) and (\ref{EX8})  can be 
 explicitly integrated over $k$ by using the techniques reported in this part of the appendix.


\newpage 
\begin{thebibliography}{99}

\bibitem{WMAP3a}  D.~N.~Spergel {\it et al.}  [WMAP Collaboration],  Astrophys.\ J.\ Suppl.\  {\bf 170}, 377 (2007).

\bibitem{WMAP3b} L.~Page {\it et al.}  [WMAP Collaboration],  Astrophys.\ J.\ Suppl.\  {\bf 170}, 335 (2007).

\bibitem{WMAP51}  E .~Komatsu {\it et al.}  [WMAP Collaboration], Astrophys.\ J.\ Suppl.\  {\bf 180}, 330 (2009).
  
\bibitem{WMAP52} R.~S.~Hill {\it et al.}  [WMAP Collaboration],  Astrophys.\ J.\ Suppl.\  {\bf 180}, 246 (2009).
 
\bibitem{WMAP53}  B.~Gold {\it et al.}  [WMAP Collaboration],  Astrophys.\ J.\ Suppl.\  {\bf 180}, 265 (2009).

\bibitem{WMAP54} J.~Dunkley {\it et al.}  [WMAP Collaboration],  Astrophys.\ J.\ Suppl.\  {\bf 180}, 306 (2009).

\bibitem{WMAP55} M.~R.~Nolta {\it et al.}  [WMAP Collaboration],  Astrophys.\ J.\ Suppl.\  {\bf 180}, 296 (2009).

\bibitem{KS1} L.~Kofman and A.~A.~Starobinsky,  Sov.\ Astron.\ Lett.\  {\bf 11}, 271 (1985) [Pisma Astron.\ Zh.\  {\bf 11}, 643 (1985)].

\bibitem{KS2} D. Yu. Pogosian, and A. A. Starobinsky, Sov.\ Astron.\ Lett.\  {\bf 12}, 175 (1987) [Pisma Astron.\ Zh.\  {\bf 12}, 419 (1986)].

\bibitem{hugor} C.~Gordon and W.~Hu, Phys.\ Rev.\  D {\bf 70}, 083003 (2004).

\bibitem{elga}   T.~Multamaki and O.~Elgaroy,  Astron.\ Astrophys.\  {\bf 423}, 811 (2004).

\bibitem{karwan} K.~Karwan,  JCAP {\bf 0707}, 009 (2007).

\bibitem{BDVS} D.~Boyanovsky, H.~J.~de Vega and N.~G.~Sanchez,  Phys.\ Rev.\  D {\bf 74}, 123006 (2006); Phys.\ Rev.\  D {\bf 74}, 123007 (2006).
  
\bibitem{maxa1}  M.~Giovannini,  Phys.\ Rev.\  D {\bf 70}, 123507 (2004); Class.\ Quant.\ Grav.\  {\bf 23}, R1 (2006).

\bibitem{maxa2} M.~Giovannini, Phys.\ Rev.\  D {\bf 73}, 101302 (2006).
  
\bibitem{max1} M. Giovannini, Phys.\ Rev.\ D {\bf 79}, 121302(R) (2009). 

\bibitem{max2}  M. Giovannini,  Phys. Rev. D {\bf 79}, 103007 (2009).

\bibitem{max3} M.~Giovannini,  PMC Phys.\  A {\bf 1}, 5 (2007); 
M.~Giovannini and K.~E.~Kunze, Phys.\ Rev.\  D {\bf 77}, 061301 (2008); 
Phys.\ Rev.\  D {\bf 77}, 063003 (2008).

\bibitem{gal1} R. Beck, A. Brandenburg, D. Moss, A. Skhurov, and D. Sokoloff
 Annu. Rev. Astron. Astrophys. {\bf 34}, 155 (1996).

\bibitem{gal2}  T.~G.~Arshakian, R.~Beck, M.~Krause and D.~Sokoloff,  arXiv:0810.3114 [astro-ph].

\bibitem{cl1}   T.E. Clarke, P.P. Kronberg and H. B\"ohringer,  Astrophys. J.  {\bf 547}, L111 (2001).

\bibitem{cl2}  H.~Xu, H.~Li, D.~C.~Collins, S.~Li and M.~L.~Norman,  Astrophys.\ J.\  {\bf 698}, L14 (2009).

\bibitem{scl1}  J.~Lee, U.~L.~Pen, A.~R.~Taylor, J.~M.~Stil and C.~Sunstrum, arXiv:0906.1631 [astro-ph.CO].

\bibitem{q1} P.~P.~Kronberg, M.~L.~Bernet, F.~Miniati, S.~J.~Lilly, M.~B.~Short and D.~M.~Higdon,   arXiv:0712.0435 [astro-ph].

\bibitem{q2} M.~L.~Bernet, F.~Miniati, S.~J.~Lilly, P.~P.~Kronberg and M.~Dessauges-Zavadsky, arXiv:0807.3347 [astro-ph].

\bibitem{maxrev}  M.~Giovannini,  Int.\ J.\ Mod.\ Phys.\  D {\bf 13}, 391 (2004).

\bibitem{pb1} H. Alfv\'en and C.-G. F\"althammer, {\it Cosmical Electrodynamics}, 2nd edn., (Clarendon press, Oxford, 1963).

\bibitem{pb2}   L. Spitzer, {\it Physics of Fully ionized plasmas} (J. Wiley and Sons, New York, 1962).

\bibitem{max4}  M. Giovannini and N. Q. Lan, {\it Ohmic currents and pre-decoupling magnetism} CERN-PH-TH/2009-062.

\bibitem{bertschinger1} C.~P.~Ma and E.~Bertschinger,  Astrophys.\ J.\  {\bf 455}, 7 (1995).

\bibitem{PV1} W. Press and E. Vishniac, Astrophys. J. {\bf 239}, 1 (1980); Astrophys. J. {\bf 236}, 323 (1980).

\bibitem{bardeen}  J. M. Bardeen, Phys. Rev. D {\bf 22}, 1882 (1980).

\bibitem{hwang1}  J. Hwang, Astrophys. J.  {\bf 375}, 443 (1990).

\bibitem{hwang2} J.~Hwang and H.~Noh,   Class.\ Quant.\ Grav.\  {\bf 19}, 527 (2002).

\bibitem{bertschinger2} U.~Seljak and M.~Zaldarriaga, Astrophys.\ J.\  {\bf 469}, 437 (1996).

\bibitem{zalda1}  U.~Seljak and M.~Zaldarriaga, Astrophys.\ J.\  {\bf 469}, 437 (1996).

\bibitem{zalda2} M.~Zaldarriaga, D.~N.~Spergel and U.~Seljak,  Astrophys.\ J.\  {\bf 488}, 1 (1997).

\bibitem{EB} G.F.R. Ellis and M. Bruni, Phys. Rev. D{\bf 40}, 1804 (1989); 
M. Bruni, G. F. R.Ellis and P. K. S. Dunsby, Class. Quantum Grav. {\bf 9}, 921 (1992).

\bibitem{cov} C.~G.~Tsagas and J.~D.~Barrow,
Class.\ Quant.\ Grav.\  {\bf 14}, 2539 (1997); {\it ibid}. {\bf 15}, 3523 (1998).

\bibitem{stress1} T.~Koivisto and D.~F.~Mota, 
Phys.\ Rev.\  D {\bf 73}, 083502 (2006).

\bibitem{stress2} W.~Hu and D.~J.~Eisenstein,
  Phys.\ Rev.\  D {\bf 59}, 083509 (1999).
 
  \bibitem{zeld1} R.~A.~Sunyaev and Y.~B.~Zeldovich,  Astrophys.\ Space Sci.\  {\bf 7}, 3 (1970).

\bibitem{wyse}  B. Jones and R. Wyse, Astron. Astrophys. {\bf 149}, 144 (1985).

\bibitem{pav1} P. Naselsky and I. Novikov, Astrophys. J. {\bf 413}, 14 (1993).

\bibitem{pav2} H. Jorgensen, E. Kotok, P. Naselsky, and I Novikov, Astron. Astrophys. {\bf 294}, 639 (1995).

\bibitem{max5}   M.~Giovannini, Phys.\ Lett.\  B {\bf 659}, 661 (2008).
  
\bibitem{grad} I. S. Gradshteyn and I. M. Ryzhik {\it Table of Integrals, Series and Poducts}, (Academic Press, San Diego, 2000), sixth edition.
  
\bibitem{abr1}  M. Abramowitz and I. A. Stegun, {\it Handbook of Mathematical Functions} (Dover, New York, 1972).

\bibitem{rip1}  M.~Giovannini,  Phys.\ Rev.\  D {\bf 72}, 083508 (2005) ;  JCAP {\bf 0509}, 009 (2005).
 
\bibitem{far1}  M.~Giovannini, Phys.\ Rev.\  D {\bf 56}, 3198 (1997);  M.~Giovannini and K.~E.~Kunze,
Phys.\ Rev.\  D {\bf 79}, 063007 (2009).
 
\bibitem{far2} M.~Giovannini, Phys.\ Rev.\  D {\bf 71}, 021301(R) (2005); 
M.~Giovannini and K.~E.~Kunze,  Phys.\ Rev.\  D {\bf 79}, 087301 (2009).

\bibitem{max6} M.~Giovannini and K. Kunze, Phys.\ Rev.\  D {\bf 77}, 061301 (2008).

\bibitem{quad1}    P.Ade {\it et al.}  [QUaD collaboration], Astrophys. J. {\bf 674}, 22 (2008).

\bibitem{quad2}   C.~Pryke {\it et al.}  [QUaD collaboration],
  Astrophys.\ J.\  {\bf 692}, 1247 (2009).

\bibitem{quad3} E.~Y.~S.~Wu {\it et al.}  [QUaD Collaboration],  Phys.\ Rev.\ Lett.\  {\bf 102}, 161302 (2009).

\bibitem{bao1} D.~J.~Eisenstein {\it et al.}  [SDSS Collaboration], Astrophys.\ J.\  {\bf 633}, 560 (2005).

\bibitem{bao2}  M.~Tegmark {\it et al.}  [SDSS Collaboration], Astrophys.\ J.\  {\bf 606}, 702 (2004).

\bibitem{bao3} W.~J.~Percival, S.~Cole, D.~J.~Eisenstein, R.~C.~Nichol, J.~A.~Peacock, A.~C.~Pope and A.~S.~Szalay,
  Mon.\ Not.\ Roy.\ Astron.\ Soc.\  {\bf 381}, 1053 (2007).
  
\bibitem{sn1} P.~Astier {\it et al.}  [The SNLS Collaboration], Astron.\ Astrophys.\  {\bf 447}, 31 (2006).

\bibitem{sn2}  A.~G.~Riess {\it et al.}  [Supernova Search Team Collaboration],  Astrophys.\ J.\  {\bf 607}, 665 (2004).

\bibitem{sn3} B.~J.~Barris {\it et al.}, Astrophys.\ J.\  {\bf 602}, 571 (2004).


\end{thebibliography}
\end{document}